\documentclass[
  aps, a4paper, twocolumn, pra, superscriptaddress, amsmath, amssymb, floatfix
]{revtex4-1}
\usepackage[utf8]{inputenc}
\usepackage[T1]{fontenc}
\usepackage{subcaption}
\usepackage{physics}
\usepackage[framemethod=tikz]{mdframed}
\usepackage{algorithm}
\usepackage{algpseudocode}
\usepackage{hyperref}
\usepackage[justification=centerlast]{caption}

\algnewcommand{\LeftComment}[1]{\Statex \(\triangleright\) #1}
\newmdtheoremenv[
    nobreak=true,leftmargin=-.3cm, usetwoside=false,innermargin =0cm]{theo}{Definition}

\graphicspath{ {./figures/} }

\begin{document}
\title{Hierarchical quantum circuit representations for neural architecture search}
\author{Matt Lourens}
\email{lourensmattj@gmail.com}
\affiliation{Physics Department,
    Stellenbosch University, Stellenbosch, South Africa}
\author{Ilya Sinayskiy}
\affiliation{School of Chemistry and Physics, University of KwaZulu-Natal, Durban, South Africa}
\affiliation{National Institute for Theoretical and Computational Sciences (NITheCS), South Africa}
\author{Daniel K. Park}
\affiliation{Department of Statistics and Data Science, Yonsei University, Seoul, Korea}
\author{Carsten Blank}
\affiliation{Data Cybernetics, Landsberg, Germany}
\author{Francesco Petruccione}
\affiliation{National Institute for Theoretical and Computational Sciences (NITheCS), South Africa}
\affiliation{School of Data Science and Computational Thinking,
    Stellenbosch University, Stellenbosch, South Africa}
\affiliation{Physics Department,
    Stellenbosch University, Stellenbosch, South Africa}


\begin{abstract}
    Machine learning with hierarchical quantum circuits, usually referred to as Quantum Convolutional Neural Networks (QCNNs), is a promising prospect for near-term quantum computing. The QCNN is a circuit model inspired by the architecture of Convolutional Neural Networks (CNNs). CNNs are successful because they do not need manual feature design and can learn high-level features from raw data. Neural Architecture Search (NAS) builds on this success by learning network architecture and achieves state-of-the-art performance. However, applying NAS to QCNNs presents unique challenges due to the lack of a well-defined search space. In this work, we propose a novel framework for representing QCNN architectures using techniques from NAS, which enables search space design and architecture search. Using this framework, we generate a family of popular QCNNs, those resembling reverse binary trees. We then evaluate this family of models on a music genre classification dataset, GTZAN, to justify the importance of circuit architecture. Furthermore, we employ a genetic algorithm to perform Quantum Phase Recognition (QPR) as an example of architecture search with our representation. This work provides a way to improve model performance without increasing complexity and to jump around the cost landscape to avoid barren plateaus. Finally, we implement the framework as an open-source Python package to enable dynamic QCNN creation and facilitate QCNN search space design for NAS.
    
\end{abstract}

\flushbottom
\maketitle
\thispagestyle{empty}
\section*{Introduction}\label{sec:introduction}
\begin{figure*}[t]
    \includegraphics[width=\linewidth]{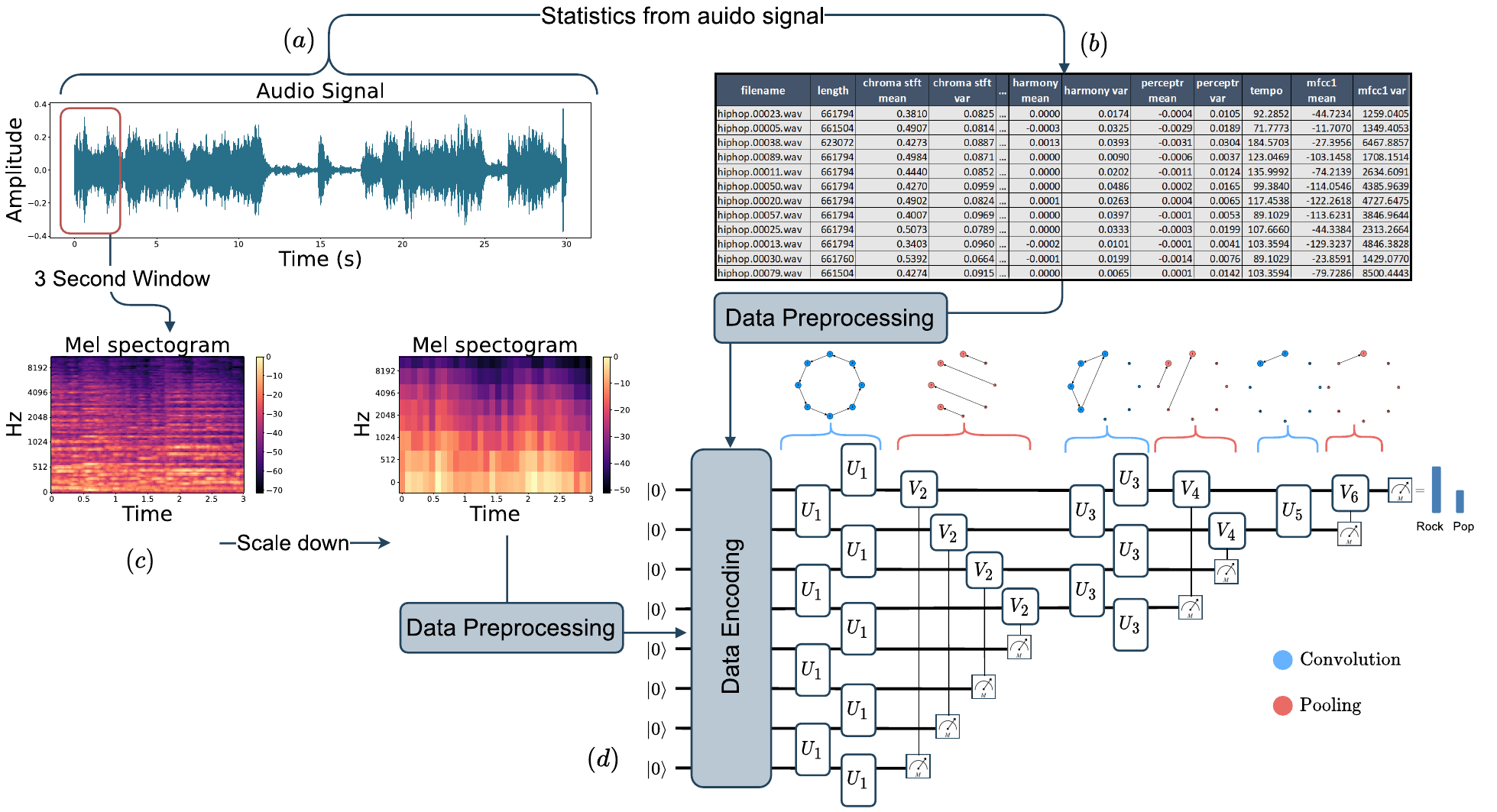}
    \caption{The machine learning pipeline we implemented for music genre classification. Given an audio signal of a song  (a), we generate two forms of data: tabular (b) and image (c). Each form has data preprocessing applied before being encoded into a quantum state (d). The QCNN circuit shown in (d) favours Principal Component Analysis (PCA) because qubits are pooled from bottom to top, and principal components are encoded from top to bottom. This architecture is an instance of the reverse binary tree family that we generated with our framework. }
    \label{fig:qcnn_pipeline}
\end{figure*}
\begin{figure*}[t]
    \includegraphics[width=\linewidth]{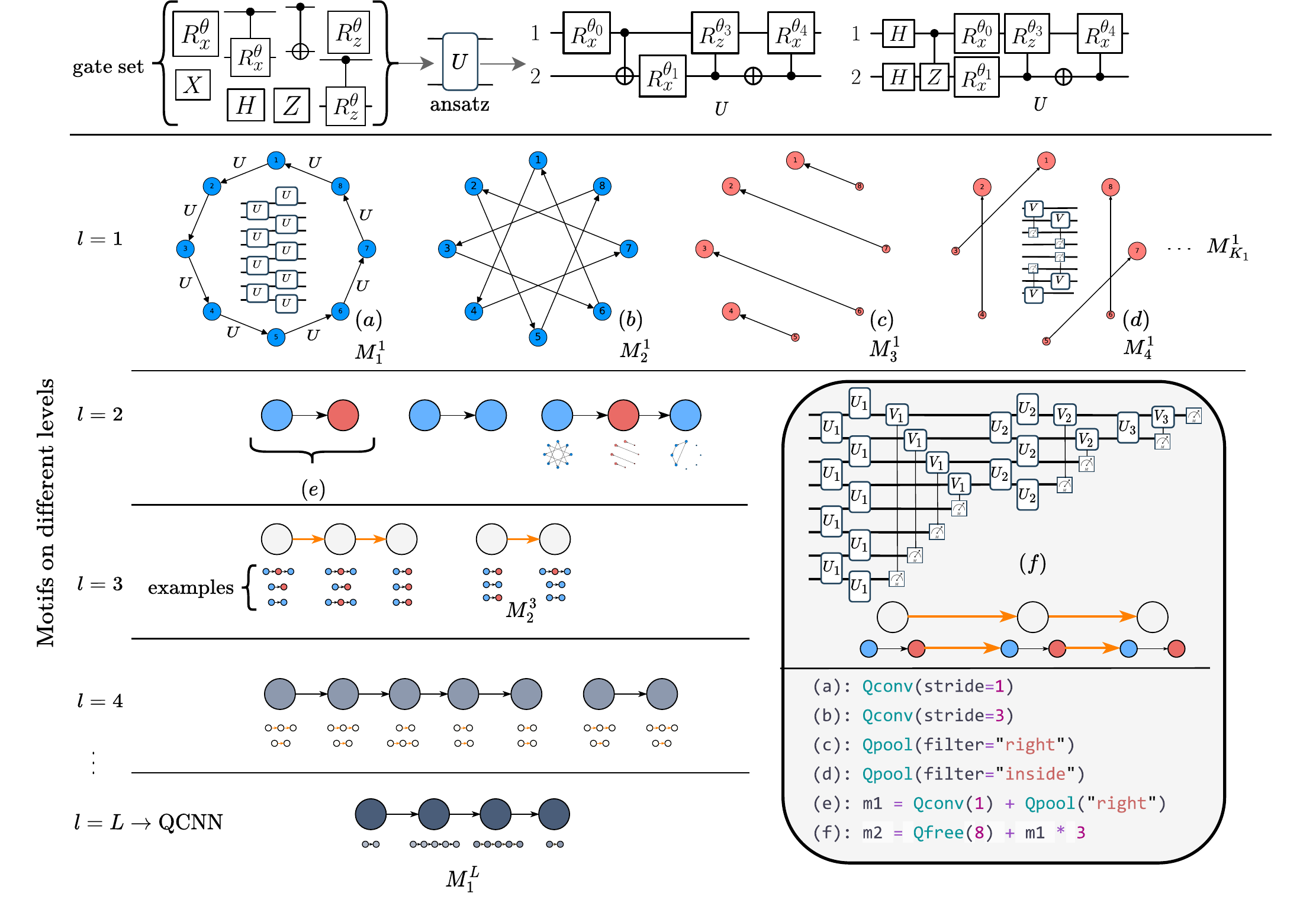}
    \caption{An overview of our architectural representation for QCNNs. From a given set of gates, we build two-qubit unitary ansatzes. The representation then captures design motifs $M^l_k$  on different levels $l$ of the hierarchy. On the lowest level $l=1$, we define primitives which act as building blocks for the architecture. For example, a convolution operation with stride one is encoded as the directed graph $M_1^1$. The directed graph $M^1_3$ is a pooling operation that measures the bottom half of the circuit. Combined, they form the level two motif (e): a convolution-pooling unit $M^2_1$. Higher-level motifs consist of combinations of lower-level motifs up until the final level $l=L$, which contains only one motif $M^L_1$, the complete QCNN architecture. $M^L_1$ is a hierarchy of directed graphs fully specifying how to spread the unitary ansatzes across the circuit. The two lines of code $(e)$ and $(f)$ show the power of this representation as it is all that is required to create the entire QCNN circuit from Figure \ref{fig:qcnn_pipeline} (d). The code comes from the Python package we implemented based on the work of this paper. It facilitates dynamic QCNN creation and search space design. }
    \label{fig:motifs}
\end{figure*}
Machine learning using trainable quantum circuits provides promising applications for quantum computing \cite{Benedetti_2019,cerezo2020variational,mangini_quantum_2021, RevModPhys.94.015004}. Among various parameterized quantum circuit (PQC) models, the Quantum Convolutional Neural Network (QCNN) introduced in Ref \cite{cong_quantum_2019} stands out for its shallow circuit depth, absence of barren plateaus \cite{pesah2020absence}, and good generalisation capabilities \cite{PRXQuantum.2.040321}. It has been implemented experimentally \cite{herrmannRealizingQuantumConvolutional2022} and combines techniques from Quantum Error Correction (QEC), Tensor Networks (TNs) and deep learning. Research at this intersection has been fruitful, yielding deep learning solutions for quantum many-body problems \cite{carleoSolvingQuantumManybody2017, carrasquillaMachineLearningPhases2017, vannieuwenburgLearningPhaseTransitions2017, dengMachineLearningTopological2017}, quantum-inspired insights for deep learning \cite{levineQuantumEntanglementDeep2019, stoudenmireSupervisedLearningTensor2016,dengQuantumEntanglementNeural2017} and equivalences between them \cite{linWhyDoesDeep2017, mehtaExactMappingVariational2014, levineDeepLearningQuantum2018}. Deep learning has been widely successful in recent years with applications spanning from content filtering and product recommendations to aided medical diagnosis and scientific research. Its main characteristic, learning features from raw data, eliminates the need for manual feature design by experts \cite{lecun_deep_2015}. AlexNet \cite{krizhevskyImageNetClassificationDeep2012} demonstrated this and marked the shift in focus from feature design to architecture design \cite{zophNeuralArchitectureSearch2017}. Naturally, the next step is learning network architecture, which Neural Architecture Search (NAS) aims to achieve \cite{elskenNeuralArchitectureSearch2019}. NAS has already produced state-of-the-art deep learning models with automatically designed architectures \cite{zophNeuralArchitectureSearch2017,realRegularizedEvolutionImage2019a, zophLearningTransferableArchitectures2018, chenSearchingEfficientMultiScale2018}. NAS consist of three main categories: search space, search strategy and performance estimation strategy \cite{elskenNeuralArchitectureSearch2019}. The search space defines the set of possible architectures that a search algorithm can consider, and carefully designed search spaces help improve search efficiency and reduce computational complexity \cite{liuHierarchicalRepresentationsEfficient2018a}. Search space design often involves encoding architectures using a cell-based representation. Usually, a set of primitive operations, such as convolutions or pooling, are combined into a cell to capture some design motif (compute graph). Different cells are then stacked to form a complete architecture. Cell-based representations are popular because they can capture repeated motifs and modular design patterns, which are often seen in successful hand-crafted architectures. Similar patterns also appear in quantum circuit designs  \cite{cong_quantum_2019,grant_hierarchical_2018,haugCapacityQuantumGeometry2021,hurQuantumConvolutionalNeural2022,ohTutorialQuantumConvolutional2020a,  frankenExplorationsQuantumNeural2020}. For example, Grant et al.  \cite{grant_hierarchical_2018} use hierarchical architectures based on tensor networks to classify classical and quantum data. Similarly,  Cong et al. \cite{cong_quantum_2019} use the multiscale entanglement renormalisation ansatz (MERA) as an instance of their proposed QCNN and discuss generalisations for quantum analogues of convolution and pooling operations. In this work, we formalise these design patterns by providing a hierarchical representation for QCNNs, thereby capturing their architecture in such a way to facilitate search space design for NAS with PQCs.
\newline
\newline
The QCNN belongs to the class of hybrid quantum-classical algorithms, in which a quantum computer executes the circuit, and a classical computer optimises its parameters. Two key factors must be considered when using PQCs for machine learning: the method of data encoding (feature map) \cite{mccleanBarrenPlateausQuantum2018a, holmesConnectingAnsatzExpressibility2022} and the choice of a quantum circuit \cite{schuldEffectDataEncoding2021, abbasPowerQuantumNeural2021, schuldSupervisedQuantumMachine2021}. Both the challenge and objective are to find a suitable quantum circuit for a given feature map that is expressive and trainable \cite{holmesConnectingAnsatzExpressibility2022}. The typical approach to finding a circuit is to keep the architecture (gates layout) fixed and to optimise continuous parameters such as rotation angles. Optimising architecture is referred to as variable structure ansatz in literature and is generally not the focus because of its computational complexity \cite{cerezo2020variational}. However, the architecture of a circuit can improve its expressive power and the effectiveness of initialisation techniques \cite{haugCapacityQuantumGeometry2021}. Also, the QCNN's defining characteristic is its architecture, which we found to impact model performance significantly. Therefore, we look towards NAS to optimise architecture in a quantum circuit setting. This approach, sometimes referred to as quantum architecture search (QAS)    \cite{zhangDifferentiableQuantumArchitecture2021,zhangNeuralPredictorBased2021}, has shown promising results for the variational quantum eigensolver (VQE) \cite{grimsleyAdaptiveVariationalAlgorithm2019,tangQubitADAPTVQEAdaptiveAlgorithm2021,yordanovQubitexcitationbasedAdaptiveVariational2021, rattewDomainagnosticNoiseresistantHardwareefficient2020}, the quantum approximate optimisation algorithm (QAOA) \cite{zhuAdaptiveQuantumApproximate2022a,liQuantumOptimizationNovel2020} and general architecture search \cite{ostaszewskiStructureOptimizationParameterized2021,zhangDifferentiableQuantumArchitecture2021,zhangNeuralPredictorBased2021,duQuantumCircuitArchitecture2022}. However, these approaches are often task-specific or impose additional constraints, such as circuit topology or allowable gates, to make them computationally feasible. To the best of the author's knowledge, there is currently no framework that can generate hierarchical architectures such as the QCNN without imposing such constraints.
\newline
\newline
One problem with the cell-based representation for NAS is that the macro architecture, the sequence of cells, is fixed and must be chosen \cite{elskenNeuralArchitectureSearch2019}. Recently, Liu et al. \cite{liuHierarchicalRepresentationsEfficient2018a} proposed a hierarchical representation as a solution, where a cell sequence acts as the third level of a multi-level hierarchy. In this representation, lower-level motifs act as building blocks for higher-level ones, allowing both macro and micro architecture to be learned. In this work, we  follow a similar approach and represent a QCNN architecture as a hierarchy of directed graphs. On the lowest level are primitive operations such as convolutions and pooling. The second level consists of sequences of these primitives, such as convolution-pooling or convolution-convolution units. Higher-level motifs then contain sequences of these lower-level motifs. For example, the third level could contain a sequence of three convolution-pooling units, as seen in Figure \ref{fig:qcnn_pipeline}d. For the primitives, we define hyperparameters such as strides and pooling filters that control their architectural effect. This way, the representation can capture design motifs on multiple levels, from the distribution of gates in a single layer to overall hierarchical patterns such as tensor tree networks. We demonstrate this by generating a family of QCNN architectures based on popular motifs in literature. We then benchmark this family of models and show that alternating architecture has a greater impact on model performance than other modelling components. By alternating architecture we mean the following: given a quantum circuit that consist of $n$ unitary gates, an altered architecture consists of the same $n$ gates rearranged in a different way on the circuit. The types of rearrangements may be changing which qubits the gates act upon, altering the order of gate occurrences, or adjusting larger architectural motifs, such as pooling specific qubits (stop using them) while leaving others available for subsequent gates and so on. We create architectural families to show the impact of alternating architecture, any two instances of the family will have the exact same unitaries, just applied in a different order on different qubits. Consider the machine learning pipeline for classifying musical genres from audio signals, seen in Figure \ref{fig:qcnn_pipeline}. We start with a 30-second recording of a song (Figure \ref{fig:qcnn_pipeline}a) and transform it in two ways. The first is tabular form (Figure \ref{fig:qcnn_pipeline}b), derived from standard digital signal processing statistics of the audio signal. The second is image form (Figure \ref{fig:qcnn_pipeline}c), constructed using a Mel frequency spectrogram. Both datasets are benchmarked separately, with their own data preprocessing and encoding techniques applied. For the tabular data, we test Principal Component Analysis (PCA) and tree-based feature selection before encoding it in a quantum state using either qubit, IQP, or amplitude encoding. Once encoded, we choose two-qubit unitary ansatzes $U_m$ and $V_m$ for the convolution and pooling primitives $m=1,2,\dots,6$, as shown in Figure \ref{fig:qcnn_pipeline}d. We show example ansatzes in Appendix \ref{appendix:circuits} and test them across different instances of an architecture family. Of all the components in this pipeline, alternating architecture, that is changing how each $U_m$ and each $V_m$ are spread across the circuit, had the greatest impact on model performance. In addition to our theoretical framework, we implement it as an open-source Python package to enable dynamic QCNN creation and facilitate search space design for NAS. It allows users to experimentally determine suitable architectures for specific modelling setups, such as finding circuits that perform well under a specific noise or hardware configuration, which is particularly relevant in the Noisy Intermediate-Scale Quantum (NISQ) \cite{Preskill2018quantumcomputingin} era. Additionally, as more qubits become available, the hierarchical nature of our framework provides a natural way to scale up the same model. In summary, our contributions are the architectural representation for QCNNs, a Python package for dynamic QCNN creation, and experimental results on the potential advantage of architecture search in a quantum setting.
\newline
\newline
The remainder of this paper is structured as follows: we begin with our main results by summarising the architectural representation for QCNNs and then show the effect of alternating architecture, justifying its importance. We then provide an example of architecture search with our representation by employing an evolutionary algorithm to perform QPR. Following this, we give details of our framework by providing a mathematical formalism for the representation and describing its use. Next, with the formalism at hand, we show how it facilitates search space design by describing the space we created for the benchmark experiments. We then discuss generalisations of the formalism and the applicability of our representation with search algorithms. After this we elaborate on our experimental setup in the Methods Section. Finally, we discuss applications and future steps.
\section*{Results}
\subsection*{Architectural Representation}
\label{sec:arc_rep}

Figure \ref{fig:motifs} shows our architectural representation for QCNNs. We define two-qubit unitary ansatzes from a given set of gates, and capture design motifs $M^l_k$ on different levels $l$ of the hierarchy. On the lowest level $l=1$, we define primitives which act as building blocks for the architecture. For example, a convolution operation with stride one is encoded as the directed graph $M_1^1$, and with stride three as $M^1_2$. The directed graph $M^1_3$ is a pooling operation that measures the bottom half of the circuit, and $M^1_4$ measures from the inside outwards. Combined, they can form higher-level motifs such as convolution-pooling units $M^2_1$ (e), convolution-convolution units $M^2_2$, or convolution-pooling-convolution units $M^2_3$. The highest level $l=L$ contains only one motif $M^L_1$, the complete QCNN architecture. $M^L_1$ is a hierarchy of directed graphs fully specifying how to spread the unitary ansatzes across the circuit. This hierarchical representation is based on the one from Liu et al. \cite{liuHierarchicalRepresentationsEfficient2018a} for deep neural networks (DNNs), and allows for the capture of modularised design patterns and repeated motifs.  The two lines of code $(e)$ and $(f)$ show the power of this representation as it is all that is required to create the entire QCNN circuit from Figure \ref{fig:qcnn_pipeline} (d). The code comes from the Python package we implemented based on the work of this paper. It facilitates dynamic QCNN creation and search space design.

\subsection*{Architectural impact}\label{sec:results}
The details regarding specific notation and representation of the framework is given after this section, first we justify it with the following experimental results. In Appendix \ref{sec:background} we also give background on QCNNs and quantum machine learning for more context. 
To illustrate the impact of architecture on model performance, we compare the fixed architecture from the experiments of Hur et al. \cite{hurQuantumConvolutionalNeural2022} to other architectures in the same family while keeping all other components the same. The only difference in each comparison is architecture (how the unitaries are spread across the circuit). The architecture in \cite{hurQuantumConvolutionalNeural2022} is represented within our framework as: $(s_c,F^*,s_p)=(1,\text{even},0)\mapsto \text{Qfree}(8) + (\text{Qconv}(1)+\text{Qpool}(0,F^{even}))\times3$, see algorithm \ref{alg:reverse_binary_tree}. To evaluate their performance, we use the country vs rock genre pair, which proved to be one of the most difficult classification tasks from the $45$ possible combinations. We compare eight unitary ansatzes with different levels of complexity, as shown in Figure \ref{fig:ansatzes_appendix}.
\begin{table}[h]
    \centering
    \begin{tabular}{|c c | c c |c c |}
        \hline
        \multicolumn{6}{|c|}{Architecture vs Ansatz}                                                                                                                \\
        \hline
        \multicolumn{2}{|c|}{Ansatz,}   & \multicolumn{2}{c|}{Architecture} &                 & Alteration                                                          \\
        \multicolumn{2}{|c|}{\# Params} & Reference                         & New alteration  & $\Delta$                 & ($s_c,F^*,s_p$)                          \\
        \hline
        \ref{fig:u_ttn},                & $6$                               & $65.37\pm{2.8}$ & $\textbf{75.14}\pm{1.7}$ & $+9.77$         & $(6,\text{left},2)$    \\
        \ref{fig:u_9},                  & $6$                               & $56.34\pm{3.2}$ & $\textbf{70.46}\pm{1.0}$ & $+14.12$        & $(1,\text{odd},3)$     \\
        \ref{fig:u_15},                 & $12$                              & $52.69\pm{3.8}$ & $\textbf{70.74}\pm{1.3}$ & $+18.05$        & $(1,\text{odd},0)$     \\
        \ref{fig:u_so4},                & $18$                              & $67.13\pm{1.5}$ & $\textbf{77.87}\pm{2.4}$ & $+9.87$         & $(1,\text{outside},2)$ \\
        \ref{fig:u_13},                 & $18$                              & $67.87\pm{2.5}$ & $\textbf{73.61}\pm{1.8}$ & $+5.74$         & $(6,\text{left},0)$    \\
        \ref{fig:u_14},                 & $18$                              & $69.21\pm{2.6}$ & $\textbf{74.80}\pm{2.8}$ & $+5.59$         & $(1,\text{left},3)$    \\
        \ref{fig:u5},                   & $30$                              & $73.24\pm{2.9}$ & $\textbf{79.47}\pm{2.2}$ & $+6.23$         & $(2,\text{left},1)$    \\
        \ref{fig:u_6},                  & $30$                              & $69.35\pm{4.1}$ & $\textbf{71.71}\pm{3.7}$ & $+2.36$         & $(2,\text{left},1)$    \\
        \hline
    \end{tabular}
    \caption{The average accuracy and standard deviation of the country vs rock genre pair on a held-out test set after 30 separate trained instances. All architectures come from the family of reverse binary trees, generated with algorithm \ref{alg:reverse_binary_tree}. The "reference" architecture is the one used in the experiments of  Ref \cite{hurQuantumConvolutionalNeural2022} and the "alteration" was found through random search within the same family. The unitary
        ansatzes also come from Ref \cite{hurQuantumConvolutionalNeural2022}, which is based on previous studies that benchmarked PQCs \cite{Sim_expressibility,PhysRevLett.122.230401,grant_hierarchical_2018}.}
    \label{tab:arc_anz}
\end{table}
\newline
Table \ref{tab:arc_anz} shows the results of the comparisons, the reference architecture is as described above and the discovered alteration found via random search. We note the first important result, we improved the performance of every ansatz, in one case, by $18.05\%$, through random search of the architecture space. Ansatz refers to the two-qubit unitary used for the convolution operation of a model. For example, the model in figure \ref{fig:qcnn_pipeline} (d) is described by $(1,\text{right},0)$ and ansatz \ref{fig:u_ttn}  corresponds to $U_1, U_2 \text{ and } U_3$ being circuit \ref{fig:u_ttn} from Appendix \ref{appendix:circuits}. Each value represents the average model accuracy and standard deviation from 30 separate trained instances on the same held-out test set.
\newline
The second important result is that alternating architecture can improve model performance without increasing complexity. For instance, the best-performing model for the reference architecture is with ansatz \ref{fig:u5}, which has an average accuracy of $73.24\%$. However, this ansatz causes the model to have $10\times3=30$ parameters. In contrast, by alternating the architecture with the simplest ansatz \Ref {fig:u_ttn}, the model outperformed the best reference model with an average accuracy of $75.14\%$ while only having $3\times2=6$ parameters. The parameter counts come from each model having $N=8$ qubits and the same number of unitaries, $3N-2 \rightarrow 3(8)-2=22$, of which 13 are for convolutions. See the search space design section and Algorithm \ref{alg:reverse_binary_tree} for more details. A model has three convolutions, and each convolution shares weights between its two-qubit unitaries. This means that the two-qubit unitary ansatz primarily determines the number of parameters to optimise for a model. For example, a model with ansatz \ref{fig:u_ttn} have $2\times3=6$ parameters to optimise because ansatz \ref{fig:u_ttn} has two parameters.
\newline
Another interesting result is for ansatz \ref{fig:u_15}, the reference architecture could only obtain an average accuracy of $52.69 \%$ indicating its inability to find any kind of local minimum during training, leading one to think it might be a barren plateu. But, the altered architecture was able to find a local minima and improve the average accuracy by $18.05 \%$.\newline\newline

\begin{table}[h!]
    \centering
    \begin{tabular}{|ccccccccc|}
        \hline
        \multicolumn{9}{|c|}{Performance across architecture search space}                                                                                                                                                                                                                                                              \\ \hline
        \multicolumn{1}{|c|}{}                 & \multicolumn{8}{c|}{Convolution stride, $s_c$}                                                                                                                                                                                                                                         \\ \hline
        \multicolumn{1}{|c|}{$F^*,s_p$}        & \multicolumn{1}{c|}{1}                         & \multicolumn{1}{c|}{2}       & \multicolumn{1}{c|}{3}       & \multicolumn{1}{c|}{4}       & \multicolumn{1}{c|}{5}       & \multicolumn{1}{c|}{6}                & \multicolumn{1}{c|}{7}                         & \textbf{Avg}     \\ \hline
        \multicolumn{1}{|c|}{\textbf{even}}    & \textbf{$\textbf{67.01}$}                      & \textbf{$63.63$}             & \textbf{$60.76$}             & \textbf{$64.93$}             & \textbf{$59.98$}             & \textbf{$63.1$}                       & \multicolumn{1}{c|}{\textbf{$59.49$}}          & \textbf{$62.81$} \\ \hline
        \multicolumn{1}{|c|}{0}                & \multicolumn{1}{c|}{$65.97$}                   & \multicolumn{1}{c|}{$58.68$} & \multicolumn{1}{c|}{$56.25$} & \multicolumn{1}{c|}{$66.67$} & \multicolumn{1}{c|}{$62.85$} & \multicolumn{1}{c|}{$59.72$}          & \multicolumn{1}{c|}{$63.43$}                   & $61.88$          \\
        \multicolumn{1}{|c|}{1}                & \multicolumn{1}{c|}{$66.32$}                   & \multicolumn{1}{c|}{$66.32$} & \multicolumn{1}{c|}{$63.54$} & \multicolumn{1}{c|}{$60.07$} & \multicolumn{1}{c|}{$61.46$} & \multicolumn{1}{c|}{$71.88$}          & \multicolumn{1}{c|}{$54.17$}                   & $63.73$          \\
        \multicolumn{1}{|c|}{2}                & \multicolumn{1}{c|}{$66.67$}                   & \multicolumn{1}{c|}{$60.76$} & \multicolumn{1}{c|}{$60.07$} & \multicolumn{1}{c|}{$68.06$} & \multicolumn{1}{c|}{$54.17$} & \multicolumn{1}{c|}{$58.8$}           & \multicolumn{1}{c|}{$63.89$}                   & $61.81$          \\
        \multicolumn{1}{|c|}{3}                & \multicolumn{1}{c|}{$69.1$}                    & \multicolumn{1}{c|}{$68.75$} & \multicolumn{1}{c|}{$63.19$} & \multicolumn{1}{c|}{$64.93$} & \multicolumn{1}{c|}{$61.46$} & \multicolumn{1}{c|}{$60.19$}          & \multicolumn{1}{c|}{$56.48$}                   & $63.84$          \\ \hline
        \multicolumn{1}{|c|}{\textbf{inside}}  & \textbf{$66.41$}                               & \textbf{$\textbf{71.96}$}    & \textbf{$58.25$}             & \textbf{$54.25$}             & \textbf{$69.27$}             & \textbf{$68.15$}                      & \multicolumn{1}{c|}{\textbf{$60.53$}}          & \textbf{$64.18$} \\ \hline
        \multicolumn{1}{|c|}{0}                & \multicolumn{1}{c|}{$65.28$}                   & \multicolumn{1}{c|}{$72.22$} & \multicolumn{1}{c|}{$60.07$} & \multicolumn{1}{c|}{$49.65$} & \multicolumn{1}{c|}{$70.49$} & \multicolumn{1}{c|}{$68.4$}           & \multicolumn{1}{c|}{$60.65$}                   & $63.94$          \\
        \multicolumn{1}{|c|}{1}                & \multicolumn{1}{c|}{$67.01$}                   & \multicolumn{1}{c|}{$71.18$} & \multicolumn{1}{c|}{$58.68$} & \multicolumn{1}{c|}{$55.9$}  & \multicolumn{1}{c|}{$66.32$} & \multicolumn{1}{c|}{$68.4$}           & \multicolumn{1}{c|}{$60.19$}                   & $64.09$          \\
        \multicolumn{1}{|c|}{2}                & \multicolumn{1}{c|}{$68.4$}                    & \multicolumn{1}{c|}{$71.53$} & \multicolumn{1}{c|}{$58.33$} & \multicolumn{1}{c|}{$51.74$} & \multicolumn{1}{c|}{$71.88$} & \multicolumn{1}{c|}{$68.98$}          & \multicolumn{1}{c|}{$58.8$}                    & $64.26$          \\
        \multicolumn{1}{|c|}{3}                & \multicolumn{1}{c|}{$64.93$}                   & \multicolumn{1}{c|}{$72.92$} & \multicolumn{1}{c|}{$55.9$}  & \multicolumn{1}{c|}{$59.72$} & \multicolumn{1}{c|}{$68.4$}  & \multicolumn{1}{c|}{$66.67$}          & \multicolumn{1}{c|}{$62.5$}                    & $64.42$          \\ \hline
        \multicolumn{1}{|c|}{\textbf{left}}    & \textbf{$62.85$}                               & \textbf{$61.63$}             & \textbf{$59.38$}             & \textbf{$59.03$}             & \textbf{$51.56$}             & \textbf{$\textbf{72.52}$}             & \multicolumn{1}{c|}{\textbf{$72.45$}}          & \textbf{$62.22$} \\ \hline
        \multicolumn{1}{|c|}{0}                & \multicolumn{1}{c|}{$66.67$}                   & \multicolumn{1}{c|}{$67.01$} & \multicolumn{1}{c|}{$56.94$} & \multicolumn{1}{c|}{$61.46$} & \multicolumn{1}{c|}{$52.08$} & \multicolumn{1}{c|}{$71.18$}          & \multicolumn{1}{c|}{$73.61$}                   & $63.79$          \\
        \multicolumn{1}{|c|}{1}                & \multicolumn{1}{c|}{$59.03$}                   & \multicolumn{1}{c|}{$62.15$} & \multicolumn{1}{c|}{$52.78$} & \multicolumn{1}{c|}{$57.99$} & \multicolumn{1}{c|}{$52.08$} & \multicolumn{1}{c|}{$71.18$}          & \multicolumn{1}{c|}{$73.61$}                   & $60.8$           \\
        \multicolumn{1}{|c|}{2}                & \multicolumn{1}{c|}{$63.19$}                   & \multicolumn{1}{c|}{$63.19$} & \multicolumn{1}{c|}{$63.19$} & \multicolumn{1}{c|}{$60.76$} & \multicolumn{1}{c|}{$51.74$} & \multicolumn{1}{c|}{$\textbf{75.93}$} & \multicolumn{1}{c|}{$71.76$}                   & $63.51$          \\
        \multicolumn{1}{|c|}{3}                & \multicolumn{1}{c|}{$62.5$}                    & \multicolumn{1}{c|}{$54.17$} & \multicolumn{1}{c|}{$64.58$} & \multicolumn{1}{c|}{$55.9$}  & \multicolumn{1}{c|}{$50.35$} & \multicolumn{1}{c|}{$72.69$}          & \multicolumn{1}{c|}{$70.83$}                   & $60.79$          \\ \hline
        \multicolumn{1}{|c|}{\textbf{odd}}     & \textbf{$61.11$}                               & \textbf{$\textbf{68.75}$}    & \textbf{$63.37$}             & \textbf{$62.76$}             & \textbf{$64.67$}             & \textbf{$60.52$}                      & \multicolumn{1}{c|}{\textbf{$57.99$}}          & \textbf{$62.96$} \\ \hline
        \multicolumn{1}{|c|}{0}                & \multicolumn{1}{c|}{$60.76$}                   & \multicolumn{1}{c|}{$71.88$} & \multicolumn{1}{c|}{$63.19$} & \multicolumn{1}{c|}{$58.33$} & \multicolumn{1}{c|}{$63.54$} & \multicolumn{1}{c|}{$59.38$}          & \multicolumn{1}{c|}{$57.87$}                   & $62.29$          \\
        \multicolumn{1}{|c|}{1}                & \multicolumn{1}{c|}{$63.54$}                   & \multicolumn{1}{c|}{$67.36$} & \multicolumn{1}{c|}{$64.58$} & \multicolumn{1}{c|}{$63.54$} & \multicolumn{1}{c|}{$64.24$} & \multicolumn{1}{c|}{$62.5$}           & \multicolumn{1}{c|}{$59.26$}                   & $63.73$          \\
        \multicolumn{1}{|c|}{2}                & \multicolumn{1}{c|}{$60.42$}                   & \multicolumn{1}{c|}{$70.14$} & \multicolumn{1}{c|}{$64.58$} & \multicolumn{1}{c|}{$65.97$} & \multicolumn{1}{c|}{$69.1$}  & \multicolumn{1}{c|}{$58.8$}           & \multicolumn{1}{c|}{$56.94$}                   & $64.16$          \\
        \multicolumn{1}{|c|}{3}                & \multicolumn{1}{c|}{$59.72$}                   & \multicolumn{1}{c|}{$65.62$} & \multicolumn{1}{c|}{$61.11$} & \multicolumn{1}{c|}{$63.19$} & \multicolumn{1}{c|}{$61.81$} & \multicolumn{1}{c|}{$61.11$}          & \multicolumn{1}{c|}{$57.87$}                   & $61.65$          \\ \hline
        \multicolumn{1}{|c|}{\textbf{outside}} & \textbf{$60.68$}                               & \textbf{$65.8$}              & \textbf{$65.54$}             & \textbf{$57.12$}             & \textbf{$62.15$}             & \textbf{$59.83$}                      & \multicolumn{1}{c|}{\textbf{$\textbf{67.13}$}} & \textbf{$62.51$} \\ \hline
        \multicolumn{1}{|c|}{0}                & \multicolumn{1}{c|}{$67.36$}                   & \multicolumn{1}{c|}{$59.72$} & \multicolumn{1}{c|}{$71.88$} & \multicolumn{1}{c|}{$54.17$} & \multicolumn{1}{c|}{$67.01$} & \multicolumn{1}{c|}{$60.07$}          & \multicolumn{1}{c|}{$70.37$}                   & $64.15$          \\
        \multicolumn{1}{|c|}{1}                & \multicolumn{1}{c|}{$53.47$}                   & \multicolumn{1}{c|}{$69.79$} & \multicolumn{1}{c|}{$62.15$} & \multicolumn{1}{c|}{$56.25$} & \multicolumn{1}{c|}{$61.11$} & \multicolumn{1}{c|}{$58.33$}          & \multicolumn{1}{c|}{$70.83$}                   & $61.49$          \\
        \multicolumn{1}{|c|}{2}                & \multicolumn{1}{c|}{$57.99$}                   & \multicolumn{1}{c|}{$70.83$} & \multicolumn{1}{c|}{$60.07$} & \multicolumn{1}{c|}{$61.11$} & \multicolumn{1}{c|}{$59.03$} & \multicolumn{1}{c|}{$59.26$}          & \multicolumn{1}{c|}{$66.67$}                   & $62.07$          \\
        \multicolumn{1}{|c|}{3}                & \multicolumn{1}{c|}{$63.89$}                   & \multicolumn{1}{c|}{$62.85$} & \multicolumn{1}{c|}{$68.06$} & \multicolumn{1}{c|}{$56.94$} & \multicolumn{1}{c|}{$61.46$} & \multicolumn{1}{c|}{$61.57$}          & \multicolumn{1}{c|}{$60.65$}                   & $62.29$          \\ \hline
        \multicolumn{1}{|c|}{\textbf{right}}   & \textbf{$\textbf{70.05}$}                      & \textbf{$65.63$}             & \textbf{$64.41$}             & \textbf{$53.65$}             & \textbf{$68.66$}             & \textbf{$63.69$}                      & \multicolumn{1}{c|}{\textbf{$60.65$}}          & \textbf{$63.94$} \\ \hline
        \multicolumn{1}{|c|}{0}                & \multicolumn{1}{c|}{$70.14$}                   & \multicolumn{1}{c|}{$63.54$} & \multicolumn{1}{c|}{$64.58$} & \multicolumn{1}{c|}{$50$}    & \multicolumn{1}{c|}{$68.4$}  & \multicolumn{1}{c|}{$61.11$}          & \multicolumn{1}{c|}{$62.96$}                   & $62.96$          \\
        \multicolumn{1}{|c|}{1}                & \multicolumn{1}{c|}{$69.79$}                   & \multicolumn{1}{c|}{$67.71$} & \multicolumn{1}{c|}{$64.58$} & \multicolumn{1}{c|}{$69.1$}  & \multicolumn{1}{c|}{$68.06$} & \multicolumn{1}{c|}{$67.01$}          & \multicolumn{1}{c|}{$57.87$}                   & $66.62$          \\
        \multicolumn{1}{|c|}{2}                & \multicolumn{1}{c|}{$70.14$}                   & \multicolumn{1}{c|}{$62.15$} & \multicolumn{1}{c|}{$63.89$} & \multicolumn{1}{c|}{$43.75$} & \multicolumn{1}{c|}{$68.75$} & \multicolumn{1}{c|}{$62.04$}          & \multicolumn{1}{c|}{$61.57$}                   & $61.75$          \\
        \multicolumn{1}{|c|}{3}                & \multicolumn{1}{c|}{$70.14$}                   & \multicolumn{1}{c|}{$69.1$}  & \multicolumn{1}{c|}{$64.58$} & \multicolumn{1}{c|}{$51.74$} & \multicolumn{1}{c|}{$69.44$} & \multicolumn{1}{c|}{$64.35$}          & \multicolumn{1}{c|}{$60.19$}                   & $64.37$          \\ \hline
        \multicolumn{1}{|c|}{\textbf{Avg}}     & \textbf{$64.68$}                               & \textbf{$66.23$}             & \textbf{$61.95$}             & \textbf{$58.62$}             & \textbf{$62.72$}             & \textbf{$64.69$}                      & \multicolumn{1}{c|}{\textbf{$63.04$}}          & \textbf{$63.11$} \\ \hline
    \end{tabular}
    \caption{Country vs Rock average accuracy within the reverse binary tree search space, all with \ref{fig:u_ttn} as ansatz. The convolution stride $s_c$ is shown on the horizontal axis and the combinations of pooling filter $F^{*}$ and stride $s_p$  on the vertical. The best pooling filter and convolution stride combinations are presented in bold along with the overall best architecture $(s_c,F^*,s_p)=(6,\text{left},2)$. }
    \label{tab:space}
\end{table}
We would like to note that our primary objective in these experiments is to demonstrate the potential for performance improvement. As such, we only conducted random search for approximately 2 hours on an i7-1165G7 processor for each ansatz. Consequently, for higher parameter ansatzes, which correspond to longer training times, the search space was less explored. This is likely the reason behind the observed decrease in performance improvement for larger parameter ansatzes. Therefore the observed improvements are all lower bounds for the potential performance increase from alternating architecture. We anticipate that significantly better architectures may still exist within the space. Table \ref{tab:space} presents the performance of the family of reverse binary trees (as described in Algorithm \ref{alg:reverse_binary_tree}) for ansatz \ref{fig:u_ttn}. Due to its quick training time, ansatz \ref{fig:u_ttn} was the only case for which we managed to exhaust the search space (168 architectures).  In the search space design section, we discuss how the size of the family can be easily increased or decreased. Each value represents the average accuracy of five trained instances on the country vs rock genre pair. The overall accuracy of the whole space is $63.11 \%$, indicating that the reference architecture from table \ref{tab:arc_anz} was close to the mean performance. The best-performing architecture in this space is $(s_c,F^{*},s_p)=(6,\text{left},2)$, with an average accuracy of $75.93\%$. This is the alteration from Table \ref{tab:arc_anz} discovered through random search within the family of reverse binary trees. It seems that the combination of $F^{\text{left}}$ and $s_c=6$ performs particularly well for this task, with an average accuracy of $72.52 \%$. In general, it appears that the convolution stride $s_c$ and pooling filter $F^{*}$ have the most significant impact on performance.  It is also worth noting that convolution strides of $s_c = 3,4,5$ performed poorly compared to the other values. The range of performance in this space goes from a minimum of $43.75\%$ to a maximum of $75.93\%$, demonstrating the potential impact of architectural choices on model performance.
\newline\newline
\begin{figure}[t]
    \centering
    \includegraphics[width=\linewidth]{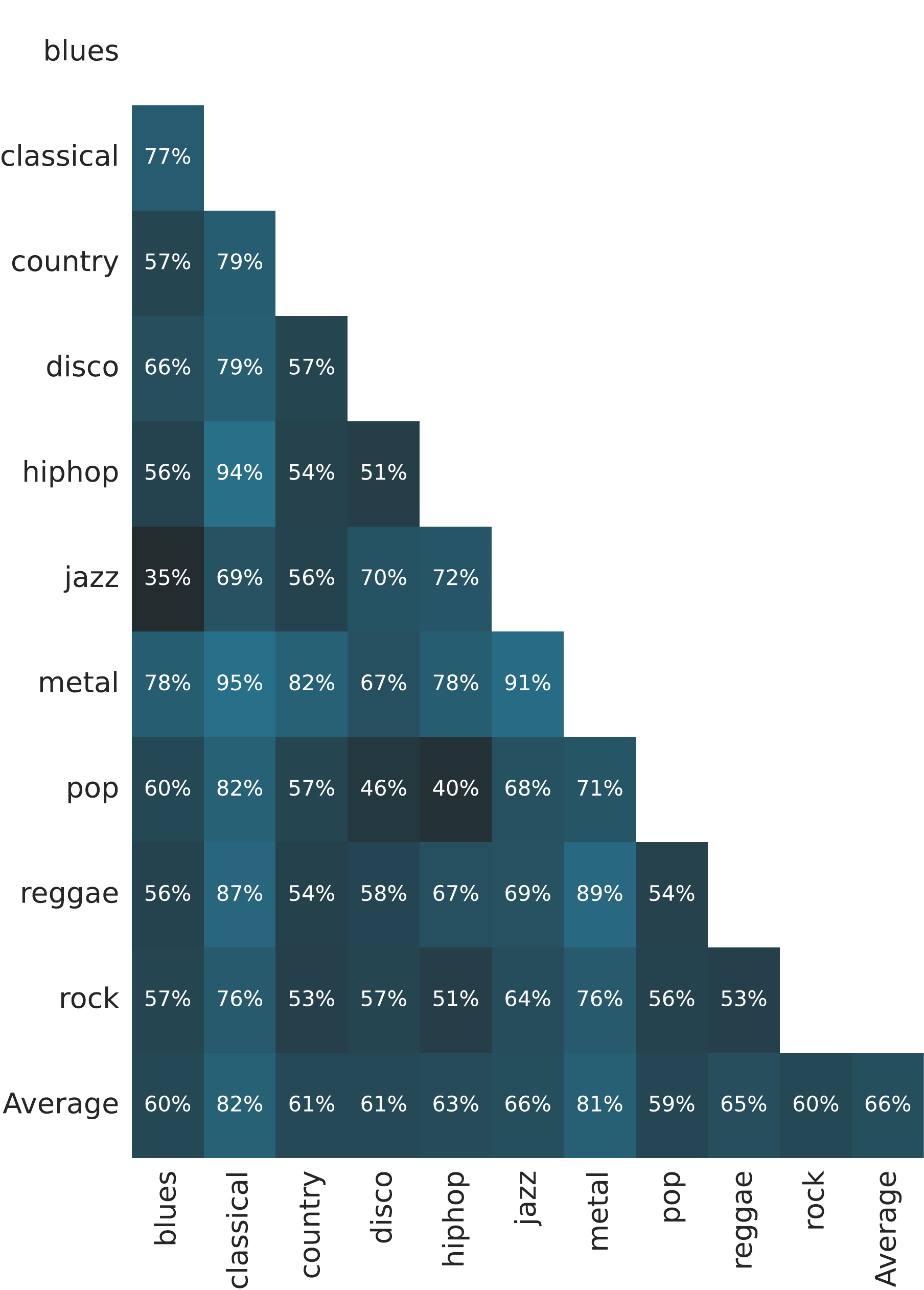}
    \caption{QCNN with the $F_m^{\rm{right}}$ pooling filter using low resolution image data. The accuracies for all genre pairs are provided.}
    \label{fig:quantum_image_right}
\end{figure}
\begin{figure}[t]
    \centering
    \includegraphics[width=\linewidth]{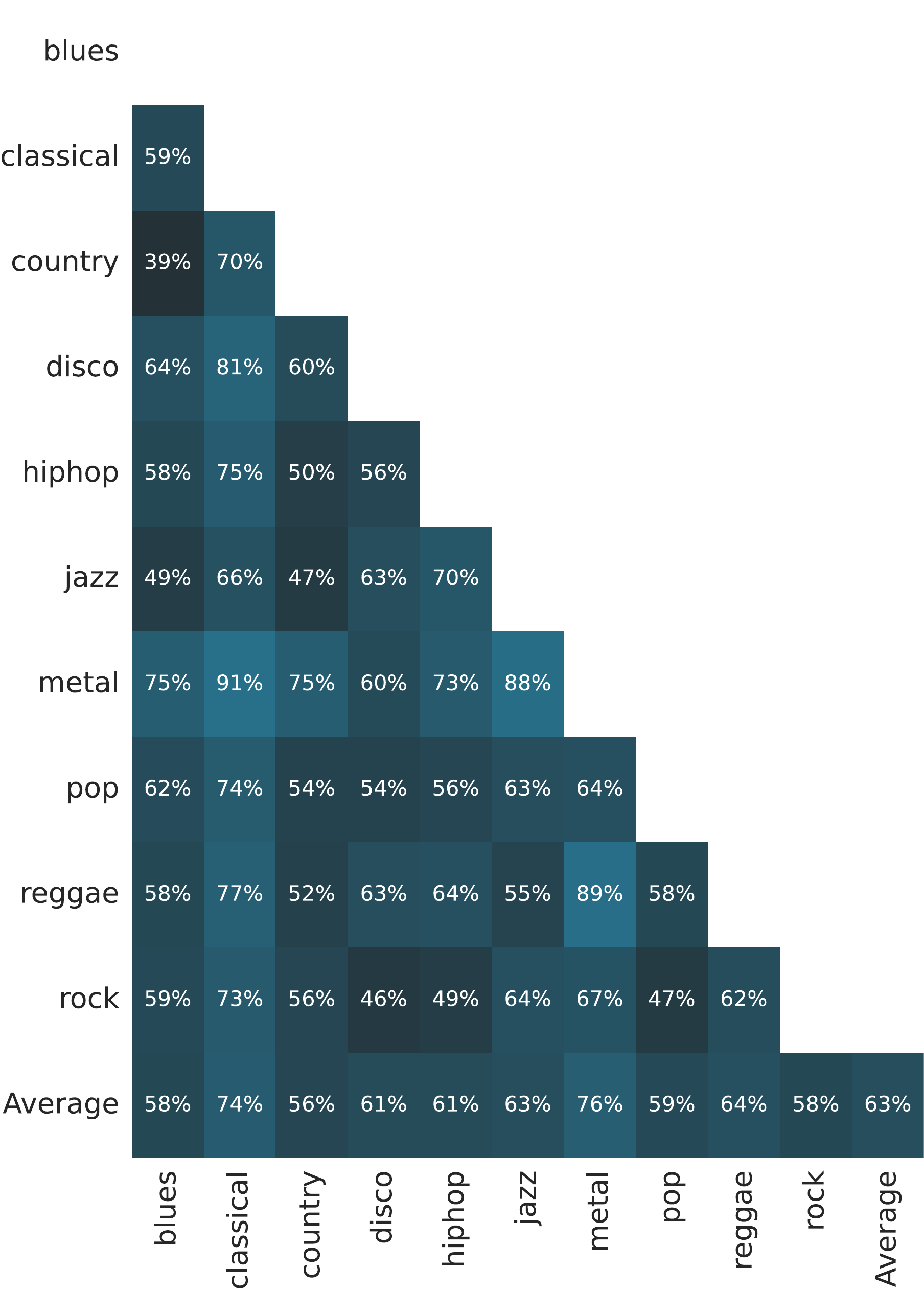}
    \caption{QCNN with the $F_m^{\rm{even}}$ pooling filter using low resolution image data. The accuracies for all genre pairs are provided. }
    \label{fig:quantum_image_even}
\end{figure}
Finally, we compared the performance of two different architectures on the image data across all genres. This time, we used ansatz \ref{fig:u5} to compare the $F^{\rm{right}}_m$ and $F^{\rm{even}}_m$ pooling filters, shown in Figures \ref{fig:quantum_image_right} and \ref{fig:quantum_image_even}. The image data is a low-resolution ($8\times32=256=2^8$ pixels) spectrogram of the audio signal. We did not expect high accuracy from this data, but were interested in the variation of performance for different architectures. Figures \ref{fig:quantum_image_right} and \ref{fig:quantum_image_even} show the difficulty of some genre pairs. Interestingly, the $F^{\rm{right}}_m$ pooling filter outperformed the $F^{\rm{even}}_m$ filter on almost all genres. If we focus on the genre pairs that the models were able to classify, we see that $F^{\rm{right}}_m$ had 14 models that achieved an accuracy above $75\%$, compared to the 5 of $F^{\rm{even}}_m$. We also note that the image data had no PCA or tree-based feature selection applied to it, and the $F^{\rm{right}}_m$ filter was still favoured. A similar result was obtained with ansatz \ref{fig:u_ttn}. This shows architecture impacts performance even on low-resolution data.\newline

\subsection*{Architectural Search}
\label{sec:results}
\begin{figure}[h]
    \includegraphics[width=\linewidth]{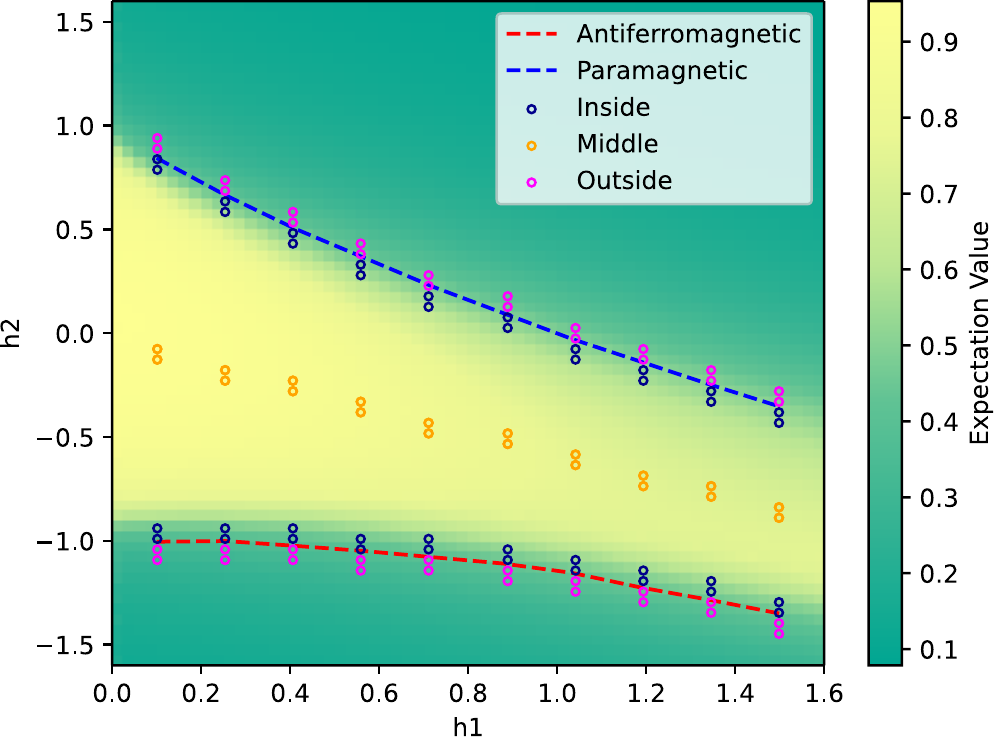}
    \caption{Expectation values for the circuit found via evolutionary search for a system of $N=15$ spins. Points represent a test set of $64\times64$ ground states for various $h_1$ and $h_2$ values of the Hamiltonian, $J=1$. The inside, middle and outside points were used to evaluate an architecture's fitness during search. The same color scale as in \cite{cong_quantum_2019} is used to facilitate comparison.}
    \label{fig:evolve_phase}
\end{figure}

In this section, we present an example of applying our architectural representation in conjunction with evolutionary search to perform Quantum Phase Recognition (QPR). The specifics of the search algorithm can be found in the Generalisation and Search section but we utilize an algorithm similar to the one employed by Liu et al. \cite{liuHierarchicalRepresentationsEfficient2018a}.  Mutations involve replacing a primitive within a motif with a randomly generated one, while crossover consists of combining two motifs end-to-end, if possible, or interweaving them otherwise. To facilitate comparison, we consider the same task and setup from the original QCNN paper \cite{cong_quantum_2019}.  The objective is to recognize a $\mathbb{Z}_2\times\mathbb{Z}_2$ symmetry-protected topological (SPT) phase for a ground state that belongs to a family of cluster-Ising Hamiltonians \cite{smacchiaStatisticalMechanicsCluster2011a}:
\begin{align}
    H & = -J\sum_{i=1}^{N-2} Z_iX_{i+1}Z_{i+2} - h_1\sum_{i=1}^{N} X_i - h_2\sum_{i=1}^{N-1} X_iX_{i+1}.\label{eq:ham}
\end{align}

Here, $X_i,Z_i$ are Pauli operators acting on the spin at site $i$ and the SPT phase contains a $S=1$ Haldane chain \cite{haldaneNonlinearFieldTheory1983}.  The ground state can belong to an SPT, paramagnetic or antiferromagnetic phase depending on the values of $h_1$, $h_2$, and $J$. Our goal is to identify a QCNN capable of distinguishing between SPT and other phases by measuring a single qubit. Following the approach in \cite{cong_quantum_2019}, we consider a system of $N=15$ spins and train a circuit on 40 equally spaced points along $h_2=0$, where the ground state is known to be in the SPT phase when $h_1\leq 1$. We also evaluate the circuit with the same sample complexity \cite{cong_quantum_2019}:
\begin{align}
    M_{\text{min}} & = \frac{1.96^{2}}{(\arcsin{\sqrt{p}}-\sqrt{\arcsin{p_0}})}\label{eq:sample}
\end{align}
where $p$ represents the probability of measuring a non-zero expectation value and $p_0=0.5$. Equation (\ref{eq:sample}) calculates the minimum number of measurements required to be $95\%$ confident that $p\neq0.5$, with $p$ being the expectation value of the circuit $U$ encoded with the ground state $\ket{\psi_g}$ transformed into a probability: $p=(\bra{\psi_g}U\ket{\psi_g}+1)/2$. Therefore a well-performing QCNN will yield low values of $M_{\text{min}}$ near the phase boundary for points within the SPT phase. We define the fitness of an architecture as a linear combination of the sample complexity values $M_{\text{in}},M_{\text{middle}}$ for points in the SPT phase, and the mean squared error $\text{MSE}_{\text{out}}$ for points outside the boundary. Figure \ref{fig:evolve_phase} illustrates the points considered for $M_{\text{in}},M_{\text{middle}}$ and $\text{MSE}_{\text{out}}$. During search we assigned the majority of the weight to $M_{\text{in}}$  as the goal is to develop a model that confidently identifies SPT phases near the boundary.  To prevent a model from classifying all points as SPT, $\text{MSE}_{\text{out}}$ is included, while $M_{\text{middle}}$ ensures overall good performance. Finally, during search we added a regularization term for the number of parameters, to find well-performing architectures with low computational complexity.\newline

\begin{table}[ht]
    \centering
    \begin{tabular}{lcc}
        \hline
        \textbf{Metric}            & \textbf{Reference} & \textbf{Found}    \\ \hline
        Number of parameters       & $1308$             & $\textbf{11}$     \\
        Sample Complexity (Inside) & $61.523$           & $\textbf{36.079}$ \\
        Sample Complexity (Middle) & $\textbf{10.992}$  & $13.253$          \\
        MSE (Outside)              & $\textbf{0.164}$   & $0.167 $          \\ \hline
    \end{tabular}
    \caption{Different performance metrics (lower is better) for the 15-qubit QCNN from \cite{cong_quantum_2019} and the architecture found via evolutionary search. Sample complexity represents the expected number of measurements required to be $95\%$ confident that the ground state is in the SPT phase (non-zero expectation value). Metrics are calculated on a set of points in the test set, where inside refers to SPT points near the phase boundary, outside to non-SPT points near to the phase boundary and middle to points in between, as shown in Figure \ref{fig:evolve_phase}.}
    \label{tab:evolve_architecture_comparison}
\end{table}

Table \ref{tab:evolve_architecture_comparison} and Figure \ref{fig:evolve_phase} show the performance of the best architecture found during search. The search algorithm identified a QCNN with only 11 parameters, in contrast to the 1308 parameters of the original reference architecture. For points in the SPT phase near the boundary, the sample complexity of the discovered architecture ($M_{\text{in}}=36.079$) is lower than that of the reference ($61.523$), resulting in $25$ fewer measurements required on average. Although the reference architecture exhibits slightly better sample complexity for points in the middle of the phase boundary ($M_{\text{middle}}=10.992$) compared to the discovered architecture ($M_{\text{middle}}=13.253$), and a marginally lower MSE for points outside the phase boundary ($\text{MSE}{\text{out}}=0.164$ compared to $\text{MSE}{\text{out}}=0.167$), the improvements in $M_{\text{in}}$ and the number of parameters are substantial and more advantageous. The discovered architecture can be found in Appendix \ref{fig:evolve_circuit}, and the phase diagram it generates is shown in Figure \ref{fig:evolve_phase}. The search was conducted on a system equipped with two Intel Xeon E5-2640 processors (2.0 GHz) and 128 GB of RAM, and it took approximately 2 hours to discover the final architecture  (over 831 generations). Although we anticipate that extending the search may yield even better architectures, the primary goal of this experiment was to demonstrate a representative example of the search process and showcase the ease of obtaining promising results. This emphasizes the potential advantages of architecture search in quantum computing tasks, where the computational cost of a circuit can be reduced while maintaining or even improving performance. We attribute this success to a well-defined search space, with our representation aiming to simplify the process of creating such spaces. Moreover, our representation allows for the incorporation of hardware constraints, facilitating the search for architectures that perform well on specific quantum devices. We believe this to be a necessary step towards the development of efficient quantum algorithms for real-world applications. By employing a well-structured representation and search space, we can streamline the process of discovering optimized quantum circuit architectures that are better suited for specific tasks and hardware.



\subsection*{Digraph Formalism}
\label{ssec:formalisation}
We represent the QCNN architecture as a sequence of directed graphs, each acting as a primitive operation such as a convolution (Qconv)     or pooling (Qpool). A primitive is the directed graph $G=(Q,E)$; its nodes $Q$ represent available qubits, and oriented edges $E$ the   connectivity of the unitary applied between a pair of them. The direction of an edge indicates the order of interaction for the unitary.  For example, a CNOT gate with qubit $i$ as control and $j$ as target is represented by the edge from qubit $i$ to qubit $j$. We also introduce other primitives, such as Qfree, that free up pooled qubits for future operations. The effect of a primitive is based on its  hyperparameters and the effect of its predecessor. This way, their individual and combined architectural effects are captured, enabling  them to be dynamically stacked one after another to form the second level $l=2$ motifs. Stacking these stacks in different ways  constitutes higher-level motifs until a final level $l=L$, where one motif constitutes the entire QCNN architecture. In the case of  pooling, controlled unitaries are used in place of measurement due to the deferred measurement principle \cite{Nielsen:2011:QCQ:1972505}. We define a QCNN architecture in Definition \ref{def:1}.
\begin{theo}
    \label{def:1}
    The $k^{th}=1,2,\dots K_l$ motif on level $l=1,2,\dots,L$ is the tuple $M^{l}_{k}=(M^{l-1}_{j}|j \in \{1,2,\dots, K_{l-1}\})$. Motifs on the lowest level $M^{1}_k$ are primitive operations, which form the set $M^{(1)} = \{M^1_1,M^1_2,\dots,M^1_{K_1}\}$. For example, $M^1_1=\text{Qconv$(2)$}, M^1_2=\text{Qpool(right)}$. At the highest level $l=L$ there is only one motif $M^L_1$ which is a hierarchy of tuples. $M^L_1$ is flattened through an assemble operation: $M=\text{assemble}(M^L_1)$ which encodes each primitive into a directed graph $G_m=(Q_m,E_m)$, the nodes $Q_m$ are available qubits and edges $E_m$ the connectevity of unitaries applied between them. $M$ describes the entire QCNN architecture, $M=({G_{1}},{G_{2}},\dots,{G_{|M|}})$.
\end{theo}
Figure \ref{fig:motifs} shows example motifs on different levels for a QCNN. Higher level motifs are tuples and the lowest level ones directed graphs. The dependence between successive motifs is specified in definition \ref{def:2}.
\begin{theo}
    \label{def:2}
    Let $x\in\{c,p,f\}$ indicate the primitive type for $\{\text{Qconv, Qpool, Qfree}\}$ and $M^L_1$ be the highest level motif for a QCNN. Then $\text{assemble}(M^L_1)$ flattens depth-wise into $M=({G_{1}},{G_{2}},\dots,{G_{|M|}})$ where $G_m=(Q^x_m,E^x_m)$. $G_1$ is always a Qfree($N_q$) primitive specifying the number of available qubits with $N_q$. For $m>1$, $G_m$ is defined as:
    \begin{flalign}
                & \text{If }G_m \text{ is a Qfree$(N_f)$ primitive then: }                           &  & \nonumber \\
        Q_{m}^f & =  \{1,2,\ldots,N_f\},                                                             &  & \nonumber \\
        E_{m}^f & = \{\}.                                                                            &  & \nonumber \\
                & \text{If } G_m \text{ is a convolution primitive: }                                &  & \nonumber \\
        Q_{m}^c & = \begin{cases}
                        Q_{m-1}^x                                          & \text{if $x\in\{c,f\}$ }, \\
                        Q_{m-1}^x \setminus \{i \in (i,j) \in  E_{m-1}^x\} & \text{if $x=p$ },
                    \end{cases}  &  & \nonumber                \\
        E_{m}^c & = \{(i,j) | (i,j) \in Q_{m}^c \times Q_{m}^c\}.                                    &  & \nonumber \\
                & \text{If } G_m \text{ is a pooling primitive: }                                    &  & \nonumber \\
        Q_m^p   & = \begin{cases}
                        Q_{m-1}^x                                          & \text{if $x\in \{c,f\}$ }, \\
                        Q_{m-1}^x \setminus \{i \in (i,j) \in  E_{m-1}^x\} & \text{if $x=p$ },
                    \end{cases} &  & \nonumber               \\
        E_m^p   & = \{(i,j) | (i,j) \in Q_m^p \times Q_m^p,i\neq j,                                  &  & \nonumber \\
                & \hspace{.5cm}  d^{-}(i)=0, d^{+}(i)=1,d^{-}(j)\geq1, d^{+}(j)=0 \}.                &  & \nonumber
    \end{flalign}
    with $d^{-}(i)$ and $d^{+}(i)$ referring to the indegree and outdegree of node $i$, respectively and $\setminus$ to set difference.
\end{theo}

\begin{figure*}[t]
    \centering
    \includegraphics[width=\linewidth]{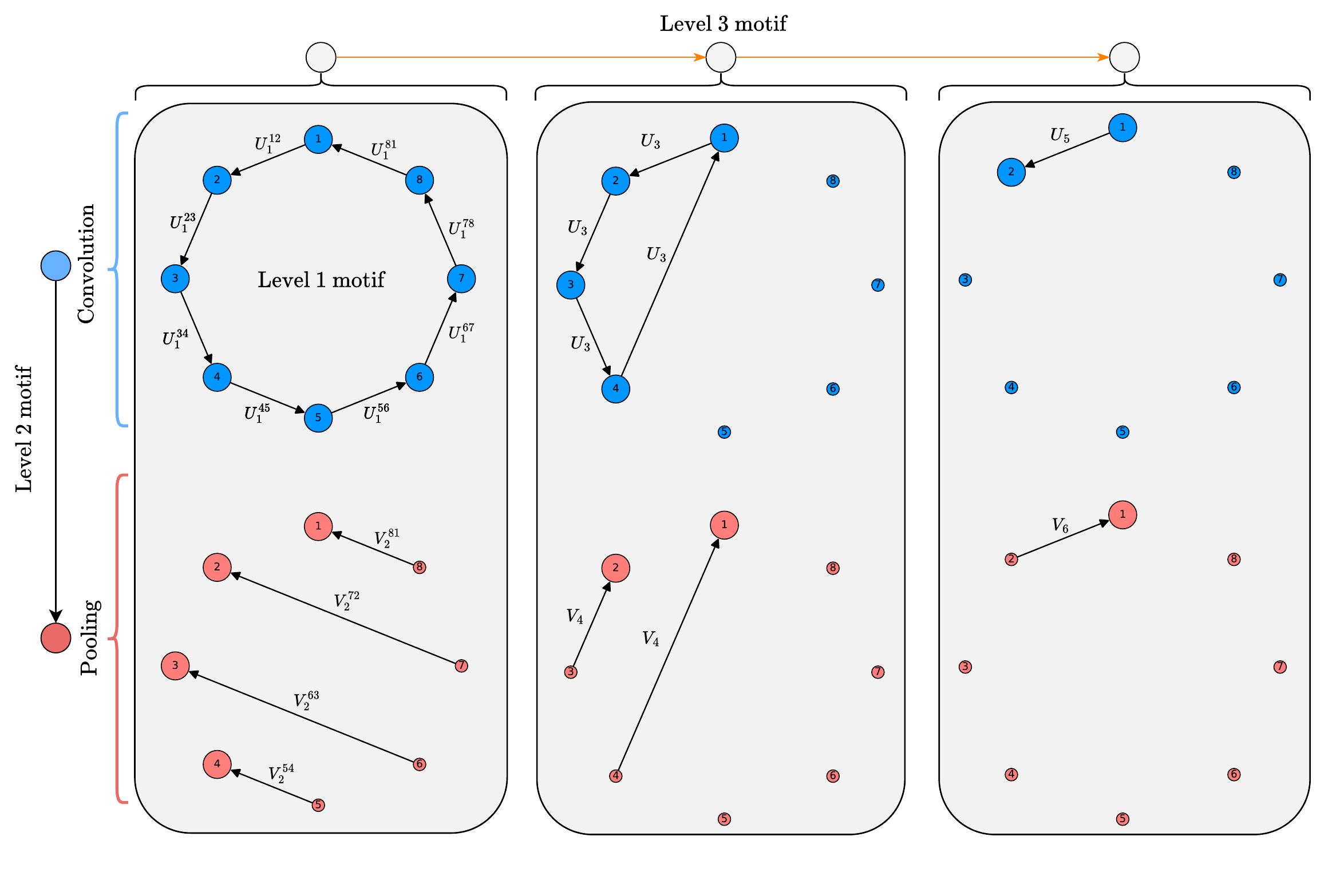}
    \caption{Graph view for the circuit architecture in Figure \ref{fig:qcnn_pipeline} (d). The same two-qubit unitary is used in all layers for the convolution operation, i.e. $U_m^{ij}=U_m.$ Similarly, in this example, we use the same two-qubit pooling unitaries $V_m^{ij}=V_m$. The top left graph is $G_1=(Q^c_1,E^c_1)$ with all eight qubits $Q_1^c$ available for the convolution operations $U^{ij}_1, (i,j) \in E^c_1$. Below $G_1$ is $G_2$ with half the qubits of $Q_2^p$ measured, indicated by the $i^{th}$ indices of $V_m^{ij}, (i,j) \in E_2^p$. For example, qubit $8\in Q^p_2$ is measured and $V_2$ applied to qubit $1\in Q_2^p$ as indicated by $V_2^{81},(8,1)\in E_2^p$. This pattern repeats until one qubit remains in $G_6$, which is measured and used to classify the music genre.}
    \label{fig:qcnn_digraph}
\end{figure*}
We show this digraph perspective in Figure \ref{fig:qcnn_digraph}, it is the data structure of the circuit in Figure \ref{fig:qcnn_pipeline}d. If the $m^{th}$ graph in $M$ is a convolution, we denote its two-qubit unitary acting on qubit $i$ and $j$ as $U_{m}^{ij}(\theta)$. Similarly, for pooling, we notate the unitary as $V_{m}^{ij}(\theta)$. The action of $V_{m}^{ij}(\theta)$ is measuring qubit $i$ (the control), which causes a unitary rotation $V$ on qubit $j$ (the target). With this figure and notational scheme in mind, Definition \ref{def:2} reads as follows:\newline

$Q_m^x$ is the set of available qubits for the $m^{th}$ primitive in $M$, where $x\in\{c, p, f\}$ for convolution, pooling or Qfree respectively. The first primitive $G_1$ is Qfree($N_q$) which specifies the number of available qubits $N_q$ for future operations. Any proceeding $m>1$ primitive $G_m$ only has access to qubits not measured up to that point. This is the previous primitive's available qubits $Q^x_{m-1}$ if its type $x\in\{c,f\}$ is a convolution or Qfree. Otherwise, for pooling, $x=p$, it's the set difference: $Q_{m-1}^x \setminus \{i \in (i,j) \in E_{m-1}^x\}$ since the $i$ indices during pooling $(i,j) \in E^p_m$ indicates measured qubits. This is visualised as small red circles in Figure \ref{fig:qcnn_digraph}. The only way to make those qubits available again is through Qfree($N_f$), which can be used to free up $N_f$ qubits. For the convolution primitive, $E_m^c$ is the set of all pairs of qubits that have $U_{m}^{ij}(\theta)$ applied to them. Finally, for the pooling primitive, $E_m^p$ is the set of pairs of qubits that have pooling unitaries $V_{m}^{ij}(\theta)$ applied to them. The restriction is that if qubit $i$ is measured, it cannot have any other rotational unitary $V$ applied to it within the same primitive $G_m$. This means the indegree $d^{-}$ of node $i$ is zero. Similarly, if qubit $i$ is measured, it may only have one corresponding target, meaning that the outdegree $d^{+}$ of node $i$ is one. In the same vein, no target qubit $j$ can be the control for another, $d^{+}(j)=0$. Every target qubit $j$ have at least one corresponding control qubit $i$, $d^{-}(j)\geq 1$. It is possible for multiple measured qubits to have the same target qubit, giving $E^{p}_{m}$ a surjective property.\newline
Following this definition, we can express a convolution or pooling operation for the $m^{th}$ graph in $M$ as:
\begin{flalign}
    \widetilde{U}_m & = \prod_{(i,j) \in E_m^c} U_{m}^{ij}(\theta), &  & \label{eq:1} \\
    \widetilde{V}_m & = \prod_{(i,j) \in E_m^p} V_{m}^{ij}(\theta). &  & \label{eq:2}
\end{flalign}

Let $\widetilde{W}_m=\widetilde{U}_m \text{ or } \widetilde{V}_m$ be the  $m^{th}$ primitive in $M$ based on whether it's a convolution or pooling and the identity $I$ if it's a Qfree primitive. Then the state of the QCNN after one training run is:
\begin{flalign}\label{eq:3}
    \ket{\psi} & =  \widetilde{W}_{|M|}\cdots \widetilde{W}_4\widetilde{W}_3\widetilde{W}_2\widetilde{W}_1U_{\text{encoding}}\ket{0}. &  & 
\end{flalign}

We note that the choice of $V$ is unrestricted, which means that within one layer each $V$ can be a different rotation. Figure \ref{fig:qcnn_pipeline}d shows a special case where the same $V$ is used per layer, which is computationally favourable compared to using different ones. To enable weight sharing, the QCNN require convolution unitaries to be the same i.e. $U_{m}^{ij}=U_m^{kh} \text{ where } (i,j) \in, (k,h) \in E_m^c$. This formulation only regards one and two qubit unitaries for convolutions, one qubit unitaries being described with $E_m^c=(i,i), i\in Q_m^c$. In the generalisation and search section we extend it to multiple qubit unitaries. \newline

After training, $\ket{\psi}$ in eq. \ref{eq:3} is measured based on the type of classification task, in this work we focus on binary classification allowing us estimate $\hat{y}$ by measuring the remaining or specified qubit in the computational basis:
\begin{flalign}\label{eq:yhat}
    \hat{y} & = P(y=1)= |\bra{1}\ket{\psi}|^2. &  & 
\end{flalign}

We note that multi-class classification is also possible by measuring the other qubits and associating each with a different class outcome. Following this, we calculate the cost of a training run with $C(y,\hat{y})$, then using numerical optimization the cost is reduced by updating the parameters from Equations \ref{eq:1} to \ref{eq:2} and repeating the whole process until some local minimum is reached. Resulting in a model alongside a set of parameters to be used for classifying unseen data.\newline
\subsection*{Controlling the primitives}
\label{ssec:control_prim}
We define basic hyperparameters that control the individual architectural effect of a primitive. There are two broad classes of primitives, special and operational. A special primitive has no operational effect on the circuit, such as Qfree. Its purpose is to make qubits available for future operational primitives and therefore has one hyperparamater $N_f$ for this specification. $N_f$ is typically an integer or set of integers corresponding to qubit numberings:
\begin{flalign}
    Q_{m}^f & =  \{1,2,\ldots,N_f\} & \text{if } N_f \text{ is an integer},        & \label{eq:4} \\
    Q_{m}^f & =  N_f                & \text{if } N_f \text{ is a set of integers}. & \label{eq:5}
\end{flalign}

Each operational primitive has its own stride parameter analogous to classical CNNs. For a given stride $s$, each qubit gets paired with the one $s$ qubits away modulo the number of available qubits. For example a stride of $1$ pairs each qubit with its neighbour. This depends on the qubit numbering used which is based on the circuit topology. For illustration purposes, we use a circular topological ordering, but any layout is possible as long as some ordering is provided for $Q^f_1$.  For the convolution primitive we define its stride $s_c \in \{1,2,3,...\}$ as:
\begin{flalign}
    E_m^c & =\{(i,(i+s_c) \bmod |Q_m^c|) | i \in Q_m^c\} & \text{if } |Q_m^c|>2, & \label{eq:6} \\
    E_m^c & =\{(i,j) | i,j \in Q_m^c, i\neq j\}          & \text{if } |Q_m^c|=2, & \label{eq:7} \\
    E_m^c & =\{(i,i) | i \in Q_m^c\}                     & \text{if } |Q_m^c|=1. & \label{eq:8}
\end{flalign}
Equation (\ref{eq:7}) captures the case where there are only two qubits available for a convolution and equation (\ref{eq:8}) when there is only one which implies the convolution unitaries only consist of single qubit gates. A stride of $s_c=1$ is a typical design motif for PQCs and the graph formalism allow for a simple way to capture and generalise it. To achieve translational invariance for all strides the two constraints:  $|E_m^c|=|Q_m^c|$ and $(i,j) \neq (k,h) \text{ where } (i,j) \in,(k,h) \in E_m^c$ are added. Another option for translational invariance is a Qdense primitive, which only differs from Qconv in that $E_m^c$ generates all possible pairwise combinations of $Q_m^c$. This primitive is available in the python package but left out from the definition (because of its similarity).
\begin{figure*}[t]
    \includegraphics[scale=.8]{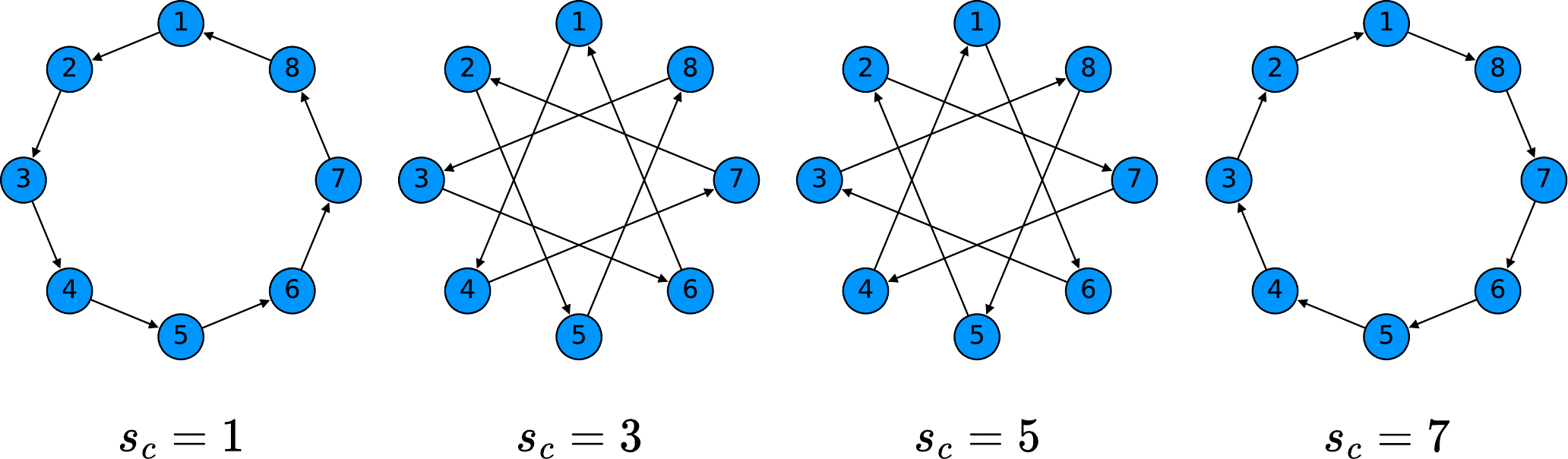}
    \caption{Diagram showing how changing the convolution stride $s_c$ generates different configurations for $E_m^c$.}
    \label{fig:convolution_step}
\end{figure*}
Figure \ref{fig:convolution_step} show different ways in which $s_c$ generate $E_m^c$ for $|Q_m^c|=8$.\newline

The pooling primitive has two hyperparameters, a stride $s_p$ and filter $F^*_m$. The filter indicates which qubits to measure and the stride how to pair them with the qubits remaining. We define the filter as a binary string:
\begin{flalign}
    F_m^* & =w_1w_2\cdots w_{|Q_m^p|} & \begin{cases}
                                            w_i=1  \text{ if qubit $i$ is measured}, \\
                                            w_i=0  \text{ otherwise}.
                                        \end{cases} & 
\end{flalign}
For $N=8$ qubits, the binary string $F_m^*=00001111$ translates to measuring the rightmost qubits, i.e. $\{i|i\in Q_m^p, i\ge5\}$. Figure \ref{fig:qcnn_digraph} is an example where the pattern $F_2^*=00001111\rightarrow F_4^*=0011\rightarrow F_6^*=01$ is used, visually the qubits are removed from bottom to top. Encoding filters as binary strings is useful since generating them becomes generating languages, enabling the use of computer scientific tools such as context free grammars and regular expressions to describe families of filters. Pooling primitives enable hierarchical architectures for QCNNs, and in the search space design section, we illustrate how they can be implemented to create a family resembling reverse binary trees. The action of the filter is expressed as: $F^{*}_m \star Q_m^p=Q_{m+1}^x$ where $\star$ slices $Q^p_m$ corresponding to the $0$ indices of $F^*_m$, i.e. $w_i=0$(not measured). For example $010\star\{4,7,2\} = \{4,2\}$. This example illustrates the case where an ordering was given to the set of available qubits to represent some specific topology of the circuit.
Let $Q_{m+1}^x=F^{*}_m \star Q_m^p$ then the pooling primitive stride $s_p=\{1,2,\dots \}$ is defined as:
\begin{flalign}
    E_m^p & =\{(i,(j+s_p) \bmod |Q_{m+1}^x|)| i \in Q_m^p \setminus Q_{m+1}^x, &  & \nonumber \\
          & \hspace{1cm} j \in Q_{m+1}^x\}.                                    &  & 
\end{flalign}
\subsection*{Search Space Design}
\label{ssec:search_space_design}

We show how the digraph formalism facilitates QCNN generation and search space design. Grant et al. \cite{grant_hierarchical_2018} exhibited the success of hierarchical designs that resemble reverse binary trees. To     create a space of these architectures, we only need three levels of motifs. The idea is to reduce the system size in half until one qubit remains while alternating between convolution and pooling operations. Given   $N$ qubits, a convolution stride $s_c$, pooling stride $s_p$ and a pooling filter $F^*$ that reduce system size in half, a reverse binary tree QCNN is generated in Algorithm \ref{alg:reverse_binary_tree}. 

\begin{figure*}[t]
    \includegraphics[width=\linewidth]{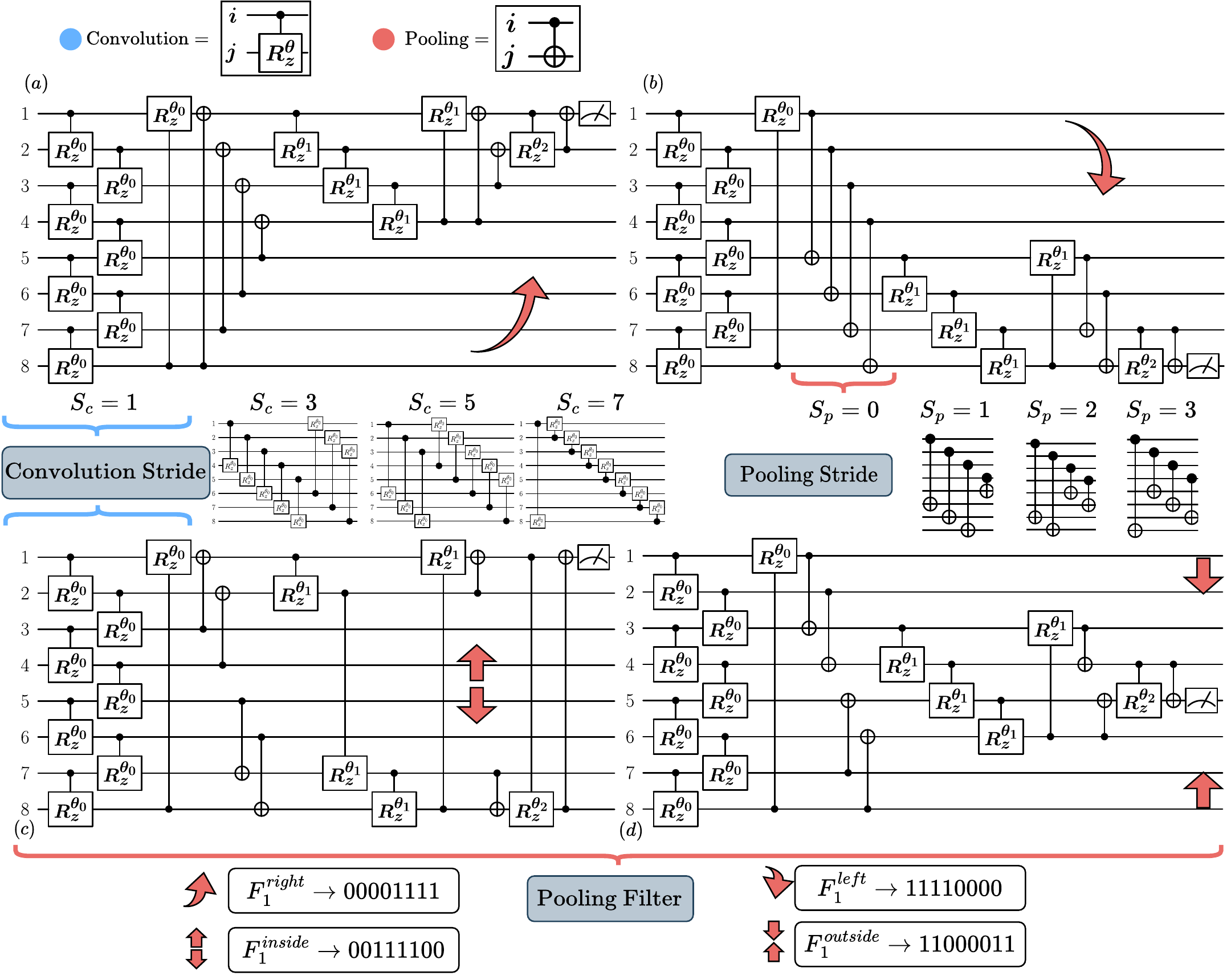}
    \caption{An example of how the hyperparameters of the primitives effect the circuit architecture of the family generated by Algorithm \ref{alg:reverse_binary_tree}. Three are shown, the convolution stride $s_c$, pooling stride $s_p$ and pooling filter $F^*$. These are specified in the controlling primitives section. Controlled-$R_z^\theta$ gates are used for convolutions and CNOTs for pooling as an example. The convolution stride $s_c$ determine how convolution unitaries are distributed across the circuit. Each convolution primitive typically consist of multiple unitaries and the QCNN requires them to be identical for weight sharing. The pooling stride $s_p$ determine how pooling unitaries are distributed, for a given pooling primitive, a portion of available qubits gets pooled via controlled unitary operations and $s_p$ dictates which controls match to which targets. The pooling filter $F^*$ dictates which qubits to pool according to some recursive pattern/mask. For example, circuit d) always pools the outside qubits during pooling primitives, resulting in the middle qubit making it to the end of the circuit. }
    \label{fig:hyperparams_summary}
\end{figure*}

\begin{algorithm}[H]
    \begin{algorithmic}
        \Require{$N,s_c,s_p,F^{*}$}
        \Ensure{QCNN$\rightarrow M=(G_1,G_2,\dots,G_{|M|})$}\\
        \LeftComment{Primitives:}
        \State $M_1^1 \gets \text{Qconv}(stride=s_c)$
        \State $M_2^1 \gets \text{Qpool}(stride=s_p,filter=F^{*})$\\
        \LeftComment{Motif: alternate convolution and pooling}
        \State $M_1^2 \gets M_1^1 + M_2^1$\\
        \LeftComment{Motif: repeat until one qubit remain}
        \State $M_1^3 \gets \text{Qfree}(N) + M_1^2\times\log_2{N}$\\
        \State $M\gets \text{assemble}(M_1^3)$
    \end{algorithmic}
    \caption{QCNN, reverse binary tree architecture.}
    \label{alg:reverse_binary_tree}
\end{algorithm}
Algorithm \ref{alg:reverse_binary_tree} shows how to create instances of this architecture family. First, two primitives are created on the first level of the hierarchy, a convolution operation $M^1_1$ and a pooling operation $M^1_2$. They are then sequentially combined on level two as $M^2_1=(M^1_1, M^1_2)$ to form a convolution-pooling unit. The third-level motif $M^3_1$ repeats this second-level motif $M^2_1$ until the system only contains one qubit. This is $\log_2(N)$ repetitions for $N$ qubits because we chose $F^*$ to reduce the system size in half during each pooling operation. The addition and multiplication symbols act as append and extend for tuples. For example $M_{1}^{l}+M_{2}^{l}=(M_{1}^{l},M_{2}^{l})$ and $M_{k}^l\times 3= (M_{k}^l,M_{k}^l,M_{k}^l)$ which allow for an intuitive way to build motifs.  It is easy to expand the algorithm for more intricate architectures, for instance, by increasing the number of motifs per level and the number of levels. A valid level four motif for algorithm \ref{alg:reverse_binary_tree} would be $M^4_1= (M^3_1 +M^3_2) \times 3$, where $M^3_2 = \text{Qfree(4)} + M^2_2 + M^2_1$ and $M^2_2=M_1^1\times 2$ which is the reverse binary tree architecture $M^3_1$ then two convolutions and one convolution-pooling unit on four qubits, all repeated three times. Motifs can also be randomly selected on each level to generate novel architectures. The python package we provide acts as a tool to facilitate architecture generation this way.
\newline
In more detail, we now analyse the family of architectures generated by algorithm \ref{alg:reverse_binary_tree}. First, we consider the possible pooling filters $F^*$ that reduce system size in half. It is equivalent to generating strings for the language $A = \{w|w \text{ has an equal number of 0s and 1s }, |w|=|Q_m^p|\}$. Let $N_{m-1}=|Q^x_{m-1}|$ indicate the number of available qubits for the filter $F^*_m$. Then based on the ${4\choose2}=6$ possible equal binary strings \footnote{These are 0011,1100,1010,0101,0110,1001, equal in the sense that they have the same number of 1s and 0s} of length four, we construct the following \textit{pooling filters}: 
\begin{flalign}
    F^{right}_m   & = \{0^{\frac{n}{2}}1^{\frac{n}{2}} | n=N_{m-1}\},                                                                                      &  & \\
    F^{left}_m    & = \{1^{\frac{n}{2}}0^{\frac{n}{2}} | n=N_{m-1}\},                                                                                      &  & \\
    F^{odd}_m     & = \{(01)^{\frac{n}{2}} | n=N_{m-1}\},                                                                                                  &  & \\
    F^{even}_m    & = \{(10)^{\frac{n}{2}} | n=N_{m-1}\},                                                                                                  &  & \\
    F^{inside}_m  & =  \begin{cases}
                           \{0^{\frac{n}{4}}1^{\frac{n}{2}}0^{\frac{n}{4}} | n=N_{m-1}\} & \text{if $N_{m-1} > 2$}, \\
                           \{01\}                                                        & \text{if $N_{m-1}=2$ },
                       \end{cases} &  &    \\
    F^{outside}_m & = \begin{cases}
                          \{1^{\frac{n}{4}}0^{\frac{n}{2}}1^{\frac{n}{4}} | n=N_{m-1}\} & \text{if $N_{m-1} > 2$}, \\
                          \{10\}                                                        & \text{if $N_{m-1}=2$ }.
                      \end{cases} &  & 
\end{flalign}
where the exponent $a^3\equiv\{a\}\circ \{a\}\circ \{a\}=aaa$ refers to the regular operation concatenation: $A \circ B = \{xy|x\in A, y\in B\}$. The pooling filter $F_{inside}$ yields 0110. Visually this pattern pools qubits from the inside (the middle of the circuit). See Figure \ref{fig:hyperparams_summary} (c). Figure \ref{fig:hyperparams_summary} (a) shows the repeated usage of $F^{right}$ for pooling. This particular pattern is useful for data preprocessing techniques such as principal component analysis (PCA) since PCA introduces an order of importance to the features used in the model. Typically, the first principal component (which explains the most variance) is encoded on the first qubit, the second principal component on the second qubit and so on. Therefore, it makes sense to pool the last qubits and leave the first qubits in the model for as long as possible.
\newline
If $N=8$, $s_c=1$, $s_p=0$ and $F^*=F^{\rm{right}}$ then Algorithm \ref{alg:reverse_binary_tree} generates the circuit in Figure \ref{fig:qcnn_pipeline} (d), Figure \ref{fig:motifs}, Figure \ref{fig:qcnn_digraph} (f) and Figure \ref{fig:hyperparams_summary} (a). Specifically, Figure \ref{fig:hyperparams_summary} shows how different values for $s_c,s_p\text{ and } F^*$ generate different instances of the family using Algorithm \ref{alg:reverse_binary_tree}. The possible combinations of $N,s_c,s_p,F^{*}$ represent the search space/family size. Since $F^{*}$ reduces system size in half, it's required that the number of available qubits $N$ is a power of two. Using integer strides causes the $|E_m^c|=|Q_m^c|$ constraint (see the controlling primitives section), which enable translational invariance. The complexity of the model(in terms of the number of unitaries used) then scales linearly with the number of qubits $N$ available. Specifically, $N$ qubits result in $3N-2$ number of unitaries \footnote{This is because of the geometric series: $N(\frac{1}{2^0}+\frac{1}{2^1} + \cdots + \frac{1}{2^{\log_2{N}-1}})+N(\frac{1}{2^1}+\frac{1}{2^2} + \cdots + \frac{1}{2^{\log_2{N}-1}})$. Where the first sum is for convolution unitaries and the second for pooling.}.
\subsection*{Generalisation and Search}
\label{ssec:general_and_search}
\begin{figure*}[t]
    \includegraphics[scale=.8]{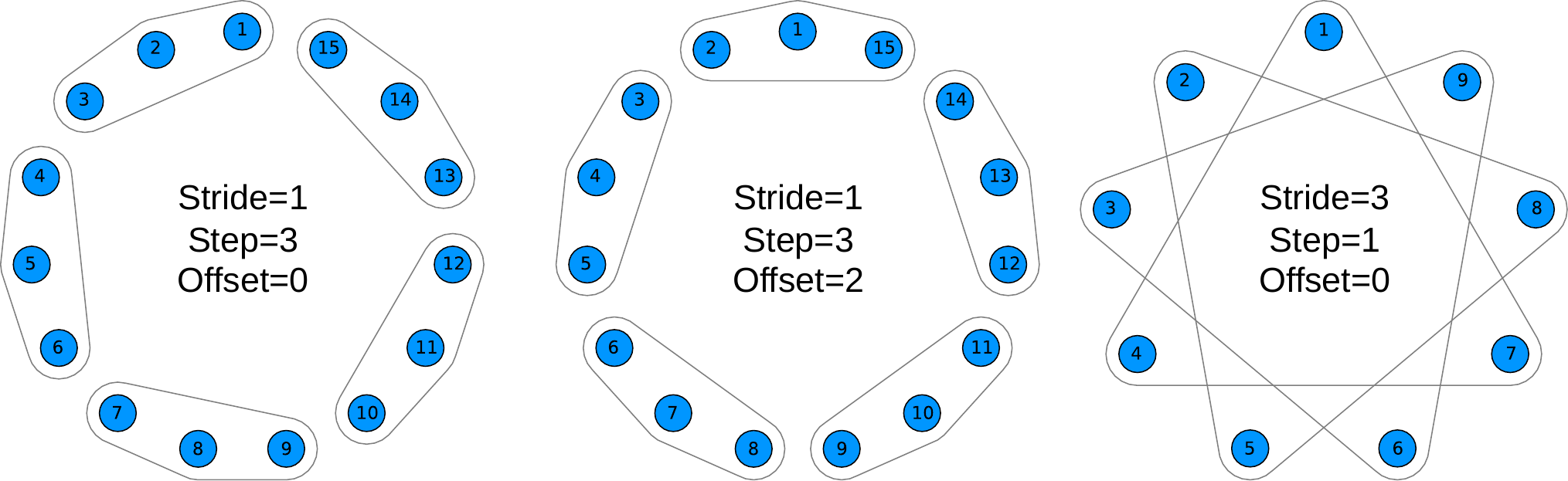}
    \caption{Examples how $3$-qubit unitaries are represented with the framework. For general $n$-qubit unitaries the graphs become hypergraphs with $n$-tuples as edges.}
    \label{fig:hypergraph}
\end{figure*}
The digraph formalism extends naturally to multi-qubit unitaries, enabling the representation of more intricate and larger scale architectures. In general, a primitive with $n$-qubit unitaries is represented as a hypergraph $G=(Q, E)$, where the edges $E$ consist of $n$-tuples. We introduce two additional hyperparameters, step and offset, which control the construction of $E$. For instance, Figure \ref{fig:hypergraph} shows three primitives, each with 3-qubit unitaries. The first two have a stride of one, meaning that each 3-qubit unitary connects to its neighbors. In contrast, the last primitive has a stride of three, connecting every third qubit within the unitary. The offset parameter determines the starting point for counting; Figure \ref{fig:hypergraph}a begins with the first qubit, while Figure \ref{fig:hypergraph}b starts with the third. The step parameter controls the position of the next unitary; for example, Figure \ref{fig:hypergraph}a and b have a step of three, skipping two qubits before creating another edge starting on the third qubit. Consequently, the primitives with 2-qubit unitaries we've been considering thus far are all special cases with a step of one and an offset of zero.  Another aspect to consider is the execution order of the unitaries, which by default is the sequence in which the edges were created for a primitive. Our package introduces an additional hyperparameter to control this order. For example to execute the third edge of Figure \ref{fig:hypergraph}a first followed by edge five, four, one and two, a value of $(3,5,4,1,2)$ can be passed to the edge order hyperparameter. Lastly, a boundary condition hyperparameter can also be specified, allowing for the definition of open or periodic boundaries for the qubits. This essentially determines whether edge creation is calculated in modulo with respect to the number of qubits or not, which in turn influences whether edge creation ceases when no further connections can be made based on the stride parameter. 

\begin{algorithm}[H]
    \begin{algorithmic}
        \Require{$n,N, F^*$}
        \Ensure{QCNN$\rightarrow M=(G_1,G_2,\dots,G_{|M|})$}\\
        \LeftComment{Primitives:}
        \State $M_1^1 \gets \text{Qfree}(n+1) + \text{Qdense}()$
        \State $M_2^1 \gets \text{Qconv}(1,n,n-1, mapping=M_1^1)$
        \State $M_3^1 \gets \sum_{i=0}^{n-1} \text{Qconv}(1,n,i)$
        \State $M_4^1 \gets  \text{Qpool}(1,n,0, filter=F^*)$
        \State $M_5^1 \gets \text{Qconv}(1,n,0)$\\
        
        \LeftComment{Motif: Apply all primitives to N qubits}
        \State $M_1^2 \gets \text{Qfree}(N)+ \sum_{i=1}^5 M^1_i$\\
        \LeftComment{Motif: repeat $d$ (depth) times}
        \State $M_1^3 \gets M_1^2*d$\\
        
        \State $M\gets \text{assemble}(M_1^3)$
    \end{algorithmic}
    \caption{Original QCNN from \cite{cong_quantum_2019}.}
    \label{alg:orignal_qcnn}
\end{algorithm}
\begin{figure*}[t]
    \includegraphics[scale=.8]{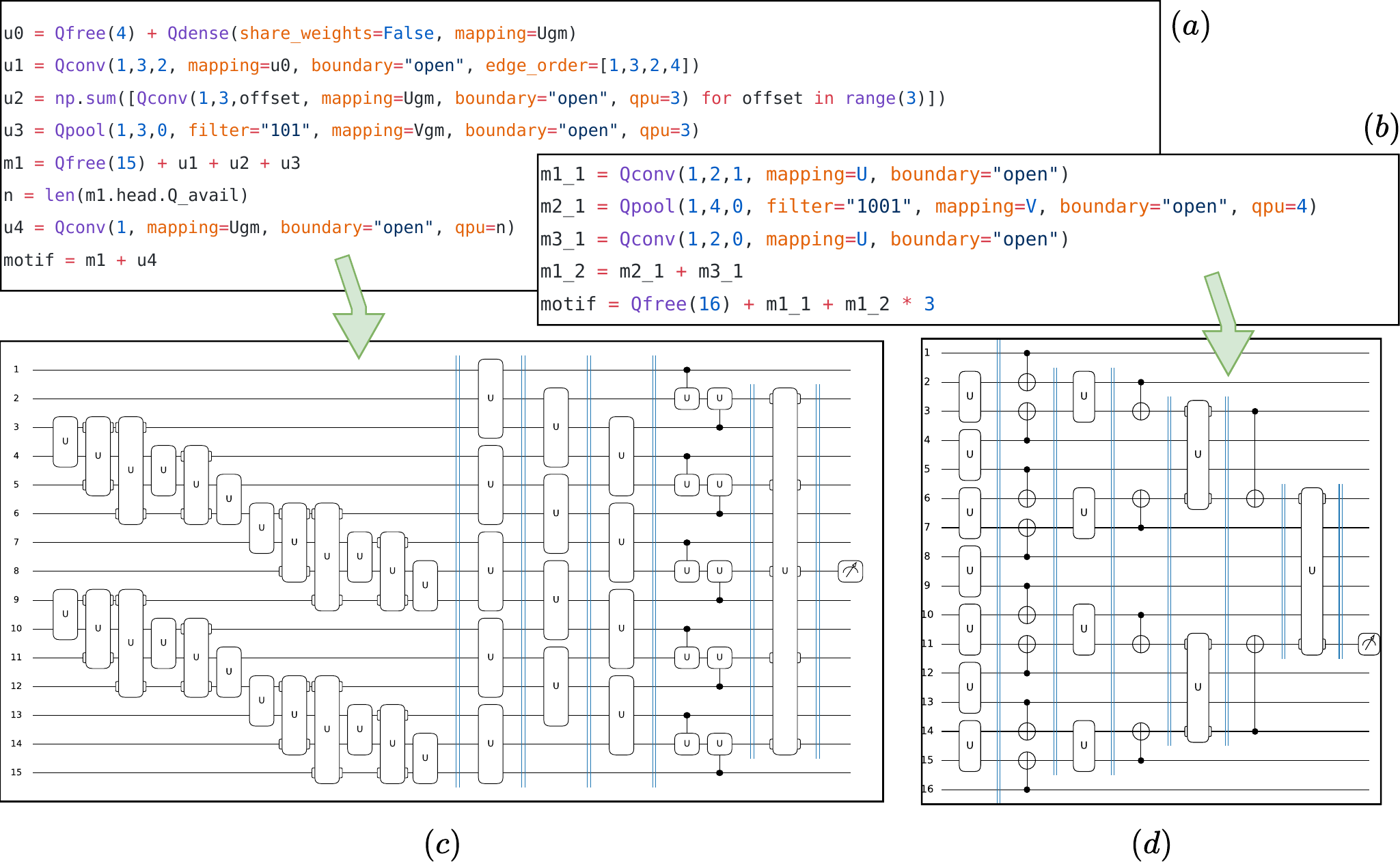}
    \caption{Example architectures from Cong et al. \cite{cong_quantum_2019} (c) and  Grant et al. \cite{grant_hierarchical_2018} (d), generated using our Python package to demonstrate its expressibility, interpretability, and scalability. In (a), the 15-qubit original QCNN is created with the first three parameters of each primitive being stride, step, and offset, respectively. The unitary $U$ mappings employ generalised Gell-Man matrices parameterised based on the number of qubits \textit{qpu} they act upon. Line 5, $\text{Qfree}(15)+m_1+m_2+m_3$, controls the system size; applying the same architecture to $N$ qubits only requires changing it to $\text{Qfree}(N)$. To introduce a depth $d$ parameter to the circuit, the last line should be modified to $m_4*d+m_5$. In (b), a 16-qubit MERA circuit is generated. For an 8-qubit MERA circuit, the last line would be changed to $\text{Qfree}(8)+m_1^1+m_2^1*2$, and in general, $\text{Qfree}(N)+m_1^1+m_2^1(\log_2{N}-1)$ produces an $N$-qubit MERA circuit. These examples highlight the representation's strengths: the essence of an architecture is captured with a few lines of code in a modular and understandable manner, and scaling up to larger systems is accomplished with minimal adjustments. }
    \label{fig:expressivity}
\end{figure*}
The hyperparameters provided are sufficient to generate a diverse array of hierarchical architectures. For example, we demonstrate how to represent the original QCNN from \cite{cong_quantum_2019} within our framework in Algorithm \ref{alg:orignal_qcnn}. The arguments for each convolution and pooling primitive are stride, step, and offset. The Qdense primitive generates 2-qubit unitaries between all pairwise combinations of $n+1$ qubits. Subsequently, the second primitive $M^1_2$ takes $M^1_1$ as its mapping, which just means it treats $M^1_1$ as a single $n+1$-qubit unitary, and distributes it across the circuit with a stride of 1, step of $n$, and offset of $n-1$. This is followed by $n$ convolutions of $n$-qubit unitaries, each having an offset incremented by one from the previous. For $n=3$ and $N=15$, the first and last convolution is illustrated in Figure \ref{fig:hypergraph}a,b. Next, a pooling layer with $n$-qubit unitaries is applied, measuring the outer $n-1$ qubits from each $n^{th}$ qubit, this corresponds to the filter $F^* =  \{1^{\frac{n-1}{2}}01^{\frac{n-1}{2}}\}$. Finally, a convolution is performed on all remaining qubits. In practice, each of these primitives is given a mapping for their corresponding unitary. The mappings of the original QCNN are based on $2^{v}\times 2^{v}$ gellman matrices, where $v$ indicates the number of qubits the unitary acts upon. For instance, the first unitary of the primitive $M^1_2$ operates on $v=n+1$ qubits, $M^1_3$ on $v=n$ qubits and pooling $M^1_3$ on $n$ qubits where $v=n-1$ to leave a qubit for the control. For $M_5^1$, $v$ equals the number of remaining qubits. It's easy to generate a family of architectures related to the original by providing the algorithm with different values of stride, step, offset, pooling filters, mappings and relaxing the dependance on $n$ based on how large we want the search space to be. \newline

Next, we discuss the applicability of search algorithms with our representation. The framework's expressiveness is demonstrated in Figure \ref{fig:qcnn_pipeline}(e,f), where only two lines of code are needed to specify a complete architecture, and in Figure \ref{fig:expressivity}, which illustrates how to capture circuits from \cite{cong_quantum_2019,grant_hierarchical_2018}. This expressiveness allows search algorithms to explore an extensive range of architectures and numerous design choices. Moreover, the modularity of the framework enables search algorithms to identify robust building blocks to combine into motifs, serving as the foundation for architectural designs. This is especially advantageous in the context of genetic algorithms, as it facilitates the definition of crossover and mutation operations in various ways. For example, mutations can involve adjusting a single hyperparameter of a primitive or replacing an entire primitive within a motif. Crossovers may include combining motifs at the same or different levels or interweaving two motifs by alternating their final sequence of primitives. In the case of reinforcement learning, the modularity allows an agent to make decisions at multiple levels of granularity, enabling it to explore and exploit different combinations of primitives and motifs. Hill climbing algorithms can also leverage this modularity in various ways. For instance, we can generate a random fixed high-level motif, such as a MERA circuit, and then iteratively optimize the hyperparameters of each primitive within the motif. In each step, we adjust a hyperparameter to neighboring values, evaluate the resulting objective values, and select the best configuration. Once we have updated all the hyperparameters of all primitives, we obtain a final motif, which can be used as a starting point for the next iteration. This approach of adjusting individual hyperparameters within a multilevel motif allows for incremental changes to the architecture. Such fine-grained modifications can be beneficial in approaches like Bayesian optimization, where smoothness in objective values is advantageous. Additionally, the hierarchical nature of the representation promotes scalability, enabling search algorithms to investigate smaller subsystems before scaling up to the full problem. This can reduce computational costs and allow for the exploration of more architectures. Lastly, the intuitive nature of the representation facilitates understanding the performance of discovered architectures, which enhances interpretability. For instance, in one experiment, we observed a spike in performance for a convolution stride of five. Upon further investigation, we discovered a strong correlation between features one and six, which was previously unknown. This insight informed future experiments and design choices for the problem at hand.\newline

Finally, we present the evolutionary algorithm used in our experiments, which is based on the approach described in \cite{liuHierarchicalRepresentationsEfficient2018a} and detailed in Algorithm \ref{alg:evolutionary_algorithm}.  We refer to an architecture as a genotype, and its fitness is determined by the sample complexity for both inside and middle points, as well as the mean squared error (MSE) for outside points in the test set (see Figure \ref{fig:evolve_phase}).  Specifically, $\text{fitness}=c_1\frac{M_{\text{in}}}{M_{\text{cap}}}+c_2\frac{M_{\text{middle}}}{M_{\text{cap}}}+c_3\text{MSE}_{\text{out}} +\lambda n_p$ where we cap $M_{\text{in}}$ and $M_{\text{middle}}$ by some large value $M_{\text{cap}}$ and $n_p$ is the number of paramaters required for the architecture. The weights $c_1$, $c_2$, and $c_3$ sum to one, assigning importance to each term. Our experiments showed that setting $c_1 = 0.7$, $c_2 = 0.05$, and $c_3 = 0.25$ led to generally well-performing architectures, we also chose $M_{\text{cap}}=500$ since fit genotypes exhibit sample complexity below $100$. We initialise the population with a pool of 100 random primitives (Qconv, Qpool, Qdense), each having random hyperparameters. Upon initialization, we perform mutation and crossover operations based on tournament selection with a 5\% selection pressure. After the selection, we mutate the fittest genotype by choosing one of its primitives and replacing it with a randomly generated one. The crossover operator acts on the two fittest individuals, attempting to combine them tail-to-head. If this is not possible, they are interleaved up to the point where they can be combined. Just like the approach in \cite{liuHierarchicalRepresentationsEfficient2018a}, we do not remove any genotypes from the pool, leading to a more diverse population.

\begin{algorithm}[H]
    \caption{Evolutionary Search Algorithm}
    \label{alg:evolutionary_algorithm}
    \begin{algorithmic}
        \State \textbf{Input:} Initial population $P$, memory table $M$ containing fitness values        
        \Function{Controller}{$P$, $M$}
        \While{True}
        \State $g_1, g_2 \gets \Call{tournament\_selection}{M}$
        \State $\text{mutated} \gets \Call{Mutate}{g_1}$
        \State $\text{combined} \gets \Call{Combine}{g_1, g_2}$
        \State Add mutated genotype to task queue
        \State Add combined genotype to task queue
        \If{idle worker available}
        \State Assign top task in task queue to the idle worker
        \EndIf
        \EndWhile
        \EndFunction
        \Function{Worker}{$\text{task}$, $M$}
        \State $\text{genotype} \gets \text{task.genotype}$
        \State $\text{fitness} \gets \Call{EvaluateFitness}{\text{genotype}}$
        \State Update memory table with the new fitness value
        \EndFunction
    \end{algorithmic}
\end{algorithm}


\section*{Discussion}

The main contribution of this paper is a framework that enables the dynamic generation of QCNNs and the creation of QCNN search spaces. The framework is provided theoretically in this paper and implemented as a Python package that is ready for use. Our numerical experiments demonstrate the importance of alternating architectures for PQCs, and illustrate a way to increase model performance without increasing its complexity. Our next step is to explore search strategies using this architectural representation to find high-performing QCNNs for different classification tasks automatically. We've already shown how the representation is useful for evolutionary algorithms, as in the classical case \cite{liuHierarchicalRepresentationsEfficient2018a} but we'd like to explore other search algorithms such as reinforcement learning or bayesian optimization.

Another interesting consideration is the theoretical analysis of QCNN architectures that generalise well across multiple data sets. Recently, it has been shown how symmetry can be used to inform the inductive biases of a model \cite{laroccaGroupInvariantQuantumMachine2022, meyerExploitingSymmetryVariational2022}, and we suspect that our numerical results stem from the search finding architectures that respect symmetries of the data. Symmetry is a natural starting point for creating primitives, the convolution primitive is already constrained by translational symmetry and additional primitives can be developed by considering other symmetries. This approach effectively narrows the search space, enabling a system to automatically discover general equivariant architectures that align well with the data. The framework also allows for the specification qubit orderings that correspond to physical hardware setups. Therefore, benchmarking the effect of noise on different architectures on NISQ devices would be a useful exploration.
\newline
\section*{Methods}

\label{sec:experiment_setup}
Figure \ref{fig:qcnn_pipeline} gives a broad view of the machine learning pipeline we implement for the benchmarks. There are various factors influencing model performance during such a pipeline. Each step, from a raw audio signal to a classified musical genre, contains various possible configurations, the influence of which propagates throughout the pipeline. For this reason, it is difficult to isolate any configuration and evaluate its effect on the model. With our goal being to analyse QCNN architectures (Figure \ref{fig:qcnn_pipeline} d) on the audio data, we perform random search in the family created by algorithm \ref{alg:reverse_binary_tree} with different choices of circuit ansatz and quantum data encoding. These are evaluated on two different datasets: Mel spectrogram data  (Figure \ref{fig:qcnn_pipeline} b) and 2D statistical data (Figure \ref{fig:qcnn_pipeline} c), both being derived from the same audio signal (Figure \ref{fig:qcnn_pipeline} a). We preprocess the data based on requirements imposed by the model implementation before encoding it into a quantum state. These configurations are expanded on below:\newline

\subsection*{Data}
\label{ssec:data}
We aimed to use a practical and widely applicable dataset for the data component and chose the well-known \cite{sturmSurveyEvaluationMusic2014} music genre dataset, \text{GTZAN}. It consists of 1000 audio tracks, each being a 30-second recording of some song. These recordings were obtained from radio, compact disks and compressed MP3 audio files \cite{tzanetakis_essl_cook_2001}. Each is given a label of one of the following ten musical genres: (\textbf{blues}, \textbf{classical}, \textbf{country}, \textbf{disco}, \textbf{hip-hop}, \textbf{jazz}, \textbf{metal}, \textbf{pop}, \textbf{reggae}, \textbf{rock}). Binary classification is used for the analysis of model performance across different architectures. Meaning there are ${10\choose2}=45$ possible genre pairs to build models from. Each pair is equally balanced since there are 100 songs for each genre. The dataset enables the comparison of 45 models per configuration within the audio domain.

\subsection*{Model Implementation}
\label{ssec:experiment_model}
\begin{figure}[h!]
    \includegraphics[width=.35\linewidth]{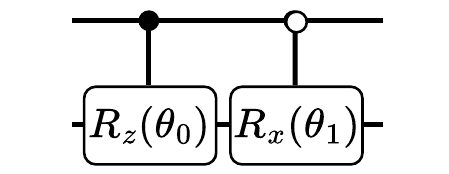}
    \caption{Pooling ansatz from the experiments of \cite{hurQuantumConvolutionalNeural2022}. A rotation is applied on the second qubit based on whether the control is one (filled circle) or zero (open circle).}
    \label{fig:psatz1}
\end{figure}

For all experiments, we evaluate instances of Algorithm \ref{alg:reverse_binary_tree} with $N=8$ qubits, resulting in $3(8)-2=22$ two-qubit unitaries. We test each model based on different combinations of model architecture, two-qubit unitary ansatz and quantum data encoding. The specific unitaries for $U_m$ are chosen from a set of eight ansatzes that were used by \cite{hurQuantumConvolutionalNeural2022}. They are based on previous studies that explore the expressibility and entangling capability of parameterised circuits \cite{Sim_expressibility}, hierarchical quantum classifiers \cite{grant_hierarchical_2018} and extensions to the VQE \cite{PhysRevLett.122.230401}. These are shown in Figure \ref{fig:ansatzes_appendix}, the ansatz for pooling also comes from \cite{hurQuantumConvolutionalNeural2022} and is shown in figure \ref{fig:psatz1}. For quantum data encoding, we compare qubit encoding \cite{schuldSupervisedQuantumMachine2021} with IQP encoding \cite{Havlicek2019} on the tabular dataset. Amplitude encoding \cite{SupervisedQML} is used for the image data.\newline
Each model configuration considers all 45 genre pairs for classification, for example, rock vs reggae. Cross entropy is used as the cost function $C(y,\hat{y})$ during training, for rock vs reggae this would be: 
\begin{flalign}
    C(y,\hat{y}) & =-(y\log(\hat{y})+(1-y)\log(1-\hat{y})). &  & 
\end{flalign}
where
\begin{flalign}
    y_i & =\begin{cases}1 \text{ if song $i$ is labelled rock,} \\
               0 \text{ if song $i$ labelled reggae}.
           \end{cases} &  & 
\end{flalign}
$\hat{y_i}$ is obtained from equation \ref{eq:yhat}, $i$ represents one observation and both $y$, $\hat{y}$ are all the observations in vector form.

\subsection*{Data Creation}
\label{ssec:experiment_data_creation}
We benchmark the model against two different forms of data, namely tabular and image. To construct the dataset in tabular form, we extract specific features from each audio signal using librosa \cite{brian_mcfee_2021_4792298} as shown in Figure \ref{fig:qcnn_pipeline} (b). Each row represents a single audio track with its features as columns. The specific features extracted are those typically used by music information retrieval systems, namely: \textit{chroma frequencies, harmonic and percussive elements, Mel-frequency cepstral coefficients, root-mean-square, spectral bandwidth, spectral centroid, spectral roll-off, tempo and the zero crossing rate}. See Appendix \ref{appendix:features} for a short description of these features. To construct the data set in image form, we extract a Mel frequency spectrogram (Figure \ref{fig:qcnn_pipeline} c) from each audio signal. The Mel scale is a non-linear transformation based on how humans perceive sound and is frequently used in speech recognition applications \cite{davisComparisonParametricRepresentations1980}. The spectrogram size depends on the number of qubits available for the QCNN. We can encode $2^N$ values with amplitude encoding into a quantum state, where $N$ is the number of available qubits. Using $N=8$ qubits, we scale the image to $8\times32=256=2^8$ pixels, normalising each pixel between 0 and 1. The downscaling is done by binning the Mel frequencies into eight groups and taking the first three seconds of each audio signal.
\newline        
\subsection*{Data Preprocessing}
\label{ssec:experiment_data_preprop}
Two primary forms of preprocessing are applied to the data before it is sent to the model: feature scaling and feature selection. The features are scaled using min-max scaling, where the range is based on the type of quantum data encoding used. For amplitude encoding, the data is scaled to the range $[0,1]$, qubit encoding to $[0,\pi/2]$ and IQP encoding to $[0,\pi]$. Feature selection is only applied to the tabular data. Using qubit encoding with $N=8$ qubits result in selecting eight features. Principal Component Analysis (PCA) and decision trees are used to perform the selection. The tree-based selection is used to compare against PCA to verify whether PCA does not heavily bias the model's results.
\newline        
\subsection*{Model Evaluation}
\label{ssec:experiment_model_eval}
The model is trained with $70\%$ of the data while $30\%$ is held out as a test set to evaluate performance. During training, five-fold cross-validation is used on each model. The average classification accuracy and standard deviation of 30 separate trained instances are calculated on the test set as performance metrics.

\bibliography{bibliography}

\begin{thebibliography}{70}%
\makeatletter
\providecommand \@ifxundefined [1]{%
 \@ifx{#1\undefined}
}%
\providecommand \@ifnum [1]{%
 \ifnum #1\expandafter \@firstoftwo
 \else \expandafter \@secondoftwo
 \fi
}%
\providecommand \@ifx [1]{%
 \ifx #1\expandafter \@firstoftwo
 \else \expandafter \@secondoftwo
 \fi
}%
\providecommand \natexlab [1]{#1}%
\providecommand \enquote  [1]{``#1''}%
\providecommand \bibnamefont  [1]{#1}%
\providecommand \bibfnamefont [1]{#1}%
\providecommand \citenamefont [1]{#1}%
\providecommand \href@noop [0]{\@secondoftwo}%
\providecommand \href [0]{\begingroup \@sanitize@url \@href}%
\providecommand \@href[1]{\@@startlink{#1}\@@href}%
\providecommand \@@href[1]{\endgroup#1\@@endlink}%
\providecommand \@sanitize@url [0]{\catcode `\\12\catcode `\$12\catcode
  `\&12\catcode `\#12\catcode `\^12\catcode `\_12\catcode `\%12\relax}%
\providecommand \@@startlink[1]{}%
\providecommand \@@endlink[0]{}%
\providecommand \url  [0]{\begingroup\@sanitize@url \@url }%
\providecommand \@url [1]{\endgroup\@href {#1}{\urlprefix }}%
\providecommand \urlprefix  [0]{URL }%
\providecommand \Eprint [0]{\href }%
\providecommand \doibase [0]{http://dx.doi.org/}%
\providecommand \selectlanguage [0]{\@gobble}%
\providecommand \bibinfo  [0]{\@secondoftwo}%
\providecommand \bibfield  [0]{\@secondoftwo}%
\providecommand \translation [1]{[#1]}%
\providecommand \BibitemOpen [0]{}%
\providecommand \bibitemStop [0]{}%
\providecommand \bibitemNoStop [0]{.\EOS\space}%
\providecommand \EOS [0]{\spacefactor3000\relax}%
\providecommand \BibitemShut  [1]{\csname bibitem#1\endcsname}%
\let\auto@bib@innerbib\@empty
\bibitem [{\citenamefont {Benedetti}\ \emph {et~al.}(2019)\citenamefont
  {Benedetti}, \citenamefont {Lloyd}, \citenamefont {Sack},\ and\ \citenamefont
  {Fiorentini}}]{Benedetti_2019}%
  \BibitemOpen
  \bibfield  {author} {\bibinfo {author} {\bibfnamefont {M.}~\bibnamefont
  {Benedetti}}, \bibinfo {author} {\bibfnamefont {E.}~\bibnamefont {Lloyd}},
  \bibinfo {author} {\bibfnamefont {S.}~\bibnamefont {Sack}}, \ and\ \bibinfo
  {author} {\bibfnamefont {M.}~\bibnamefont {Fiorentini}},\ }\href {\doibase
  10.1088/2058-9565/ab4eb5} {\bibfield  {journal} {\bibinfo  {journal} {Quantum
  Science and Technology}\ }\textbf {\bibinfo {volume} {4}},\ \bibinfo {pages}
  {043001} (\bibinfo {year} {2019})}\BibitemShut {NoStop}%
\bibitem [{\citenamefont {Cerezo}\ \emph {et~al.}(2021)\citenamefont {Cerezo},
  \citenamefont {Arrasmith}, \citenamefont {Babbush}, \citenamefont {Benjamin},
  \citenamefont {Endo}, \citenamefont {Fujii}, \citenamefont {{McClean}},
  \citenamefont {Mitarai}, \citenamefont {Yuan}, \citenamefont {Cincio},\ and\
  \citenamefont {Coles}}]{cerezo2020variational}%
  \BibitemOpen
  \bibfield  {author} {\bibinfo {author} {\bibfnamefont {M.}~\bibnamefont
  {Cerezo}}, \bibinfo {author} {\bibfnamefont {A.}~\bibnamefont {Arrasmith}},
  \bibinfo {author} {\bibfnamefont {R.}~\bibnamefont {Babbush}}, \bibinfo
  {author} {\bibfnamefont {S.~C.}\ \bibnamefont {Benjamin}}, \bibinfo {author}
  {\bibfnamefont {S.}~\bibnamefont {Endo}}, \bibinfo {author} {\bibfnamefont
  {K.}~\bibnamefont {Fujii}}, \bibinfo {author} {\bibfnamefont {J.~R.}\
  \bibnamefont {{McClean}}}, \bibinfo {author} {\bibfnamefont {K.}~\bibnamefont
  {Mitarai}}, \bibinfo {author} {\bibfnamefont {X.}~\bibnamefont {Yuan}},
  \bibinfo {author} {\bibfnamefont {L.}~\bibnamefont {Cincio}}, \ and\ \bibinfo
  {author} {\bibfnamefont {P.~J.}\ \bibnamefont {Coles}},\ }\href {\doibase
  10.1038/s42254-021-00348-9} {\bibfield  {journal} {\bibinfo  {journal}
  {Nature Reviews Physics}\ }\textbf {\bibinfo {volume} {3}},\ \bibinfo {pages}
  {625} (\bibinfo {year} {2021})}\BibitemShut {NoStop}%
\bibitem [{\citenamefont {Mangini}\ \emph {et~al.}(2021)\citenamefont
  {Mangini}, \citenamefont {Tacchino}, \citenamefont {Gerace}, \citenamefont
  {Bajoni},\ and\ \citenamefont {Macchiavello}}]{mangini_quantum_2021}%
  \BibitemOpen
  \bibfield  {author} {\bibinfo {author} {\bibfnamefont {S.}~\bibnamefont
  {Mangini}}, \bibinfo {author} {\bibfnamefont {F.}~\bibnamefont {Tacchino}},
  \bibinfo {author} {\bibfnamefont {D.}~\bibnamefont {Gerace}}, \bibinfo
  {author} {\bibfnamefont {D.}~\bibnamefont {Bajoni}}, \ and\ \bibinfo {author}
  {\bibfnamefont {C.}~\bibnamefont {Macchiavello}},\ }\href {\doibase
  10.1209/0295-5075/134/10002} {\bibfield  {journal} {\bibinfo  {journal} {EPL
  (Europhysics Letters)}\ }\textbf {\bibinfo {volume} {134}},\ \bibinfo {pages}
  {10002} (\bibinfo {year} {2021})}\BibitemShut {NoStop}%
\bibitem [{\citenamefont {Bharti}\ \emph {et~al.}(2022)\citenamefont {Bharti},
  \citenamefont {Cervera-Lierta}, \citenamefont {Kyaw}, \citenamefont {Haug},
  \citenamefont {Alperin-Lea}, \citenamefont {Anand}, \citenamefont {Degroote},
  \citenamefont {Heimonen}, \citenamefont {Kottmann}, \citenamefont {Menke},
  \citenamefont {Mok}, \citenamefont {Sim}, \citenamefont {Kwek},\ and\
  \citenamefont {Aspuru-Guzik}}]{RevModPhys.94.015004}%
  \BibitemOpen
  \bibfield  {author} {\bibinfo {author} {\bibfnamefont {K.}~\bibnamefont
  {Bharti}}, \bibinfo {author} {\bibfnamefont {A.}~\bibnamefont
  {Cervera-Lierta}}, \bibinfo {author} {\bibfnamefont {T.~H.}\ \bibnamefont
  {Kyaw}}, \bibinfo {author} {\bibfnamefont {T.}~\bibnamefont {Haug}}, \bibinfo
  {author} {\bibfnamefont {S.}~\bibnamefont {Alperin-Lea}}, \bibinfo {author}
  {\bibfnamefont {A.}~\bibnamefont {Anand}}, \bibinfo {author} {\bibfnamefont
  {M.}~\bibnamefont {Degroote}}, \bibinfo {author} {\bibfnamefont
  {H.}~\bibnamefont {Heimonen}}, \bibinfo {author} {\bibfnamefont {J.~S.}\
  \bibnamefont {Kottmann}}, \bibinfo {author} {\bibfnamefont {T.}~\bibnamefont
  {Menke}}, \bibinfo {author} {\bibfnamefont {W.-K.}\ \bibnamefont {Mok}},
  \bibinfo {author} {\bibfnamefont {S.}~\bibnamefont {Sim}}, \bibinfo {author}
  {\bibfnamefont {L.-C.}\ \bibnamefont {Kwek}}, \ and\ \bibinfo {author}
  {\bibfnamefont {A.}~\bibnamefont {Aspuru-Guzik}},\ }\href {\doibase
  10.1103/RevModPhys.94.015004} {\bibfield  {journal} {\bibinfo  {journal}
  {Rev. Mod. Phys.}\ }\textbf {\bibinfo {volume} {94}},\ \bibinfo {pages}
  {015004} (\bibinfo {year} {2022})}\BibitemShut {NoStop}%
\bibitem [{\citenamefont {Cong}\ \emph {et~al.}(2019)\citenamefont {Cong},
  \citenamefont {Choi},\ and\ \citenamefont {Lukin}}]{cong_quantum_2019}%
  \BibitemOpen
  \bibfield  {author} {\bibinfo {author} {\bibfnamefont {I.}~\bibnamefont
  {Cong}}, \bibinfo {author} {\bibfnamefont {S.}~\bibnamefont {Choi}}, \ and\
  \bibinfo {author} {\bibfnamefont {M.~D.}\ \bibnamefont {Lukin}},\ }\href
  {\doibase 10.1038/s41567-019-0648-8} {\bibfield  {journal} {\bibinfo
  {journal} {Nature Physics}\ }\textbf {\bibinfo {volume} {15}},\ \bibinfo
  {pages} {1273} (\bibinfo {year} {2019})}\BibitemShut {NoStop}%
\bibitem [{\citenamefont {Pesah}\ \emph {et~al.}(2021)\citenamefont {Pesah},
  \citenamefont {Cerezo}, \citenamefont {Wang}, \citenamefont {Volkoff},
  \citenamefont {Sornborger},\ and\ \citenamefont {Coles}}]{pesah2020absence}%
  \BibitemOpen
  \bibfield  {author} {\bibinfo {author} {\bibfnamefont {A.}~\bibnamefont
  {Pesah}}, \bibinfo {author} {\bibfnamefont {M.}~\bibnamefont {Cerezo}},
  \bibinfo {author} {\bibfnamefont {S.}~\bibnamefont {Wang}}, \bibinfo {author}
  {\bibfnamefont {T.}~\bibnamefont {Volkoff}}, \bibinfo {author} {\bibfnamefont
  {A.~T.}\ \bibnamefont {Sornborger}}, \ and\ \bibinfo {author} {\bibfnamefont
  {P.~J.}\ \bibnamefont {Coles}},\ }\href {\doibase 10.1103/PhysRevX.11.041011}
  {\bibfield  {journal} {\bibinfo  {journal} {Phys. Rev. X}\ }\textbf {\bibinfo
  {volume} {11}},\ \bibinfo {pages} {041011} (\bibinfo {year}
  {2021})}\BibitemShut {NoStop}%
\bibitem [{\citenamefont {Banchi}\ \emph {et~al.}(2021)\citenamefont {Banchi},
  \citenamefont {Pereira},\ and\ \citenamefont
  {Pirandola}}]{PRXQuantum.2.040321}%
  \BibitemOpen
  \bibfield  {author} {\bibinfo {author} {\bibfnamefont {L.}~\bibnamefont
  {Banchi}}, \bibinfo {author} {\bibfnamefont {J.}~\bibnamefont {Pereira}}, \
  and\ \bibinfo {author} {\bibfnamefont {S.}~\bibnamefont {Pirandola}},\ }\href
  {\doibase 10.1103/PRXQuantum.2.040321} {\bibfield  {journal} {\bibinfo
  {journal} {PRX Quantum}\ }\textbf {\bibinfo {volume} {2}},\ \bibinfo {pages}
  {040321} (\bibinfo {year} {2021})}\BibitemShut {NoStop}%
\bibitem [{\citenamefont {Herrmann}\ \emph {et~al.}()\citenamefont {Herrmann},
  \citenamefont {Llima}, \citenamefont {Remm}, \citenamefont {Zapletal},
  \citenamefont {McMahon}, \citenamefont {Scarato}, \citenamefont {Swiadek},
  \citenamefont {Andersen}, \citenamefont {Hellings}, \citenamefont {Krinner},
  \citenamefont {Lacroix}, \citenamefont {Lazar}, \citenamefont {Kerschbaum},
  \citenamefont {Zanuz}, \citenamefont {Norris}, \citenamefont {Hartmann},
  \citenamefont {Wallraff},\ and\ \citenamefont
  {Eichler}}]{herrmannRealizingQuantumConvolutional2022}%
  \BibitemOpen
  \bibfield  {author} {\bibinfo {author} {\bibfnamefont {J.}~\bibnamefont
  {Herrmann}}, \bibinfo {author} {\bibfnamefont {S.~M.}\ \bibnamefont {Llima}},
  \bibinfo {author} {\bibfnamefont {A.}~\bibnamefont {Remm}}, \bibinfo {author}
  {\bibfnamefont {P.}~\bibnamefont {Zapletal}}, \bibinfo {author}
  {\bibfnamefont {N.~A.}\ \bibnamefont {McMahon}}, \bibinfo {author}
  {\bibfnamefont {C.}~\bibnamefont {Scarato}}, \bibinfo {author} {\bibfnamefont
  {F.}~\bibnamefont {Swiadek}}, \bibinfo {author} {\bibfnamefont {C.~K.}\
  \bibnamefont {Andersen}}, \bibinfo {author} {\bibfnamefont {C.}~\bibnamefont
  {Hellings}}, \bibinfo {author} {\bibfnamefont {S.}~\bibnamefont {Krinner}},
  \bibinfo {author} {\bibfnamefont {N.}~\bibnamefont {Lacroix}}, \bibinfo
  {author} {\bibfnamefont {S.}~\bibnamefont {Lazar}}, \bibinfo {author}
  {\bibfnamefont {M.}~\bibnamefont {Kerschbaum}}, \bibinfo {author}
  {\bibfnamefont {D.~C.}\ \bibnamefont {Zanuz}}, \bibinfo {author}
  {\bibfnamefont {G.~J.}\ \bibnamefont {Norris}}, \bibinfo {author}
  {\bibfnamefont {M.~J.}\ \bibnamefont {Hartmann}}, \bibinfo {author}
  {\bibfnamefont {A.}~\bibnamefont {Wallraff}}, \ and\ \bibinfo {author}
  {\bibfnamefont {C.}~\bibnamefont {Eichler}},\ }\href {\doibase
  10.1038/s41467-022-31679-5} {\ \textbf {\bibinfo {volume} {13}},\ \bibinfo
  {pages} {4144}}\BibitemShut {NoStop}%
\bibitem [{\citenamefont {Carleo}\ and\ \citenamefont
  {Troyer}(2017)}]{carleoSolvingQuantumManybody2017}%
  \BibitemOpen
  \bibfield  {author} {\bibinfo {author} {\bibfnamefont {G.}~\bibnamefont
  {Carleo}}\ and\ \bibinfo {author} {\bibfnamefont {M.}~\bibnamefont
  {Troyer}},\ }\href {\doibase 10.1126/science.aag2302} {\bibfield  {journal}
  {\bibinfo  {journal} {Science}\ }\textbf {\bibinfo {volume} {355}},\ \bibinfo
  {pages} {602} (\bibinfo {year} {2017})},\ \bibinfo {note} {publisher:
  American Association for the Advancement of Science}\BibitemShut {NoStop}%
\bibitem [{\citenamefont {Carrasquilla}\ and\ \citenamefont
  {Melko}(2017)}]{carrasquillaMachineLearningPhases2017}%
  \BibitemOpen
  \bibfield  {author} {\bibinfo {author} {\bibfnamefont {J.}~\bibnamefont
  {Carrasquilla}}\ and\ \bibinfo {author} {\bibfnamefont {R.~G.}\ \bibnamefont
  {Melko}},\ }\href {\doibase 10.1038/nphys4035} {\bibfield  {journal}
  {\bibinfo  {journal} {Nature Physics}\ }\textbf {\bibinfo {volume} {13}},\
  \bibinfo {pages} {431} (\bibinfo {year} {2017})}\BibitemShut {NoStop}%
\bibitem [{\citenamefont {van Nieuwenburg}\ \emph {et~al.}(2017)\citenamefont
  {van Nieuwenburg}, \citenamefont {Liu},\ and\ \citenamefont
  {Huber}}]{vannieuwenburgLearningPhaseTransitions2017}%
  \BibitemOpen
  \bibfield  {author} {\bibinfo {author} {\bibfnamefont {E.~P.~L.}\
  \bibnamefont {van Nieuwenburg}}, \bibinfo {author} {\bibfnamefont {Y.-H.}\
  \bibnamefont {Liu}}, \ and\ \bibinfo {author} {\bibfnamefont {S.~D.}\
  \bibnamefont {Huber}},\ }\href {\doibase 10.1038/nphys4037} {\bibfield
  {journal} {\bibinfo  {journal} {Nature Physics}\ }\textbf {\bibinfo {volume}
  {13}},\ \bibinfo {pages} {435} (\bibinfo {year} {2017})}\BibitemShut
  {NoStop}%
\bibitem [{\citenamefont {Deng}\ \emph
  {et~al.}(2017{\natexlab{a}})\citenamefont {Deng}, \citenamefont {Li},\ and\
  \citenamefont {Sarma}}]{dengMachineLearningTopological2017}%
  \BibitemOpen
  \bibfield  {author} {\bibinfo {author} {\bibfnamefont {D.-L.}\ \bibnamefont
  {Deng}}, \bibinfo {author} {\bibfnamefont {X.}~\bibnamefont {Li}}, \ and\
  \bibinfo {author} {\bibfnamefont {S.~D.}\ \bibnamefont {Sarma}},\ }\href
  {\doibase 10.1103/PhysRevB.96.195145} {\bibfield  {journal} {\bibinfo
  {journal} {Physical Review B}\ }\textbf {\bibinfo {volume} {96}},\ \bibinfo
  {pages} {195145} (\bibinfo {year} {2017}{\natexlab{a}})},\ \Eprint
  {http://arxiv.org/abs/1609.09060} {arXiv:1609.09060 [cond-mat,
  physics:quant-ph]} \BibitemShut {NoStop}%
\bibitem [{\citenamefont {Levine}\ \emph {et~al.}(2019)\citenamefont {Levine},
  \citenamefont {Sharir}, \citenamefont {Cohen},\ and\ \citenamefont
  {Shashua}}]{levineQuantumEntanglementDeep2019}%
  \BibitemOpen
  \bibfield  {author} {\bibinfo {author} {\bibfnamefont {Y.}~\bibnamefont
  {Levine}}, \bibinfo {author} {\bibfnamefont {O.}~\bibnamefont {Sharir}},
  \bibinfo {author} {\bibfnamefont {N.}~\bibnamefont {Cohen}}, \ and\ \bibinfo
  {author} {\bibfnamefont {A.}~\bibnamefont {Shashua}},\ }\href {\doibase
  10.1103/PhysRevLett.122.065301} {\bibfield  {journal} {\bibinfo  {journal}
  {Physical Review Letters}\ }\textbf {\bibinfo {volume} {122}},\ \bibinfo
  {pages} {065301} (\bibinfo {year} {2019})},\ \Eprint
  {http://arxiv.org/abs/1803.09780} {arXiv:1803.09780 [quant-ph]} \BibitemShut
  {NoStop}%
\bibitem [{\citenamefont {Stoudenmire}\ and\ \citenamefont
  {Schwab}(2016)}]{stoudenmireSupervisedLearningTensor2016}%
  \BibitemOpen
  \bibfield  {author} {\bibinfo {author} {\bibfnamefont {E.}~\bibnamefont
  {Stoudenmire}}\ and\ \bibinfo {author} {\bibfnamefont {D.~J.}\ \bibnamefont
  {Schwab}},\ }in\ \href
  {https://papers.nips.cc/paper/2016/hash/5314b9674c86e3f9d1ba25ef9bb32895-Abstract.html}
  {\emph {\bibinfo {booktitle} {Advances in Neural Information Processing
  Systems}}},\ Vol.~\bibinfo {volume} {29}\ (\bibinfo  {publisher} {Curran
  Associates, Inc.},\ \bibinfo {year} {2016})\BibitemShut {NoStop}%
\bibitem [{\citenamefont {Deng}\ \emph
  {et~al.}(2017{\natexlab{b}})\citenamefont {Deng}, \citenamefont {Li},\ and\
  \citenamefont {Sarma}}]{dengQuantumEntanglementNeural2017}%
  \BibitemOpen
  \bibfield  {author} {\bibinfo {author} {\bibfnamefont {D.-L.}\ \bibnamefont
  {Deng}}, \bibinfo {author} {\bibfnamefont {X.}~\bibnamefont {Li}}, \ and\
  \bibinfo {author} {\bibfnamefont {S.~D.}\ \bibnamefont {Sarma}},\ }\href
  {\doibase 10.1103/PhysRevX.7.021021} {\bibfield  {journal} {\bibinfo
  {journal} {Physical Review X}\ }\textbf {\bibinfo {volume} {7}},\ \bibinfo
  {pages} {021021} (\bibinfo {year} {2017}{\natexlab{b}})},\ \Eprint
  {http://arxiv.org/abs/1701.04844} {arXiv:1701.04844 [cond-mat,
  physics:quant-ph]} \BibitemShut {NoStop}%
\bibitem [{\citenamefont {Lin}\ \emph {et~al.}(2017)\citenamefont {Lin},
  \citenamefont {Tegmark},\ and\ \citenamefont {Rolnick}}]{linWhyDoesDeep2017}%
  \BibitemOpen
  \bibfield  {author} {\bibinfo {author} {\bibfnamefont {H.~W.}\ \bibnamefont
  {Lin}}, \bibinfo {author} {\bibfnamefont {M.}~\bibnamefont {Tegmark}}, \ and\
  \bibinfo {author} {\bibfnamefont {D.}~\bibnamefont {Rolnick}},\ }\href
  {\doibase 10.1007/s10955-017-1836-5} {\bibfield  {journal} {\bibinfo
  {journal} {Journal of Statistical Physics}\ }\textbf {\bibinfo {volume}
  {168}},\ \bibinfo {pages} {1223} (\bibinfo {year} {2017})},\ \Eprint
  {http://arxiv.org/abs/1608.08225} {arXiv:1608.08225 [cond-mat, stat]}
  \BibitemShut {NoStop}%
\bibitem [{\citenamefont {Mehta}\ and\ \citenamefont
  {Schwab}(2014)}]{mehtaExactMappingVariational2014}%
  \BibitemOpen
  \bibfield  {author} {\bibinfo {author} {\bibfnamefont {P.}~\bibnamefont
  {Mehta}}\ and\ \bibinfo {author} {\bibfnamefont {D.~J.}\ \bibnamefont
  {Schwab}},\ }\href {\doibase 10.48550/arXiv.1410.3831} {\bibfield  {journal}
  {\bibinfo  {journal} {arXiv:1410.3831[cond-mat]}\ } (\bibinfo {year}
  {2014}),\ 10.48550/arXiv.1410.3831},\ \Eprint
  {http://arxiv.org/abs/1410.3831} {arXiv:1410.3831 [cond-mat, stat]}
  \BibitemShut {NoStop}%
\bibitem [{\citenamefont {Levine}\ \emph {et~al.}(2018)\citenamefont {Levine},
  \citenamefont {Yakira}, \citenamefont {Cohen},\ and\ \citenamefont
  {Shashua}}]{levineDeepLearningQuantum2018}%
  \BibitemOpen
  \bibfield  {author} {\bibinfo {author} {\bibfnamefont {Y.}~\bibnamefont
  {Levine}}, \bibinfo {author} {\bibfnamefont {D.}~\bibnamefont {Yakira}},
  \bibinfo {author} {\bibfnamefont {N.}~\bibnamefont {Cohen}}, \ and\ \bibinfo
  {author} {\bibfnamefont {A.}~\bibnamefont {Shashua}},\ }in\ \href
  {https://openreview.net/forum?id=SywXXwJAb} {\emph {\bibinfo {booktitle}
  {International Conference on Learning Representations}}}\ (\bibinfo {year}
  {2018})\BibitemShut {NoStop}%
\bibitem [{\citenamefont {LeCun}\ \emph {et~al.}(2015)\citenamefont {LeCun},
  \citenamefont {Bengio},\ and\ \citenamefont {Hinton}}]{lecun_deep_2015}%
  \BibitemOpen
  \bibfield  {author} {\bibinfo {author} {\bibfnamefont {Y.}~\bibnamefont
  {LeCun}}, \bibinfo {author} {\bibfnamefont {Y.}~\bibnamefont {Bengio}}, \
  and\ \bibinfo {author} {\bibfnamefont {G.}~\bibnamefont {Hinton}},\ }\href
  {\doibase 10.1038/nature14539} {\bibfield  {journal} {\bibinfo  {journal}
  {Nature}\ }\textbf {\bibinfo {volume} {521}},\ \bibinfo {pages} {436}
  (\bibinfo {year} {2015})}\BibitemShut {NoStop}%
\bibitem [{\citenamefont {Krizhevsky}\ \emph {et~al.}(2012)\citenamefont
  {Krizhevsky}, \citenamefont {Sutskever},\ and\ \citenamefont
  {Hinton}}]{krizhevskyImageNetClassificationDeep2012}%
  \BibitemOpen
  \bibfield  {author} {\bibinfo {author} {\bibfnamefont {A.}~\bibnamefont
  {Krizhevsky}}, \bibinfo {author} {\bibfnamefont {I.}~\bibnamefont
  {Sutskever}}, \ and\ \bibinfo {author} {\bibfnamefont {G.~E.}\ \bibnamefont
  {Hinton}},\ }in\ \href
  {https://papers.nips.cc/paper/2012/hash/c399862d3b9d6b76c8436e924a68c45b-Abstract.html}
  {\emph {\bibinfo {booktitle} {Advances in Neural Information Processing
  Systems}}},\ Vol.~\bibinfo {volume} {25}\ (\bibinfo  {publisher} {Curran
  Associates, Inc.},\ \bibinfo {year} {2012})\BibitemShut {NoStop}%
\bibitem [{\citenamefont {Zoph}\ and\ \citenamefont
  {Le}(2017)}]{zophNeuralArchitectureSearch2017}%
  \BibitemOpen
  \bibfield  {author} {\bibinfo {author} {\bibfnamefont {B.}~\bibnamefont
  {Zoph}}\ and\ \bibinfo {author} {\bibfnamefont {Q.~V.}\ \bibnamefont {Le}},\
  }in\ \href {https://openreview.net/forum?id=r1Ue8Hcxg} {\emph {\bibinfo
  {booktitle} {International Conference on Learning Representations}}}\
  (\bibinfo {year} {2017})\BibitemShut {NoStop}%
\bibitem [{\citenamefont {Elsken}\ \emph {et~al.}(2019)\citenamefont {Elsken},
  \citenamefont {Metzen},\ and\ \citenamefont
  {Hutter}}]{elskenNeuralArchitectureSearch2019}%
  \BibitemOpen
  \bibfield  {author} {\bibinfo {author} {\bibfnamefont {T.}~\bibnamefont
  {Elsken}}, \bibinfo {author} {\bibfnamefont {J.~H.}\ \bibnamefont {Metzen}},
  \ and\ \bibinfo {author} {\bibfnamefont {F.}~\bibnamefont {Hutter}},\
  }\href@noop {} {\bibfield  {journal} {\bibinfo  {journal} {The Journal of
  Machine Learning Research}\ }\textbf {\bibinfo {volume} {20}},\ \bibinfo
  {pages} {1997} (\bibinfo {year} {2019})}\BibitemShut {NoStop}%
\bibitem [{\citenamefont {Real}\ \emph {et~al.}(2019)\citenamefont {Real},
  \citenamefont {Aggarwal}, \citenamefont {Huang},\ and\ \citenamefont
  {Le}}]{realRegularizedEvolutionImage2019a}%
  \BibitemOpen
  \bibfield  {author} {\bibinfo {author} {\bibfnamefont {E.}~\bibnamefont
  {Real}}, \bibinfo {author} {\bibfnamefont {A.}~\bibnamefont {Aggarwal}},
  \bibinfo {author} {\bibfnamefont {Y.}~\bibnamefont {Huang}}, \ and\ \bibinfo
  {author} {\bibfnamefont {Q.~V.}\ \bibnamefont {Le}},\ }\href {\doibase
  10.1609/aaai.v33i01.33014780} {\bibfield  {journal} {\bibinfo  {journal}
  {Proceedings of the AAAI Conference on Artificial Intelligence}\ }\textbf
  {\bibinfo {volume} {33}},\ \bibinfo {pages} {4780} (\bibinfo {year}
  {2019})}\BibitemShut {NoStop}%
\bibitem [{\citenamefont {Zoph}\ \emph {et~al.}(2018)\citenamefont {Zoph},
  \citenamefont {Vasudevan}, \citenamefont {Shlens},\ and\ \citenamefont
  {Le}}]{zophLearningTransferableArchitectures2018}%
  \BibitemOpen
  \bibfield  {author} {\bibinfo {author} {\bibfnamefont {B.}~\bibnamefont
  {Zoph}}, \bibinfo {author} {\bibfnamefont {V.}~\bibnamefont {Vasudevan}},
  \bibinfo {author} {\bibfnamefont {J.}~\bibnamefont {Shlens}}, \ and\ \bibinfo
  {author} {\bibfnamefont {Q.~V.}\ \bibnamefont {Le}},\ }in\ \href@noop {}
  {\emph {\bibinfo {booktitle} {Proceedings of the IEEE Conference on Computer
  Vision and Pattern Recognition}}}\ (\bibinfo {year} {2018})\ pp.\ \bibinfo
  {pages} {8697--8710}\BibitemShut {NoStop}%
\bibitem [{\citenamefont {Chen}\ \emph {et~al.}(2018)\citenamefont {Chen},
  \citenamefont {Collins}, \citenamefont {Zhu}, \citenamefont {Papandreou},
  \citenamefont {Zoph}, \citenamefont {Schroff}, \citenamefont {Adam},\ and\
  \citenamefont {Shlens}}]{chenSearchingEfficientMultiScale2018}%
  \BibitemOpen
  \bibfield  {author} {\bibinfo {author} {\bibfnamefont {L.-C.}\ \bibnamefont
  {Chen}}, \bibinfo {author} {\bibfnamefont {M.}~\bibnamefont {Collins}},
  \bibinfo {author} {\bibfnamefont {Y.}~\bibnamefont {Zhu}}, \bibinfo {author}
  {\bibfnamefont {G.}~\bibnamefont {Papandreou}}, \bibinfo {author}
  {\bibfnamefont {B.}~\bibnamefont {Zoph}}, \bibinfo {author} {\bibfnamefont
  {F.}~\bibnamefont {Schroff}}, \bibinfo {author} {\bibfnamefont
  {H.}~\bibnamefont {Adam}}, \ and\ \bibinfo {author} {\bibfnamefont
  {J.}~\bibnamefont {Shlens}},\ }in\ \href
  {https://proceedings.neurips.cc/paper/2018/hash/c90070e1f03e982448983975a0f52d57-Abstract.html}
  {\emph {\bibinfo {booktitle} {Advances in Neural Information Processing
  Systems}}},\ Vol.~\bibinfo {volume} {31}\ (\bibinfo  {publisher} {Curran
  Associates, Inc.},\ \bibinfo {year} {2018})\BibitemShut {NoStop}%
\bibitem [{\citenamefont {Liu}\ \emph {et~al.}(2018)\citenamefont {Liu},
  \citenamefont {Simonyan}, \citenamefont {Vinyals}, \citenamefont {Fernando},\
  and\ \citenamefont
  {Kavukcuoglu}}]{liuHierarchicalRepresentationsEfficient2018a}%
  \BibitemOpen
  \bibfield  {author} {\bibinfo {author} {\bibfnamefont {H.}~\bibnamefont
  {Liu}}, \bibinfo {author} {\bibfnamefont {K.}~\bibnamefont {Simonyan}},
  \bibinfo {author} {\bibfnamefont {O.}~\bibnamefont {Vinyals}}, \bibinfo
  {author} {\bibfnamefont {C.}~\bibnamefont {Fernando}}, \ and\ \bibinfo
  {author} {\bibfnamefont {K.}~\bibnamefont {Kavukcuoglu}},\ }in\ \href
  {https://openreview.net/forum?id=BJQRKzbA-} {\emph {\bibinfo {booktitle}
  {International Conference on Learning Representations}}}\ (\bibinfo {year}
  {2018})\BibitemShut {NoStop}%
\bibitem [{\citenamefont {Grant}\ \emph {et~al.}(2018)\citenamefont {Grant},
  \citenamefont {Benedetti}, \citenamefont {Cao}, \citenamefont {Hallam},
  \citenamefont {Lockhart}, \citenamefont {Stojevic}, \citenamefont {Green},\
  and\ \citenamefont {Severini}}]{grant_hierarchical_2018}%
  \BibitemOpen
  \bibfield  {author} {\bibinfo {author} {\bibfnamefont {E.}~\bibnamefont
  {Grant}}, \bibinfo {author} {\bibfnamefont {M.}~\bibnamefont {Benedetti}},
  \bibinfo {author} {\bibfnamefont {S.}~\bibnamefont {Cao}}, \bibinfo {author}
  {\bibfnamefont {A.}~\bibnamefont {Hallam}}, \bibinfo {author} {\bibfnamefont
  {J.}~\bibnamefont {Lockhart}}, \bibinfo {author} {\bibfnamefont
  {V.}~\bibnamefont {Stojevic}}, \bibinfo {author} {\bibfnamefont {A.~G.}\
  \bibnamefont {Green}}, \ and\ \bibinfo {author} {\bibfnamefont
  {S.}~\bibnamefont {Severini}},\ }\href {\doibase 10.1038/s41534-018-0116-9}
  {\bibfield  {journal} {\bibinfo  {journal} {npj Quantum Information}\
  }\textbf {\bibinfo {volume} {4}},\ \bibinfo {pages} {65} (\bibinfo {year}
  {2018})}\BibitemShut {NoStop}%
\bibitem [{\citenamefont {Haug}\ \emph {et~al.}(2022)\citenamefont {Haug},
  \citenamefont {Bharti},\ and\ \citenamefont
  {Kim}}]{haugCapacityQuantumGeometry2021}%
  \BibitemOpen
  \bibfield  {author} {\bibinfo {author} {\bibfnamefont {T.}~\bibnamefont
  {Haug}}, \bibinfo {author} {\bibfnamefont {K.}~\bibnamefont {Bharti}}, \ and\
  \bibinfo {author} {\bibfnamefont {M.}~\bibnamefont {Kim}},\ }\href {\doibase
  10.1103/PRXQuantum.2.040309} {\bibfield  {journal} {\bibinfo  {journal}
  {{PRX} Quantum}\ }\textbf {\bibinfo {volume} {2}},\ \bibinfo {pages} {040309}
  (\bibinfo {year} {2022})},\ \bibinfo {note} {publisher: American Physical
  Society}\BibitemShut {NoStop}%
\bibitem [{\citenamefont {Hur}\ \emph {et~al.}(2022)\citenamefont {Hur},
  \citenamefont {Kim},\ and\ \citenamefont
  {Park}}]{hurQuantumConvolutionalNeural2022}%
  \BibitemOpen
  \bibfield  {author} {\bibinfo {author} {\bibfnamefont {T.}~\bibnamefont
  {Hur}}, \bibinfo {author} {\bibfnamefont {L.}~\bibnamefont {Kim}}, \ and\
  \bibinfo {author} {\bibfnamefont {D.~K.}\ \bibnamefont {Park}},\ }\href
  {\doibase 10.1007/s42484-021-00061-x} {\bibfield  {journal} {\bibinfo
  {journal} {Quantum Machine Intelligence}\ }\textbf {\bibinfo {volume} {4}},\
  \bibinfo {pages} {3} (\bibinfo {year} {2022})}\BibitemShut {NoStop}%
\bibitem [{\citenamefont {Oh}\ \emph {et~al.}(2020)\citenamefont {Oh},
  \citenamefont {Choi},\ and\ \citenamefont
  {Kim}}]{ohTutorialQuantumConvolutional2020a}%
  \BibitemOpen
  \bibfield  {author} {\bibinfo {author} {\bibfnamefont {S.}~\bibnamefont
  {Oh}}, \bibinfo {author} {\bibfnamefont {J.}~\bibnamefont {Choi}}, \ and\
  \bibinfo {author} {\bibfnamefont {J.}~\bibnamefont {Kim}},\ }in\ \href
  {\doibase 10.1109/ICTC49870.2020.9289439} {\emph {\bibinfo {booktitle} {2020
  International Conference on Information and Communication Technology
  Convergence (ICTC)}}}\ (\bibinfo {year} {2020})\ pp.\ \bibinfo {pages}
  {236--239}\BibitemShut {NoStop}%
\bibitem [{\citenamefont {Franken}\ and\ \citenamefont
  {Georgiev}(2020)}]{frankenExplorationsQuantumNeural2020}%
  \BibitemOpen
  \bibfield  {author} {\bibinfo {author} {\bibfnamefont {L.}~\bibnamefont
  {Franken}}\ and\ \bibinfo {author} {\bibfnamefont {B.}~\bibnamefont
  {Georgiev}},\ }in\ \href@noop {} {\emph {\bibinfo {booktitle} {ESANN}}}\
  (\bibinfo {year} {2020})\ pp.\ \bibinfo {pages} {297--302}\BibitemShut
  {NoStop}%
\bibitem [{\citenamefont {McClean}\ \emph {et~al.}(2018)\citenamefont
  {McClean}, \citenamefont {Boixo}, \citenamefont {Smelyanskiy}, \citenamefont
  {Babbush},\ and\ \citenamefont {Neven}}]{mccleanBarrenPlateausQuantum2018a}%
  \BibitemOpen
  \bibfield  {author} {\bibinfo {author} {\bibfnamefont {J.~R.}\ \bibnamefont
  {McClean}}, \bibinfo {author} {\bibfnamefont {S.}~\bibnamefont {Boixo}},
  \bibinfo {author} {\bibfnamefont {V.~N.}\ \bibnamefont {Smelyanskiy}},
  \bibinfo {author} {\bibfnamefont {R.}~\bibnamefont {Babbush}}, \ and\
  \bibinfo {author} {\bibfnamefont {H.}~\bibnamefont {Neven}},\ }\href
  {\doibase 10.1038/s41467-018-07090-4} {\bibfield  {journal} {\bibinfo
  {journal} {Nature Communications}\ }\textbf {\bibinfo {volume} {9}},\
  \bibinfo {pages} {4812} (\bibinfo {year} {2018})}\BibitemShut {NoStop}%
\bibitem [{\citenamefont {Holmes}\ \emph {et~al.}(2022)\citenamefont {Holmes},
  \citenamefont {Sharma}, \citenamefont {Cerezo},\ and\ \citenamefont
  {Coles}}]{holmesConnectingAnsatzExpressibility2022}%
  \BibitemOpen
  \bibfield  {author} {\bibinfo {author} {\bibfnamefont {Z.}~\bibnamefont
  {Holmes}}, \bibinfo {author} {\bibfnamefont {K.}~\bibnamefont {Sharma}},
  \bibinfo {author} {\bibfnamefont {M.}~\bibnamefont {Cerezo}}, \ and\ \bibinfo
  {author} {\bibfnamefont {P.~J.}\ \bibnamefont {Coles}},\ }\href {\doibase
  10.1103/PRXQuantum.3.010313} {\bibfield  {journal} {\bibinfo  {journal} {PRX
  Quantum}\ }\textbf {\bibinfo {volume} {3}},\ \bibinfo {pages} {010313}
  (\bibinfo {year} {2022})}\BibitemShut {NoStop}%
\bibitem [{\citenamefont {Schuld}\ \emph {et~al.}(2021)\citenamefont {Schuld},
  \citenamefont {Sweke},\ and\ \citenamefont
  {Meyer}}]{schuldEffectDataEncoding2021}%
  \BibitemOpen
  \bibfield  {author} {\bibinfo {author} {\bibfnamefont {M.}~\bibnamefont
  {Schuld}}, \bibinfo {author} {\bibfnamefont {R.}~\bibnamefont {Sweke}}, \
  and\ \bibinfo {author} {\bibfnamefont {J.~J.}\ \bibnamefont {Meyer}},\ }\href
  {\doibase 10.1103/PhysRevA.103.032430} {\bibfield  {journal} {\bibinfo
  {journal} {Physical Review A}\ }\textbf {\bibinfo {volume} {103}},\ \bibinfo
  {pages} {032430} (\bibinfo {year} {2021})}\BibitemShut {NoStop}%
\bibitem [{\citenamefont {Abbas}\ \emph {et~al.}(2021)\citenamefont {Abbas},
  \citenamefont {Sutter}, \citenamefont {Zoufal}, \citenamefont {Lucchi},
  \citenamefont {Figalli},\ and\ \citenamefont
  {Woerner}}]{abbasPowerQuantumNeural2021}%
  \BibitemOpen
  \bibfield  {author} {\bibinfo {author} {\bibfnamefont {A.}~\bibnamefont
  {Abbas}}, \bibinfo {author} {\bibfnamefont {D.}~\bibnamefont {Sutter}},
  \bibinfo {author} {\bibfnamefont {C.}~\bibnamefont {Zoufal}}, \bibinfo
  {author} {\bibfnamefont {A.}~\bibnamefont {Lucchi}}, \bibinfo {author}
  {\bibfnamefont {A.}~\bibnamefont {Figalli}}, \ and\ \bibinfo {author}
  {\bibfnamefont {S.}~\bibnamefont {Woerner}},\ }\href {\doibase
  10.1038/s43588-021-00084-1} {\bibfield  {journal} {\bibinfo  {journal}
  {Nature Computational Science}\ }\textbf {\bibinfo {volume} {1}},\ \bibinfo
  {pages} {403} (\bibinfo {year} {2021})},\ \Eprint
  {http://arxiv.org/abs/2011.00027} {arXiv:2011.00027 [quant-ph]} \BibitemShut
  {NoStop}%
\bibitem [{\citenamefont {Schuld}(2021)}]{schuldSupervisedQuantumMachine2021}%
  \BibitemOpen
  \bibfield  {author} {\bibinfo {author} {\bibfnamefont {M.}~\bibnamefont
  {Schuld}},\ }\href {http://arxiv.org/abs/2101.11020} {\bibfield  {journal}
  {\bibinfo  {journal} {arxiv:2101.11020[quant-ph]}\ } (\bibinfo {year}
  {2021})},\ \Eprint {http://arxiv.org/abs/2101.11020} {arXiv:2101.11020
  [quant-ph, stat]} \BibitemShut {NoStop}%
\bibitem [{\citenamefont {Zhang}\ \emph {et~al.}(2022)\citenamefont {Zhang},
  \citenamefont {Hsieh}, \citenamefont {Zhang},\ and\ \citenamefont
  {Yao}}]{zhangDifferentiableQuantumArchitecture2021}%
  \BibitemOpen
  \bibfield  {author} {\bibinfo {author} {\bibfnamefont {S.-X.}\ \bibnamefont
  {Zhang}}, \bibinfo {author} {\bibfnamefont {C.-Y.}\ \bibnamefont {Hsieh}},
  \bibinfo {author} {\bibfnamefont {S.}~\bibnamefont {Zhang}}, \ and\ \bibinfo
  {author} {\bibfnamefont {H.}~\bibnamefont {Yao}},\ }\href {\doibase
  10.1088/2058-9565/ac87cd} {\bibfield  {journal} {\bibinfo  {journal} {Quantum
  Science and Technology}\ }\textbf {\bibinfo {volume} {7}},\ \bibinfo {pages}
  {045023} (\bibinfo {year} {2022})}\BibitemShut {NoStop}%
\bibitem [{\citenamefont {Zhang}\ \emph {et~al.}(2021)\citenamefont {Zhang},
  \citenamefont {Hsieh}, \citenamefont {Zhang},\ and\ \citenamefont
  {Yao}}]{zhangNeuralPredictorBased2021}%
  \BibitemOpen
  \bibfield  {author} {\bibinfo {author} {\bibfnamefont {S.-X.}\ \bibnamefont
  {Zhang}}, \bibinfo {author} {\bibfnamefont {C.-Y.}\ \bibnamefont {Hsieh}},
  \bibinfo {author} {\bibfnamefont {S.}~\bibnamefont {Zhang}}, \ and\ \bibinfo
  {author} {\bibfnamefont {H.}~\bibnamefont {Yao}},\ }\href {\doibase
  10.1088/2632-2153/ac28dd} {\bibfield  {journal} {\bibinfo  {journal} {Machine
  Learning: Science and Technology}\ }\textbf {\bibinfo {volume} {2}},\
  \bibinfo {pages} {045027} (\bibinfo {year} {2021})},\ \Eprint
  {http://arxiv.org/abs/2103.06524} {arXiv:2103.06524 [quant-ph]} \BibitemShut
  {NoStop}%
\bibitem [{\citenamefont {Grimsley}\ \emph {et~al.}(2019)\citenamefont
  {Grimsley}, \citenamefont {Economou}, \citenamefont {Barnes},\ and\
  \citenamefont {Mayhall}}]{grimsleyAdaptiveVariationalAlgorithm2019}%
  \BibitemOpen
  \bibfield  {author} {\bibinfo {author} {\bibfnamefont {H.~R.}\ \bibnamefont
  {Grimsley}}, \bibinfo {author} {\bibfnamefont {S.~E.}\ \bibnamefont
  {Economou}}, \bibinfo {author} {\bibfnamefont {E.}~\bibnamefont {Barnes}}, \
  and\ \bibinfo {author} {\bibfnamefont {N.~J.}\ \bibnamefont {Mayhall}},\
  }\href {\doibase 10.1038/s41467-019-10988-2} {\bibfield  {journal} {\bibinfo
  {journal} {Nature Communications}\ }\textbf {\bibinfo {volume} {10}},\
  \bibinfo {pages} {3007} (\bibinfo {year} {2019})}\BibitemShut {NoStop}%
\bibitem [{\citenamefont {Tang}\ \emph {et~al.}(2021)\citenamefont {Tang},
  \citenamefont {Shkolnikov}, \citenamefont {Barron}, \citenamefont {Grimsley},
  \citenamefont {Mayhall}, \citenamefont {Barnes},\ and\ \citenamefont
  {Economou}}]{tangQubitADAPTVQEAdaptiveAlgorithm2021}%
  \BibitemOpen
  \bibfield  {author} {\bibinfo {author} {\bibfnamefont {H.~L.}\ \bibnamefont
  {Tang}}, \bibinfo {author} {\bibfnamefont {V.}~\bibnamefont {Shkolnikov}},
  \bibinfo {author} {\bibfnamefont {G.~S.}\ \bibnamefont {Barron}}, \bibinfo
  {author} {\bibfnamefont {H.~R.}\ \bibnamefont {Grimsley}}, \bibinfo {author}
  {\bibfnamefont {N.~J.}\ \bibnamefont {Mayhall}}, \bibinfo {author}
  {\bibfnamefont {E.}~\bibnamefont {Barnes}}, \ and\ \bibinfo {author}
  {\bibfnamefont {S.~E.}\ \bibnamefont {Economou}},\ }\href {\doibase
  10.1103/PRXQuantum.2.020310} {\bibfield  {journal} {\bibinfo  {journal} {PRX
  Quantum}\ }\textbf {\bibinfo {volume} {2}},\ \bibinfo {pages} {020310}
  (\bibinfo {year} {2021})}\BibitemShut {NoStop}%
\bibitem [{\citenamefont {Yordanov}\ \emph {et~al.}(2021)\citenamefont
  {Yordanov}, \citenamefont {Armaos}, \citenamefont {Barnes},\ and\
  \citenamefont
  {Arvidsson-Shukur}}]{yordanovQubitexcitationbasedAdaptiveVariational2021}%
  \BibitemOpen
  \bibfield  {author} {\bibinfo {author} {\bibfnamefont {Y.~S.}\ \bibnamefont
  {Yordanov}}, \bibinfo {author} {\bibfnamefont {V.}~\bibnamefont {Armaos}},
  \bibinfo {author} {\bibfnamefont {C.~H.~W.}\ \bibnamefont {Barnes}}, \ and\
  \bibinfo {author} {\bibfnamefont {D.~R.~M.}\ \bibnamefont
  {Arvidsson-Shukur}},\ }\href {\doibase 10.1038/s42005-021-00730-0} {\bibfield
   {journal} {\bibinfo  {journal} {Communications Physics}\ }\textbf {\bibinfo
  {volume} {4}},\ \bibinfo {pages} {228} (\bibinfo {year} {2021})},\ \Eprint
  {http://arxiv.org/abs/2011.10540} {arXiv:2011.10540 [quant-ph]} \BibitemShut
  {NoStop}%
\bibitem [{\citenamefont {Rattew}\ \emph {et~al.}(2020)\citenamefont {Rattew},
  \citenamefont {Hu}, \citenamefont {Pistoia}, \citenamefont {Chen},\ and\
  \citenamefont
  {Wood}}]{rattewDomainagnosticNoiseresistantHardwareefficient2020}%
  \BibitemOpen
  \bibfield  {author} {\bibinfo {author} {\bibfnamefont {A.~G.}\ \bibnamefont
  {Rattew}}, \bibinfo {author} {\bibfnamefont {S.}~\bibnamefont {Hu}}, \bibinfo
  {author} {\bibfnamefont {M.}~\bibnamefont {Pistoia}}, \bibinfo {author}
  {\bibfnamefont {R.}~\bibnamefont {Chen}}, \ and\ \bibinfo {author}
  {\bibfnamefont {S.}~\bibnamefont {Wood}},\ }\href
  {http://arxiv.org/abs/1910.09694} {\enquote {\bibinfo {title} {A
  {{Domain-agnostic}}, {{Noise-resistant}}, {{Hardware-efficient Evolutionary
  Variational Quantum Eigensolver}}},}\ } (\bibinfo {year} {2020}),\ \Eprint
  {http://arxiv.org/abs/1910.09694} {arXiv:1910.09694 [quant-ph]} \BibitemShut
  {NoStop}%
\bibitem [{\citenamefont {Zhu}\ \emph {et~al.}(2022)\citenamefont {Zhu},
  \citenamefont {Tang}, \citenamefont {Barron}, \citenamefont
  {Calderon-Vargas}, \citenamefont {Mayhall}, \citenamefont {Barnes},\ and\
  \citenamefont {Economou}}]{zhuAdaptiveQuantumApproximate2022a}%
  \BibitemOpen
  \bibfield  {author} {\bibinfo {author} {\bibfnamefont {L.}~\bibnamefont
  {Zhu}}, \bibinfo {author} {\bibfnamefont {H.~L.}\ \bibnamefont {Tang}},
  \bibinfo {author} {\bibfnamefont {G.~S.}\ \bibnamefont {Barron}}, \bibinfo
  {author} {\bibfnamefont {F.~A.}\ \bibnamefont {Calderon-Vargas}}, \bibinfo
  {author} {\bibfnamefont {N.~J.}\ \bibnamefont {Mayhall}}, \bibinfo {author}
  {\bibfnamefont {E.}~\bibnamefont {Barnes}}, \ and\ \bibinfo {author}
  {\bibfnamefont {S.~E.}\ \bibnamefont {Economou}},\ }\href {\doibase
  10.1103/PhysRevResearch.4.033029} {\bibfield  {journal} {\bibinfo  {journal}
  {Physical Review Research}\ }\textbf {\bibinfo {volume} {4}},\ \bibinfo
  {pages} {033029} (\bibinfo {year} {2022})}\BibitemShut {NoStop}%
\bibitem [{\citenamefont {Li}\ \emph {et~al.}(2020{\natexlab{a}})\citenamefont
  {Li}, \citenamefont {Fan}, \citenamefont {Coram}, \citenamefont {Riley},\
  and\ \citenamefont {Leichenauer}}]{liQuantumOptimizationNovel2020}%
  \BibitemOpen
  \bibfield  {author} {\bibinfo {author} {\bibfnamefont {L.}~\bibnamefont
  {Li}}, \bibinfo {author} {\bibfnamefont {M.}~\bibnamefont {Fan}}, \bibinfo
  {author} {\bibfnamefont {M.}~\bibnamefont {Coram}}, \bibinfo {author}
  {\bibfnamefont {P.}~\bibnamefont {Riley}}, \ and\ \bibinfo {author}
  {\bibfnamefont {S.}~\bibnamefont {Leichenauer}},\ }\href {\doibase
  10.1103/PhysRevResearch.2.023074} {\bibfield  {journal} {\bibinfo  {journal}
  {Physical Review Research}\ }\textbf {\bibinfo {volume} {2}},\ \bibinfo
  {pages} {023074} (\bibinfo {year} {2020}{\natexlab{a}})}\BibitemShut
  {NoStop}%
\bibitem [{\citenamefont {Ostaszewski}\ \emph {et~al.}(2021)\citenamefont
  {Ostaszewski}, \citenamefont {Grant},\ and\ \citenamefont
  {Benedetti}}]{ostaszewskiStructureOptimizationParameterized2021}%
  \BibitemOpen
  \bibfield  {author} {\bibinfo {author} {\bibfnamefont {M.}~\bibnamefont
  {Ostaszewski}}, \bibinfo {author} {\bibfnamefont {E.}~\bibnamefont {Grant}},
  \ and\ \bibinfo {author} {\bibfnamefont {M.}~\bibnamefont {Benedetti}},\
  }\href {\doibase 10.22331/q-2021-01-28-391} {\bibfield  {journal} {\bibinfo
  {journal} {Quantum}\ }\textbf {\bibinfo {volume} {5}},\ \bibinfo {pages}
  {391} (\bibinfo {year} {2021})}\BibitemShut {NoStop}%
\bibitem [{\citenamefont {Du}\ \emph {et~al.}(2022)\citenamefont {Du},
  \citenamefont {Huang}, \citenamefont {You}, \citenamefont {Hsieh},\ and\
  \citenamefont {Tao}}]{duQuantumCircuitArchitecture2022}%
  \BibitemOpen
  \bibfield  {author} {\bibinfo {author} {\bibfnamefont {Y.}~\bibnamefont
  {Du}}, \bibinfo {author} {\bibfnamefont {T.}~\bibnamefont {Huang}}, \bibinfo
  {author} {\bibfnamefont {S.}~\bibnamefont {You}}, \bibinfo {author}
  {\bibfnamefont {M.-H.}\ \bibnamefont {Hsieh}}, \ and\ \bibinfo {author}
  {\bibfnamefont {D.}~\bibnamefont {Tao}},\ }\href {\doibase
  10.1038/s41534-022-00570-y} {\bibfield  {journal} {\bibinfo  {journal} {npj
  Quantum Information}\ }\textbf {\bibinfo {volume} {8}},\ \bibinfo {pages} {1}
  (\bibinfo {year} {2022})}\BibitemShut {NoStop}%
\bibitem [{\citenamefont {Preskill}(2018)}]{Preskill2018quantumcomputingin}%
  \BibitemOpen
  \bibfield  {author} {\bibinfo {author} {\bibfnamefont {J.}~\bibnamefont
  {Preskill}},\ }\href {\doibase 10.22331/q-2018-08-06-79} {\bibfield
  {journal} {\bibinfo  {journal} {{Quantum}}\ }\textbf {\bibinfo {volume}
  {2}},\ \bibinfo {pages} {79} (\bibinfo {year} {2018})}\BibitemShut {NoStop}%
\bibitem [{\citenamefont {Sim}\ \emph {et~al.}(2019)\citenamefont {Sim},
  \citenamefont {Johnson},\ and\ \citenamefont
  {Aspuru-Guzik}}]{Sim_expressibility}%
  \BibitemOpen
  \bibfield  {author} {\bibinfo {author} {\bibfnamefont {S.}~\bibnamefont
  {Sim}}, \bibinfo {author} {\bibfnamefont {P.~D.}\ \bibnamefont {Johnson}}, \
  and\ \bibinfo {author} {\bibfnamefont {A.}~\bibnamefont {Aspuru-Guzik}},\
  }\href {\doibase https://doi.org/10.1002/qute.201900070} {\bibfield
  {journal} {\bibinfo  {journal} {Advanced Quantum Technologies}\ }\textbf
  {\bibinfo {volume} {2}},\ \bibinfo {pages} {1900070} (\bibinfo {year}
  {2019})}\BibitemShut {NoStop}%
\bibitem [{\citenamefont {Parrish}\ \emph {et~al.}(2019)\citenamefont
  {Parrish}, \citenamefont {Hohenstein}, \citenamefont {McMahon},\ and\
  \citenamefont {Mart\'{\i}nez}}]{PhysRevLett.122.230401}%
  \BibitemOpen
  \bibfield  {author} {\bibinfo {author} {\bibfnamefont {R.~M.}\ \bibnamefont
  {Parrish}}, \bibinfo {author} {\bibfnamefont {E.~G.}\ \bibnamefont
  {Hohenstein}}, \bibinfo {author} {\bibfnamefont {P.~L.}\ \bibnamefont
  {McMahon}}, \ and\ \bibinfo {author} {\bibfnamefont {T.~J.}\ \bibnamefont
  {Mart\'{\i}nez}},\ }\href {\doibase 10.1103/PhysRevLett.122.230401}
  {\bibfield  {journal} {\bibinfo  {journal} {Phys. Rev. Lett.}\ }\textbf
  {\bibinfo {volume} {122}},\ \bibinfo {pages} {230401} (\bibinfo {year}
  {2019})}\BibitemShut {NoStop}%
\bibitem [{\citenamefont {Smacchia}\ \emph {et~al.}()\citenamefont {Smacchia},
  \citenamefont {Amico}, \citenamefont {Facchi}, \citenamefont {Fazio},
  \citenamefont {Florio}, \citenamefont {Pascazio},\ and\ \citenamefont
  {Vedral}}]{smacchiaStatisticalMechanicsCluster2011a}%
  \BibitemOpen
  \bibfield  {author} {\bibinfo {author} {\bibfnamefont {P.}~\bibnamefont
  {Smacchia}}, \bibinfo {author} {\bibfnamefont {L.}~\bibnamefont {Amico}},
  \bibinfo {author} {\bibfnamefont {P.}~\bibnamefont {Facchi}}, \bibinfo
  {author} {\bibfnamefont {R.}~\bibnamefont {Fazio}}, \bibinfo {author}
  {\bibfnamefont {G.}~\bibnamefont {Florio}}, \bibinfo {author} {\bibfnamefont
  {S.}~\bibnamefont {Pascazio}}, \ and\ \bibinfo {author} {\bibfnamefont
  {V.}~\bibnamefont {Vedral}},\ }\href {\doibase 10.1103/PhysRevA.84.022304} {\
  \textbf {\bibinfo {volume} {84}},\ \bibinfo {pages} {022304}}\BibitemShut
  {NoStop}%
\bibitem [{\citenamefont {Haldane}()}]{haldaneNonlinearFieldTheory1983}%
  \BibitemOpen
  \bibfield  {author} {\bibinfo {author} {\bibfnamefont {F.~D.~M.}\
  \bibnamefont {Haldane}},\ }\href {\doibase 10.1103/PhysRevLett.50.1153} {\
  \textbf {\bibinfo {volume} {50}},\ \bibinfo {pages} {1153}}\BibitemShut
  {NoStop}%
\bibitem [{\citenamefont {Nielsen}\ and\ \citenamefont
  {Chuang}(2011)}]{Nielsen:2011:QCQ:1972505}%
  \BibitemOpen
  \bibfield  {author} {\bibinfo {author} {\bibfnamefont {M.~A.}\ \bibnamefont
  {Nielsen}}\ and\ \bibinfo {author} {\bibfnamefont {I.~L.}\ \bibnamefont
  {Chuang}},\ }\href@noop {} {\emph {\bibinfo {title} {Quantum Computation and
  Quantum Information: 10th Anniversary Edition}}},\ \bibinfo {edition} {10th}\
  ed.\ (\bibinfo  {publisher} {Cambridge University Press},\ \bibinfo {address}
  {New York, NY, USA},\ \bibinfo {year} {2011})\BibitemShut {NoStop}%
\bibitem [{Note1()}]{Note1}%
  \BibitemOpen
  \bibinfo {note} {These are 0011,1100,1010,0101,0110,1001, equal in the sense
  that they have the same number of 1s and 0s}\BibitemShut {NoStop}%
\bibitem [{Note2()}]{Note2}%
  \BibitemOpen
  \bibinfo {note} {This is because of the geometric series: $N(\protect \frac
  {1}{2^0}+\protect \frac {1}{2^1} + \protect \cdots + \protect \frac
  {1}{2^{\log _2{N}-1}})+N(\protect \frac {1}{2^1}+\protect \frac {1}{2^2} +
  \protect \cdots + \protect \frac {1}{2^{\log _2{N}-1}})$. Where the first sum
  is for convolution unitaries and the second for pooling.}\BibitemShut {Stop}%
\bibitem [{\citenamefont {Larocca}\ \emph {et~al.}(2022)\citenamefont
  {Larocca}, \citenamefont {Sauvage}, \citenamefont {Sbahi}, \citenamefont
  {Verdon}, \citenamefont {Coles},\ and\ \citenamefont
  {Cerezo}}]{laroccaGroupInvariantQuantumMachine2022}%
  \BibitemOpen
  \bibfield  {author} {\bibinfo {author} {\bibfnamefont {M.}~\bibnamefont
  {Larocca}}, \bibinfo {author} {\bibfnamefont {F.}~\bibnamefont {Sauvage}},
  \bibinfo {author} {\bibfnamefont {F.~M.}\ \bibnamefont {Sbahi}}, \bibinfo
  {author} {\bibfnamefont {G.}~\bibnamefont {Verdon}}, \bibinfo {author}
  {\bibfnamefont {P.~J.}\ \bibnamefont {Coles}}, \ and\ \bibinfo {author}
  {\bibfnamefont {M.}~\bibnamefont {Cerezo}},\ }\href {\doibase
  10.1103/PRXQuantum.3.030341} {\bibfield  {journal} {\bibinfo  {journal} {PRX
  Quantum}\ }\textbf {\bibinfo {volume} {3}},\ \bibinfo {pages} {030341}
  (\bibinfo {year} {2022})},\ \Eprint {http://arxiv.org/abs/2205.02261}
  {arXiv:2205.02261 [quant-ph, stat]} \BibitemShut {NoStop}%
\bibitem [{\citenamefont {Meyer}\ \emph {et~al.}(2022)\citenamefont {Meyer},
  \citenamefont {Mularski}, \citenamefont {{Gil-Fuster}}, \citenamefont {Mele},
  \citenamefont {Arzani}, \citenamefont {Wilms},\ and\ \citenamefont
  {Eisert}}]{meyerExploitingSymmetryVariational2022}%
  \BibitemOpen
  \bibfield  {author} {\bibinfo {author} {\bibfnamefont {J.~J.}\ \bibnamefont
  {Meyer}}, \bibinfo {author} {\bibfnamefont {M.}~\bibnamefont {Mularski}},
  \bibinfo {author} {\bibfnamefont {E.}~\bibnamefont {{Gil-Fuster}}}, \bibinfo
  {author} {\bibfnamefont {A.~A.}\ \bibnamefont {Mele}}, \bibinfo {author}
  {\bibfnamefont {F.}~\bibnamefont {Arzani}}, \bibinfo {author} {\bibfnamefont
  {A.}~\bibnamefont {Wilms}}, \ and\ \bibinfo {author} {\bibfnamefont
  {J.}~\bibnamefont {Eisert}},\ }\href@noop {} {\enquote {\bibinfo {title}
  {Exploiting symmetry in variational quantum machine learning},}\ } (\bibinfo
  {year} {2022}),\ \Eprint {http://arxiv.org/abs/2205.06217} {arXiv:2205.06217
  [quant-ph]} \BibitemShut {NoStop}%
\bibitem [{\citenamefont {Sturm}(2014)}]{sturmSurveyEvaluationMusic2014}%
  \BibitemOpen
  \bibfield  {author} {\bibinfo {author} {\bibfnamefont {B.~L.}\ \bibnamefont
  {Sturm}},\ }in\ \href {\doibase 10.1007/978-3-319-12093-5_2} {\emph {\bibinfo
  {booktitle} {Adaptive Multimedia Retrieval: Semantics, Context, and
  Adaptation}}},\ \bibinfo {series} {Lecture Notes in Computer Science}, Vol.\
  \bibinfo {volume} {8382},\ \bibinfo {editor} {edited by\ \bibinfo {editor}
  {\bibfnamefont {A.}~\bibnamefont {Nürnberger}}, \bibinfo {editor}
  {\bibfnamefont {S.}~\bibnamefont {Stober}}, \bibinfo {editor} {\bibfnamefont
  {B.}~\bibnamefont {Larsen}}, \ and\ \bibinfo {editor} {\bibfnamefont
  {M.}~\bibnamefont {Detyniecki}}}\ (\bibinfo  {publisher} {Springer
  International Publishing},\ \bibinfo {year} {2014})\ pp.\ \bibinfo {pages}
  {29--66}\BibitemShut {NoStop}%
\bibitem [{\citenamefont {George}\ \emph {et~al.}(2001)\citenamefont {George},
  \citenamefont {Georg},\ and\ \citenamefont
  {Perry}}]{tzanetakis_essl_cook_2001}%
  \BibitemOpen
  \bibfield  {author} {\bibinfo {author} {\bibfnamefont {T.}~\bibnamefont
  {George}}, \bibinfo {author} {\bibfnamefont {E.}~\bibnamefont {Georg}}, \
  and\ \bibinfo {author} {\bibfnamefont {C.}~\bibnamefont {Perry}},\ }in\
  \href@noop {} {\emph {\bibinfo {booktitle} {Proceedings of the 2nd
  international symposium on music information retrieval, Indiana}}},\ Vol.\
  \bibinfo {volume} {144}\ (\bibinfo {year} {2001})\BibitemShut {NoStop}%
\bibitem [{\citenamefont {Havl{\'i}cek}\ \emph {et~al.}(2019)\citenamefont
  {Havl{\'i}cek}, \citenamefont {C{\'o}rcoles}, \citenamefont {Temme},
  \citenamefont {Harrow}, \citenamefont {Kandala}, \citenamefont {Chow},\ and\
  \citenamefont {Gambetta}}]{Havlicek2019}%
  \BibitemOpen
  \bibfield  {author} {\bibinfo {author} {\bibfnamefont {V.}~\bibnamefont
  {Havl{\'i}cek}}, \bibinfo {author} {\bibfnamefont {A.~D.}\ \bibnamefont
  {C{\'o}rcoles}}, \bibinfo {author} {\bibfnamefont {K.}~\bibnamefont {Temme}},
  \bibinfo {author} {\bibfnamefont {A.~W.}\ \bibnamefont {Harrow}}, \bibinfo
  {author} {\bibfnamefont {A.}~\bibnamefont {Kandala}}, \bibinfo {author}
  {\bibfnamefont {J.~M.}\ \bibnamefont {Chow}}, \ and\ \bibinfo {author}
  {\bibfnamefont {J.~M.}\ \bibnamefont {Gambetta}},\ }\href {\doibase
  10.1038/s41586-019-0980-2} {\bibfield  {journal} {\bibinfo  {journal}
  {Nature}\ }\textbf {\bibinfo {volume} {567}},\ \bibinfo {pages} {209}
  (\bibinfo {year} {2019})}\BibitemShut {NoStop}%
\bibitem [{\citenamefont {Schuld}\ and\ \citenamefont {Petruccione}(1
  01)}]{SupervisedQML}%
  \BibitemOpen
  \bibfield  {author} {\bibinfo {author} {\bibfnamefont {M.}~\bibnamefont
  {Schuld}}\ and\ \bibinfo {author} {\bibfnamefont {F.}~\bibnamefont
  {Petruccione}},\ }\href {\doibase 10.1007/978-3-030-83098-4} {\emph {\bibinfo
  {title} {Machine {{Learning}} with {{Quantum Computers}}}}}\ (\bibinfo
  {publisher} {Springer International Publishing},\ \bibinfo {year}
  {2021-01-01})\BibitemShut {NoStop}%
\bibitem [{\citenamefont {McFee}\ \emph {et~al.}(2021)\citenamefont {McFee},
  \citenamefont {Metsai}, \citenamefont {McVicar}, \citenamefont {Balke},
  \citenamefont {Thomé}, \citenamefont {Raffel}, \citenamefont {Zalkow},
  \citenamefont {Malek}, \citenamefont {Dana}, \citenamefont {Lee},
  \citenamefont {Nieto}, \citenamefont {Ellis}, \citenamefont {Mason},
  \citenamefont {Battenberg}, \citenamefont {Seyfarth}, \citenamefont
  {Yamamoto}, \citenamefont {viktorandreevichmorozov}, \citenamefont {Choi},
  \citenamefont {Moore}, \citenamefont {Bittner}, \citenamefont {Hidaka},
  \citenamefont {Wei}, \citenamefont {nullmightybofo}, \citenamefont
  {Hereñú}, \citenamefont {Stöter}, \citenamefont {Friesch}, \citenamefont
  {Weiss}, \citenamefont {Vollrath}, \citenamefont {Kim},\ and\ \citenamefont
  {Thassilo}}]{brian_mcfee_2021_4792298}%
  \BibitemOpen
  \bibfield  {author} {\bibinfo {author} {\bibfnamefont {B.}~\bibnamefont
  {McFee}}, \bibinfo {author} {\bibfnamefont {A.}~\bibnamefont {Metsai}},
  \bibinfo {author} {\bibfnamefont {M.}~\bibnamefont {McVicar}}, \bibinfo
  {author} {\bibfnamefont {S.}~\bibnamefont {Balke}}, \bibinfo {author}
  {\bibfnamefont {C.}~\bibnamefont {Thomé}}, \bibinfo {author} {\bibfnamefont
  {C.}~\bibnamefont {Raffel}}, \bibinfo {author} {\bibfnamefont
  {F.}~\bibnamefont {Zalkow}}, \bibinfo {author} {\bibfnamefont
  {A.}~\bibnamefont {Malek}}, \bibinfo {author} {\bibnamefont {Dana}}, \bibinfo
  {author} {\bibfnamefont {K.}~\bibnamefont {Lee}}, \bibinfo {author}
  {\bibfnamefont {O.}~\bibnamefont {Nieto}}, \bibinfo {author} {\bibfnamefont
  {D.}~\bibnamefont {Ellis}}, \bibinfo {author} {\bibfnamefont
  {J.}~\bibnamefont {Mason}}, \bibinfo {author} {\bibfnamefont
  {E.}~\bibnamefont {Battenberg}}, \bibinfo {author} {\bibfnamefont
  {S.}~\bibnamefont {Seyfarth}}, \bibinfo {author} {\bibfnamefont
  {R.}~\bibnamefont {Yamamoto}}, \bibinfo {author} {\bibnamefont
  {viktorandreevichmorozov}}, \bibinfo {author} {\bibfnamefont
  {K.}~\bibnamefont {Choi}}, \bibinfo {author} {\bibfnamefont {J.}~\bibnamefont
  {Moore}}, \bibinfo {author} {\bibfnamefont {R.}~\bibnamefont {Bittner}},
  \bibinfo {author} {\bibfnamefont {S.}~\bibnamefont {Hidaka}}, \bibinfo
  {author} {\bibfnamefont {Z.}~\bibnamefont {Wei}}, \bibinfo {author}
  {\bibnamefont {nullmightybofo}}, \bibinfo {author} {\bibfnamefont
  {D.}~\bibnamefont {Hereñú}}, \bibinfo {author} {\bibfnamefont {F.-R.}\
  \bibnamefont {Stöter}}, \bibinfo {author} {\bibfnamefont {P.}~\bibnamefont
  {Friesch}}, \bibinfo {author} {\bibfnamefont {A.}~\bibnamefont {Weiss}},
  \bibinfo {author} {\bibfnamefont {M.}~\bibnamefont {Vollrath}}, \bibinfo
  {author} {\bibfnamefont {T.}~\bibnamefont {Kim}}, \ and\ \bibinfo {author}
  {\bibnamefont {Thassilo}},\ }\href {\doibase 10.5281/zenodo.4792298}
  {\enquote {\bibinfo {title} {librosa/librosa: 0.8.1rc2},}\ } (\bibinfo {year}
  {2021})\BibitemShut {NoStop}%
\bibitem [{\citenamefont {Davis}\ and\ \citenamefont
  {Mermelstein}(1980)}]{davisComparisonParametricRepresentations1980}%
  \BibitemOpen
  \bibfield  {author} {\bibinfo {author} {\bibfnamefont {S.}~\bibnamefont
  {Davis}}\ and\ \bibinfo {author} {\bibfnamefont {P.}~\bibnamefont
  {Mermelstein}},\ }\href {\doibase 10.1109/TASSP.1980.1163420} {\bibfield
  {journal} {\bibinfo  {journal} {IEEE Transactions on Acoustics, Speech, and
  Signal Processing}\ }\textbf {\bibinfo {volume} {28}},\ \bibinfo {pages}
  {357} (\bibinfo {year} {1980})}\BibitemShut {NoStop}%
\bibitem [{\citenamefont {McClean}\ \emph {et~al.}(2016)\citenamefont
  {McClean}, \citenamefont {Romero}, \citenamefont {Babbush},\ and\
  \citenamefont {Aspuru-Guzik}}]{mccleanTheoryVariationalHybrid2016}%
  \BibitemOpen
  \bibfield  {author} {\bibinfo {author} {\bibfnamefont {J.~R.}\ \bibnamefont
  {McClean}}, \bibinfo {author} {\bibfnamefont {J.}~\bibnamefont {Romero}},
  \bibinfo {author} {\bibfnamefont {R.}~\bibnamefont {Babbush}}, \ and\
  \bibinfo {author} {\bibfnamefont {A.}~\bibnamefont {Aspuru-Guzik}},\ }\href
  {\doibase 10.1088/1367-2630/18/2/023023} {\bibfield  {journal} {\bibinfo
  {journal} {New Journal of Physics}\ }\textbf {\bibinfo {volume} {18}},\
  \bibinfo {pages} {023023} (\bibinfo {year} {2016})}\BibitemShut {NoStop}%
\bibitem [{\citenamefont {Orús}(2014)}]{orusPracticalIntroductionTensor2014}%
  \BibitemOpen
  \bibfield  {author} {\bibinfo {author} {\bibfnamefont {R.}~\bibnamefont
  {Orús}},\ }\href {\doibase 10.1016/j.aop.2014.06.013} {\bibfield  {journal}
  {\bibinfo  {journal} {Annals of Physics}\ }\textbf {\bibinfo {volume}
  {349}},\ \bibinfo {pages} {117} (\bibinfo {year} {2014})}\BibitemShut
  {NoStop}%
\bibitem [{\citenamefont {Huggins}\ \emph {et~al.}(2019)\citenamefont
  {Huggins}, \citenamefont {Patel}, \citenamefont {Whaley},\ and\ \citenamefont
  {Stoudenmire}}]{hugginsQuantumMachineLearning2019}%
  \BibitemOpen
  \bibfield  {author} {\bibinfo {author} {\bibfnamefont {W.}~\bibnamefont
  {Huggins}}, \bibinfo {author} {\bibfnamefont {P.}~\bibnamefont {Patel}},
  \bibinfo {author} {\bibfnamefont {K.~B.}\ \bibnamefont {Whaley}}, \ and\
  \bibinfo {author} {\bibfnamefont {E.~M.}\ \bibnamefont {Stoudenmire}},\
  }\href {\doibase 10.1088/2058-9565/aaea94} {\bibfield  {journal} {\bibinfo
  {journal} {Quantum Science and Technology}\ }\textbf {\bibinfo {volume}
  {4}},\ \bibinfo {pages} {024001} (\bibinfo {year} {2019})},\ \Eprint
  {http://arxiv.org/abs/1803.11537} {arXiv:1803.11537 [cond-mat,
  physics:quant-ph]} \BibitemShut {NoStop}%
\bibitem [{\citenamefont {Goodfellow}\ \emph {et~al.}(2016)\citenamefont
  {Goodfellow}, \citenamefont {Bengio},\ and\ \citenamefont
  {Courville}}]{Goodfellow-et-al-2016}%
  \BibitemOpen
  \bibfield  {author} {\bibinfo {author} {\bibfnamefont {I.}~\bibnamefont
  {Goodfellow}}, \bibinfo {author} {\bibfnamefont {Y.}~\bibnamefont {Bengio}},
  \ and\ \bibinfo {author} {\bibfnamefont {A.}~\bibnamefont {Courville}},\
  }\href@noop {} {\emph {\bibinfo {title} {Deep Learning}}}\ (\bibinfo
  {publisher} {MIT Press},\ \bibinfo {year} {2016})\ \bibinfo {note}
  {\url{http://www.deeplearningbook.org}}\BibitemShut {NoStop}%
\bibitem [{\citenamefont {Kerenidis}\ \emph {et~al.}(2019)\citenamefont
  {Kerenidis}, \citenamefont {Landman},\ and\ \citenamefont
  {Prakash}}]{kerenidis2019quantum}%
  \BibitemOpen
  \bibfield  {author} {\bibinfo {author} {\bibfnamefont {I.}~\bibnamefont
  {Kerenidis}}, \bibinfo {author} {\bibfnamefont {J.}~\bibnamefont {Landman}},
  \ and\ \bibinfo {author} {\bibfnamefont {A.}~\bibnamefont {Prakash}},\
  }\href@noop {} {\bibfield  {journal} {\bibinfo  {journal}
  {arXiv:1911.01117[quant-ph]}\ } (\bibinfo {year} {2019})},\ \Eprint
  {http://arxiv.org/abs/1911.01117} {arXiv:1911.01117 [quant-ph]} \BibitemShut
  {NoStop}%
\bibitem [{\citenamefont {Li}\ \emph {et~al.}(2020{\natexlab{b}})\citenamefont
  {Li}, \citenamefont {Zhou}, \citenamefont {Xu}, \citenamefont {Luo},\ and\
  \citenamefont {Hu}}]{li_quantum_2020}%
  \BibitemOpen
  \bibfield  {author} {\bibinfo {author} {\bibfnamefont {Y.}~\bibnamefont
  {Li}}, \bibinfo {author} {\bibfnamefont {R.-G.}\ \bibnamefont {Zhou}},
  \bibinfo {author} {\bibfnamefont {R.}~\bibnamefont {Xu}}, \bibinfo {author}
  {\bibfnamefont {J.}~\bibnamefont {Luo}}, \ and\ \bibinfo {author}
  {\bibfnamefont {W.}~\bibnamefont {Hu}},\ }\href {\doibase
  10.1088/2058-9565/ab9f93} {\bibfield  {journal} {\bibinfo  {journal} {Quantum
  Science and Technology}\ }\textbf {\bibinfo {volume} {5}},\ \bibinfo {pages}
  {044003} (\bibinfo {year} {2020}{\natexlab{b}})}\BibitemShut {NoStop}%
\bibitem [{\citenamefont {Henderson}\ \emph {et~al.}(2020)\citenamefont
  {Henderson}, \citenamefont {Shakya}, \citenamefont {Pradhan},\ and\
  \citenamefont {Cook}}]{henderson_quanvolutional_2020}%
  \BibitemOpen
  \bibfield  {author} {\bibinfo {author} {\bibfnamefont {M.}~\bibnamefont
  {Henderson}}, \bibinfo {author} {\bibfnamefont {S.}~\bibnamefont {Shakya}},
  \bibinfo {author} {\bibfnamefont {S.}~\bibnamefont {Pradhan}}, \ and\
  \bibinfo {author} {\bibfnamefont {T.}~\bibnamefont {Cook}},\ }\href {\doibase
  10.1007/s42484-020-00012-y} {\bibfield  {journal} {\bibinfo  {journal}
  {Quantum Machine Intelligence}\ }\textbf {\bibinfo {volume} {2}},\ \bibinfo
  {pages} {2} (\bibinfo {year} {2020})}\BibitemShut {NoStop}%
\bibitem [{\citenamefont {Wei}\ \emph {et~al.}(2021)\citenamefont {Wei},
  \citenamefont {Chen}, \citenamefont {Zhou},\ and\ \citenamefont
  {Long}}]{wei_quantum_2021}%
  \BibitemOpen
  \bibfield  {author} {\bibinfo {author} {\bibfnamefont {S.}~\bibnamefont
  {Wei}}, \bibinfo {author} {\bibfnamefont {Y.}~\bibnamefont {Chen}}, \bibinfo
  {author} {\bibfnamefont {Z.}~\bibnamefont {Zhou}}, \ and\ \bibinfo {author}
  {\bibfnamefont {G.}~\bibnamefont {Long}},\ }\href
  {http://arxiv.org/abs/2104.06918} {\bibfield  {journal} {\bibinfo  {journal}
  {arXiv:2104.06918 [quant-ph]}\ } (\bibinfo {year} {2021})},\ \bibinfo {note}
  {arXiv: 2104.06918}\BibitemShut {NoStop}%
\end{thebibliography}%

\section*{Acknowledgements}

The development of the python package was funded by Unitary Fund (https://unitary.fund/). This research was supported by the Yonsei University Research Fund of 2022 (2022-22-0124), by the National Research Foundation of Korea (Grant Nos. 2019M3E4A1079666 and 2022M3E4A1074591), and by the KIST Institutional Program (2E31531-22-076). Support from the NICIS (National Integrated Cyber Infrastructure System) e-research grant QICSI7 and the South African Research Chair Initiative, Grant No. 64812 of the Department of Science and Innovation and the National Research Foundation of the Republic of South Africa is kindly acknowledged.

\section*{Author information}
\textbf{Contributions}
M.L. designed the theoretical framework, performed the numerical experiments, and implemented the software. I.S. and F.P. supervised the research. D.P. and C.B. conceived the experiment and provided insights on the reference architecture. All authors reviewed and discussed the analyses and results and contributed towards writing the manuscript.

\textbf{Corresponding author}
Correspondence to: Matt Lourens.

\section*{Data availability}
The dataset analysed during the current study is available on TensorFlow Datasets, https://www.tensorflow.org/datasets/catalog/gtzan. 

\section*{Code availability}
The theoretical framework discussed in this paper has been implemented as an open-source Python package, which is available on GitHub at https://github.com/matt-lourens/hierarqcal. This package was used to generate all the circuits in the paper.

\section*{Ethics declarations}

Carsten Blank is the Co-Founder of Data cybernetics and Francesco Petruccione the Chair of Scientific Board and Co-Founder of QUNOVA computing. The authors declare no other competing interests.

\clearpage
\appendix
\renewcommand\thefigure{\thesection.\arabic{figure}}
\onecolumngrid
\section{Circuit ansatz}
\label{appendix:circuits}
\setcounter{figure}{0}
\begin{figure}[h]
    \centering
    \begin{subfigure}[h]{0.32\linewidth}
        \centering
        \includegraphics[width=.45\linewidth]{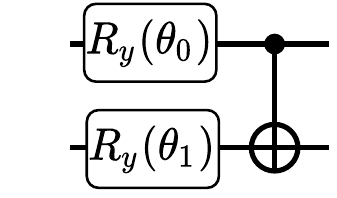}
        \caption{Two parameter ansatz.}
        \label{fig:u_ttn}
    \end{subfigure}
    \begin{subfigure}[h]{0.32\linewidth}
        \centering
        \includegraphics[width=.6\linewidth]{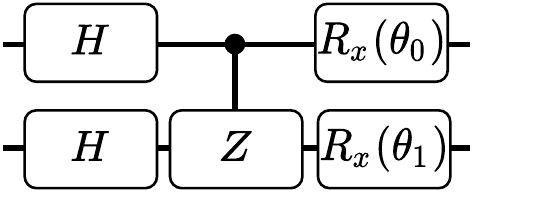}
        \caption{Two parameter ansatz.}
        \label{fig:u_9}
    \end{subfigure}
    \begin{subfigure}[h]{0.32\linewidth}
        \centering
        \includegraphics[width=.68\linewidth]{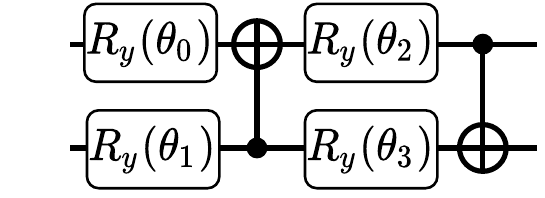}
        \caption{Four parameter ansatz.}
        \label{fig:u_15}
    \end{subfigure}
    
    \begin{subfigure}[h]{0.32\linewidth}
        \centering
        \includegraphics[width=.9\linewidth]{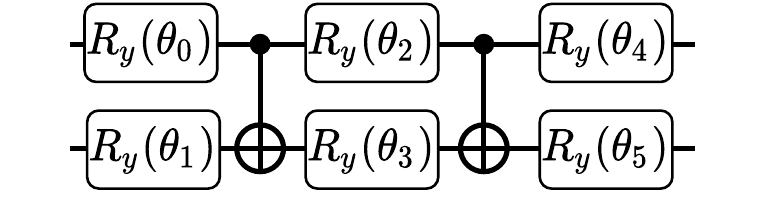}
        \caption{Six parameter ansatz.}
        \label{fig:u_so4}
    \end{subfigure}
    \begin{subfigure}[h]{0.32\linewidth}
        \centering
        \includegraphics[width=.96\linewidth]{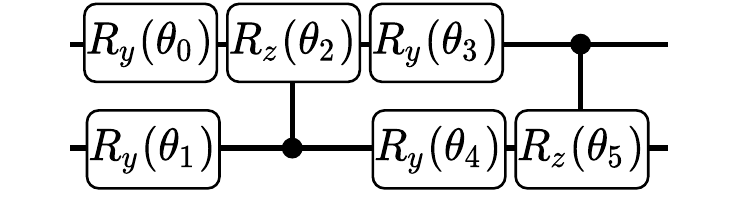}
        \caption{Six parameter ansatz.}
        \label{fig:u_13}
    \end{subfigure}
    \begin{subfigure}[h]{0.32\linewidth}
        \centering
        \includegraphics[width=\linewidth]{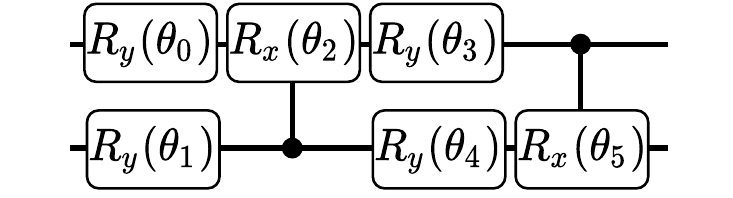}
        \caption{Six parameter ansatz.}
        \label{fig:u_14}
    \end{subfigure}
    \begin{subfigure}[h]{0.48\linewidth}
        \centering
        \includegraphics[width=.96\linewidth]{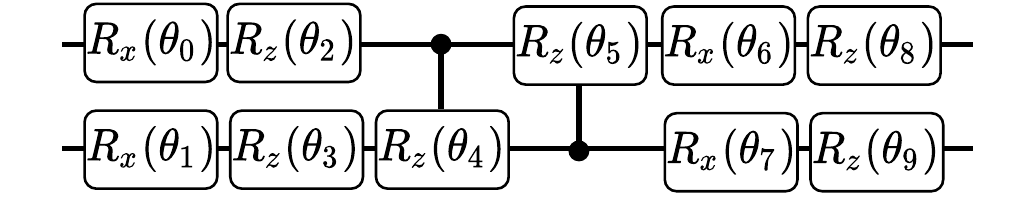}
        \caption{Ten parameter ansatz.}
        \label{fig:u5}
    \end{subfigure}
    \begin{subfigure}[h]{0.49\linewidth}
        \centering
        \includegraphics[width=\linewidth]{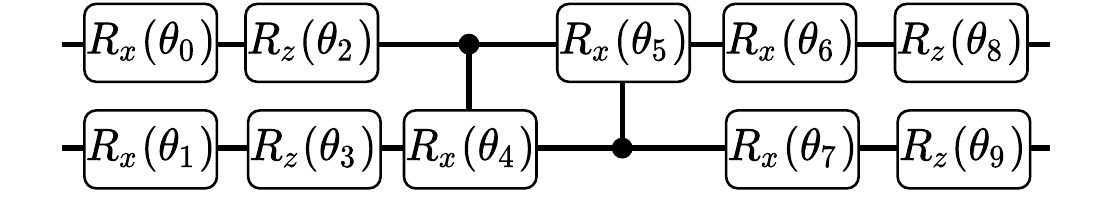}
        \caption{Ten parameter ansatz.}
        \label{fig:u_6}
    \end{subfigure}
    \caption{The different unitary ansatzes used for the convolution operation $U_m$ across all experiments. The same ansatzes were used in the benchmarks of \cite{hurQuantumConvolutionalNeural2022}. They are based on previous studies that explore the expressibility and entangling capability of parameterised circuits \cite{Sim_expressibility}, hierarchical quantum classifiers \cite{grant_hierarchical_2018}, and extensions to the VQE \cite{PhysRevLett.122.230401}.}
    \label{fig:ansatzes_appendix}
\end{figure}
\section{Discovered architecture through evolutionary search}
\begin{figure*}[h]
    \includegraphics[width=.9\linewidth]{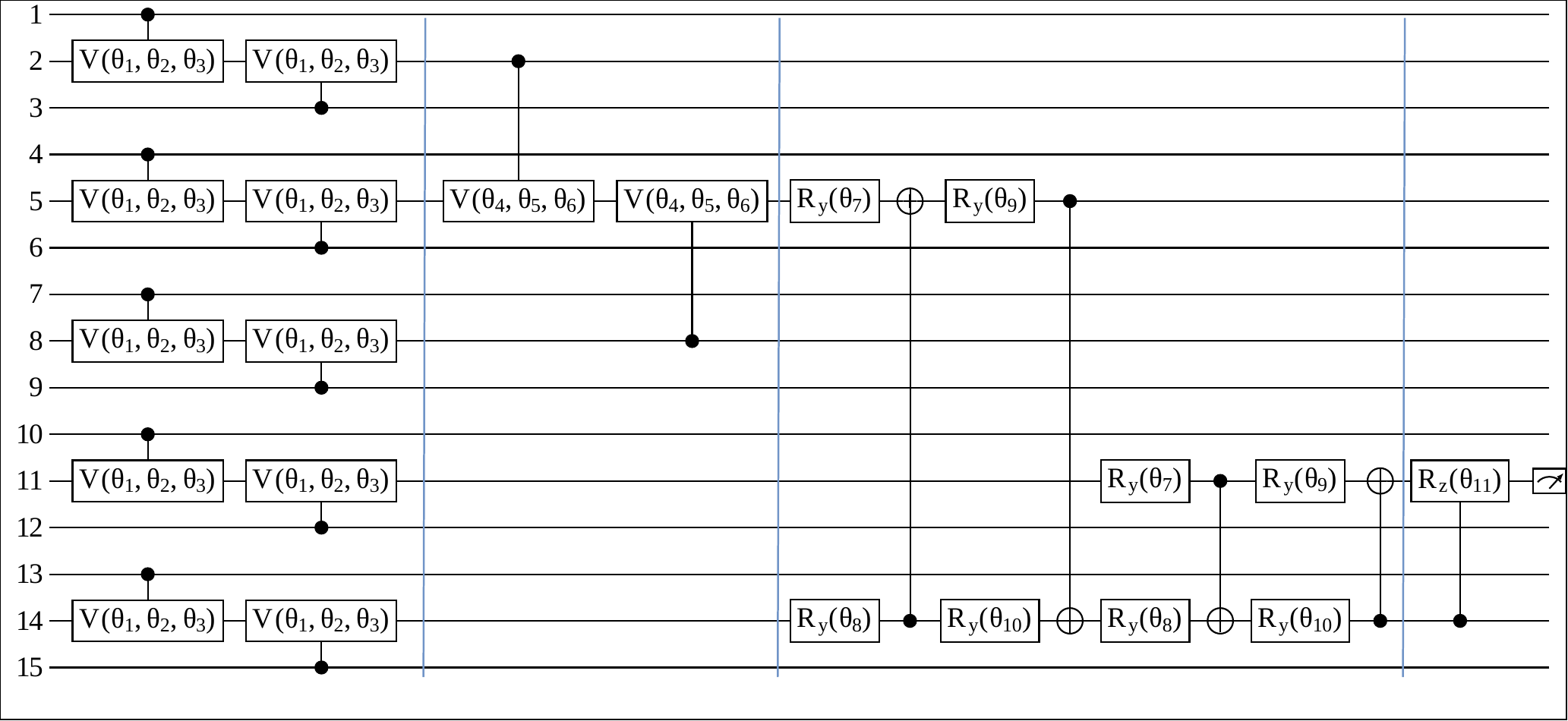}
    \caption{The architecture discovered through evolutionary search after 831 generations. It begins with a pooling layer that applies a "101" filter and a mapping of two-qubit Gell-Mann unitaries. Interestingly, the pooling in Figure \ref{fig:expressivity}c, which comes from \cite{cong_quantum_2019}, shares similarities with the one in our discovered architecture. However, differences exist, such as the control being in the computational basis and the weight sharing applied to all $V$'s, not just those moving in the same direction. In the subsequent layer, the same pooling layer with different weights is employed, further coarse-graining the circuit. This is followed by a convolution with a unitary from \ref{fig:ansatzes_appendix}(c) that operates on all four remaining qubits. The circuit concludes with a single pooling gate—a simple CRZ—before measurement.}
    \label{fig:evolve_circuit}
\end{figure*}
\section{Background}
\label{sec:background}

\subsection*{Quantum Machine Learning}
\label{ssec:learning}

The goal of classification is to utilise some data $X$ alongside a function $f_m$ (model) to accurately represent a discrete categorisation $y$, i.e. $f_m(X, \theta)=\hat{y}\approx y$. The data is utilised by iteratively changing the model $f_m$ parameters $\theta$  based on the disparity between the current representation $\hat{y}$ and the actual categorisation $y$, measured with a cost function $C(y,\hat{y})$. Minimising this function or learning is done until some specified critical point is reached, resulting in a set of parameters $\theta^*$ that can be used alongside the model $f_m$ and some new data $X^*$ to estimate their categories. This describes a supervised type of learning since some actual categorisations $y$ are known beforehand. It is achieved with the aid of computers and forms part of the broader field of machine learning, whose technology is ubiquitous in modern society. One interesting realisation of this procedure is with quantum computers, where the function $f_m$ is constructed as a variational quantum circuit that acts on a quantum state $\ket{\psi}$. Learning $\theta$ still uses classical (i.e., non-quantum) computation, resulting in a hybrid quantum-classical algorithm \cite{mccleanTheoryVariationalHybrid2016}. The hope is that the exploration of quantum circuits $f_m$ may lead to new approaches in machine learning that would be difficult to achieve classically \cite{schuldSupervisedQuantumMachine2021}. Variational quantum algorithms are also applicable in the NISQ era, making its exploration a step forward in developing future quantum technologies \cite{Preskill2018quantumcomputingin}. \newline 

The goal is then to find a quantum circuit (often called circuit ansatz) $f_m(X,\theta)$ that estimates $y$ accurately while keeping the number of required parameters $|\theta|$ as small as possible. A popular candidate for exploring and constructing different quantum circuits is tensor networks (TNs). This is because they may be used to represent quantum states and have had great theoretical and numerical success in the field of quantum many-body systems \cite{orusPracticalIntroductionTensor2014}. Within this context, tensors can be considered as multidimensional arrays, where the rank of a tensor indicates the array's dimension. For example, scalars, vectors and matrices correspond to rank-$0$, rank-$1$ and rank-$2$ tensors, respectively. A tensor network is also a tensor but composed of other, typically lower-rank, tensors through contraction operations. Being able to describe high-rank tensors through low-rank tensors in a network is, in part, what makes TNs powerful (see \cite{orusPracticalIntroductionTensor2014} for a more rigorous explanation). Experiments applying the structure of successful TNs from quantum many-body systems to quantum circuit design for machine learning show promising results. These include structures such as matrix product states (MPS) \cite{hugginsQuantumMachineLearning2019}, tensor tree networks (TTN) \cite{hugginsQuantumMachineLearning2019, grant_hierarchical_2018} and the multiscale entanglement renormalisation ansatz (MERA) \cite{grant_hierarchical_2018, cong_quantum_2019}. Specifically, the MERA tensor network overlaps with CNNs in terms of architecture \cite{cong_quantum_2019, grant_hierarchical_2018} and with the combination of QEC give rise to the QCNN presented in \cite{cong_quantum_2019}.
\subsection*{Quantum Convolutional Neural Networks}
\label{ssec:model}
In the classical CNN setting, a convolution refers to an operation that produces some feature map by cross-correlating a kernel with a given input. The input is the previous layer, and having the same kernel applied to all of its values results in weights being shared to the following layer. Sharing of weights is an important characteristic of a CNN since it shapes feature maps to be translational equivariant representations of the previous layer \cite{Goodfellow-et-al-2016}. After the convolution operation, non-linearity is introduced through an activation function. This is typically followed by a pooling operation, which down-samples the feature map to introduce local translational invariance and reduce model complexity.\newline

While there have been various proposals for the quantum analogue of convolutional neural networks \cite{cong_quantum_2019, hurQuantumConvolutionalNeural2022, kerenidis2019quantum,li_quantum_2020, frankenExplorationsQuantumNeural2020, henderson_quanvolutional_2020, wei_quantum_2021}, our work focuses on the framework proposed by Cong et al. \cite{cong_quantum_2019} and the findings of Grant et al. \cite{grant_hierarchical_2018}. As with many of these proposals, the key components are weight sharing, sequential reduction of system size via pooling and translational invariance of convolutions. This way,, the QCNN (Figure \ref{fig:qcnn_pipeline}d) implements analogous convolution and pooling operations in a quantum circuit setting. These operations are applied on a circuit architectural level, where a convolution consists of unitary operations $U_i$ being applied to all available qubits in a given layer. It is applied to all available qubits in order to achieve a type of translational invariance, and being identical unitaries allows the sharing of their weights. This relates to a CNN applying a single kernel to all input neurons in a given layer. Weight sharing is an important characteristic of the QCNN, as it causes the magnitude of its cost function gradients to increase, which is desirable in the face of barren plateaus since it counteracts vanishing gradients \cite{pesah2020absence}. Pooling consists of measuring a portion of the available qubits within a layer and then applying unitary rotations $V_i$ to the remaining ones based on the measurement outcomes. This leads to a reduction in parameters to optimise, which introduces non-linearity to the model while also reducing its computational overhead \cite{cong_quantum_2019}. Convolution and pooling operations are repeated until the system size is sufficiently small. For binary classification, one of the qubits is measured, and the expectation value is defined as the probability of binary class membership.

The MERA structure in reverse satisfies the above description, giving rise to a valid QCNN architecture. The QCNN circuit architecture has been successfully applied to problems surrounding quantum phase recognition (QPR) and quantum error correction (QEC). The partial measurement performed during pooling relates to syndrome measurements in QEC, giving the intuition that a QCNN is viewed as some combination of MERA and QEC \cite{cong_quantum_2019}.
\section{Feature Summary}
\label{appendix:features}
\def\arraystretch{2}
\begin{table}[h]
    \begin{tabular}{|p{0.28\linewidth}|p{0.68\linewidth}|}
        \hline
        Chroma frequencies                  & Bins the different pitches of a song into the equal tempered 12-tone scale commonly used in western music.                                                                                                                                                                \\
        \hline
        Harmonic and percussive elements    & The harmonic and percussive components present in the signal separated via median filtering.                                                                                                                                                                              \\
        \hline
        Mel-frequency cepstral coefficients & Coefficients that make up the mel frequency cepstrum, where mel frequency is the transformation of a signal to the mel scale which characterizes human audio perception. It's commonly used for speech recognition, mobile phone identification and genre classification. \\
        \hline
        Root-mean-square                    & The square root of the average of the square of the signal, $\sqrt{\frac{1}{T_2-T_1}\int_{T_1}^{T_2} x(t)^2 \,dt}$ where $x(t)$ is the amplitude of the signal at time $t$.                                                                                               \\
        \hline
        Spectral centroid                   & The expected value of the frequency spectrum in a time interval. A type of centre of mass which can be used as an indication of tone brightness.                                                                                                                          \\
        \hline
        Spectral bandwidth                  & The standard deviation of the frequency spectrum around its centroid in a time interval.                                                                                                                                                                                  \\
        \hline
        Spectral rolloff                    & The frequency bin where the cumulative spectral energy is a specified percentage.                                                                                                                                                                                         \\
        \hline
        Tempo                               & The speed of the music, estimated in beats per minute.                                                                                                                                                                                                                    \\
        \hline
        Zero crossing rate                  & The rate at which the amplitude of the signal crosses zero or changes sign.                                                                                                                                                                                               \\
        \hline
        
    \end{tabular}
    \caption{The information gathered from audio signals to produce the tabular form data set for genre classification benchmarks.}
\end{table}
\end{document}